\documentclass[fleqn,usenatbib]{mnras}
\usepackage{amsmath}
\usepackage{amssymb}
\usepackage{listings}
\usepackage[T1]{fontenc}
\usepackage{ae,aecompl}
\usepackage{float}
\usepackage{caption}
\usepackage{placeins}
\usepackage{subcaption}
\usepackage[dvipsnames]{xcolor}
\usepackage{longtable}
\usepackage{lineno}
\usepackage[colorinlistoftodos]{todonotes}
\usepackage{tablefootnote}
\usepackage{multirow}

\usepackage[normalem]{ulem}

\title{A Unified Catalog-level Reanalysis of Stage-III Cosmic Shear Surveys}

\begin{document}
\hbadness=10002
\vbadness=10002

\author[Emily P. Longley et al.]{
Emily P. Longley,$^{1}$
Chihway Chang,$^{2,3}$
Christopher W. Walter,$^{1}$
Joe Zuntz,$^{4}$
\newauthor
Mustapha Ishak,$^{5}$
Rachel Mandelbaum,$^{6}$
Hironao Miyatake,$^{7,8}$ \newauthor
Andrina Nicola,$^{9}$
Eske M. Pedersen,$^{10}$
Maria E.\ S.\ Pereira\,$^{11}$ 
Judit Prat,$^{2,3}$ \newauthor
J. S\'{a}nchez,$^{12}$
Lucas F. Secco,$^{3}$
Tilman Tr{\"o}ster,$^{13}$ 
Michael Troxel,$^{1}$ 
Angus Wright,$^{14}$\newauthor
The LSST Dark Energy Science Collaboration
\\
$^{1}$Department of Physics, Duke University, Durham NC 27708, USA\\
$^{2}$Department of Astronomy and Astrophysics, University of Chicago, Chicago, IL 60637, USA \\
$^{3}$Kavli Institute for Cosmological Physics, University of Chicago, Chicago, IL 60637, USA \\
$^{4}$Institute for Astronomy, University of Edinburgh, Edinburgh EH9 3HJ, United Kingdom \\
$^{5}$Department of Physics, The University of Texas at Dallas, Richardson, TX 75080, USA \\
$^{6}$Department of Physics, Carnegie Mellon University, Pittsburgh, PA 15213, USA \\
$^{7}$Kobayashi-Maskawa Institute for the Origin of Particles and the Universe (KMI),
Nagoya University, Nagoya, 464-8602, Japan \\
$^{8}$Kavli Institute for the Physics and Mathematics of the Universe (WPI), The University of Tokyo Institutes for Advanced Study (UTIAS), \\ The University of Tokyo, Chiba 277-8583, Japan \\
$^{9}$Department of Astrophysical Sciences, Princeton University, Peyton Hall, Princeton, NJ
08544, USA \\
$^{10}$Department of Physics, Harvard University, 17 Oxford street, Cambridge, MA 02138, USA \\
$^{11}$Hamburger Sternwarte, Universit{\"a}t Hamburg, Gojenbergsweg 112, 21029 Hamburg, Germany \\
$^{12}$Space Telescope Science Institute, 3700 San Martin Drive, Baltimore, MD 21218, USA \\
$^{13}$Institute for Particle Physics and Astrophysics, ETH Zürich, Wolfgang-Pauli-Strasse 27, 8093 Zürich, Switzerland \\
$^{14}$Ruhr-University Bochum, Faculty of Physics and Astronomy, Astronomical Institute (AIRUB),\\ German Centre for Cosmological Lensing, 44780 Bochum, Germany 
}

\maketitle

\begin{abstract}
    Cosmological parameter constraints from recent galaxy imaging surveys are reaching $2-3\%$-level accuracy. 
    The upcoming Legacy Survey of Space and Time (LSST) of the Vera C. Rubin Observatory will produce sub-percent level measurements of cosmological parameters, providing a milestone test of the $\Lambda$CDM model. To supply guidance to the upcoming LSST analysis, it is important to understand thoroughly the results from different recent galaxy imaging surveys and assess their consistencies. In this work we perform a unified catalog-level reanalysis of three cosmic shear datasets: the first year data from the Dark Energy Survey (DES-Y1), the 1,000 deg$^{2}$ dataset from the Kilo-Degree Survey (KiDS-1000), and the first year data from the Hyper Suprime-Cam Subaru Strategic Program (HSC-Y1).
    We utilize a pipeline developed and rigorously tested by the LSST Dark Energy Science Collaboration to perform the reanalysis and assess the robustness of the results to analysis choices.  
    We find the $S_{8}$ constraint to be robust to two different small-scale modeling approaches, and varying choices of cosmological priors. Our unified analysis allows the consistency of the surveys to be rigorously tested and we find the three surveys to be statistically consistent.  Due to the partially overlapping footprint, we model the cross-covariance between KiDS-1000 and HSC-Y1 approximately when combining all three datasets, resulting in a 1.6-1.9\% constraint on $S_8$ given different assumptions on the cross-covariance.
\end{abstract}

\section{Introduction}

Weak (gravitational) lensing refers to the subtle coherent distortion of galaxy shapes due to the bending of light in the gravitational fields sourced by the mass distribution between the galaxy and the observer. Tomographic cosmic shear measures these distortions in bins of redshift and gives a picture of the Universe's growth of structure and expansion with time. Cosmic shear is particularly sensitive to the total matter density today, $\Omega_{\rm m}$, and the normalization of the matter fluctuations on $8h^{-1}$Mpc scales, $\sigma_{8}$.  It is usually quoted via the quantity $S_{8}\equiv \sigma_{8}\sqrt{\Omega_{\rm m}/0.3}$, which approximates the most constrained direction in this parameter space using the weak lensing power spectrum \citep{Jain1997}. 

Cosmology from weak lensing with galaxy imaging surveys has reached an exciting milestone. The ``Stage-III'' and ``Stage-IV'' classification was introduced in the Dark Energy Task Force report \citep{Albrecht2006} as different phases of dark energy experiments. Stage-III refers to the dark energy experiments that started in the 2010s and Stage-IV refers to those that start in the 2020s, taking full advantage of further technological advances. Current surveys have been publishing their intermediate results, showing that the cosmological constraint on the $S_{8}$ parameter from weak lensing is reaching a similar level as that from cosmic microwave background (CMB) experiments, and will reach this in Stage-IV. In particular, the three largest surveys today -- the Dark Energy Survey \citep[DES,][]{Flaugher2015}, the Kilo-Degree Survey \citep[KiDS,][]{deJong2015} and the Hyper Suprime-Cam Subaru Strategic Program \citep[HSC-SSP,][]{Aihara2018a} -- have all shown exquisite measurements and cosmological constraints from just using a subset of their final data, conducted through blinded analyses.  The results have shown an intriguing difference is seen in the parameter $S_{8}$ when comparing all the galaxy survey and the CMB experiments -- galaxy imaging surveys tend to prefer a lower $S_{8}$ value compared to CMB experiments \citep{Hikage2019,Hamana2018,Planck2018,Asgari2021,Amon2021,Secco2021,HamanaRevision}.\footnote{We note that in recent work by \cite{Galli2022} the experiments are less discrepant when adopting extension model $\Lambda\textrm{CDM}+b$ which includes modeling of baryon clumping.}  However, this difference is not at a level of significance to indicate a distinct tension.
Thus far, the commonly adopted $\Lambda$CDM cosmological model, in which the universe is dominated by cold dark matter (CDM), and the accelerated expansion is driven by a cosmological constant $\Lambda$, has been successful in explaining growth of structure.  However, if the tension between low and high redshift experiments is found to be statistically significant, it could be an indication of new exciting physics, and evidence of a breakdown of the model.  But, on the other hand, it could also be a sign of unknown systematic errors in either the galaxy or the CMB results. 

Given the complementarity of the DES, KiDS and HSC-SSP data characteristics, and the nearly independent analysis pipelines, comparing the three surveys under unified code and analysis choices is an extremely strong test for any systematic effects. The redundancy of data products and analysis approaches provides one of the most powerful ways to check the robustness of a cosmology result. For example, \citet{Joudaki2020} showed that the different approach used in KiDS and DES for photometric redshift calibration could shift either surveys' $S_{8}$ constraints by a small amount. In \citet{Asgari2020}, they explore mitigating baryon feedback uncertainty in a joint KiDS and DES analysis.  Also, \citet{Troxel2018} showed that correcting for the survey geometry in the covariance matrix could improve the goodness-of-fit for both KiDS and DES.  Finally, as shown in \citet{Doux2021} and \citet{Asgari2021}, the use of different cosmic shear estimators and scale cuts could explain the difference seen in two analyses from the same dataset in both HSC and KiDS. Most of these analyses focus on a few specific effects and datasets, making it rather challenging to state conclusively whether cosmic shear results are consistent with CMB constraints from {\it Planck} \citep{Planck2018}. 

The LSST Dark Energy Science Collaboration (DESC) has developed a program over the years re-analyzing Stage-III data with DESC pipelines. \citet{Chang2019} was the pilot work of reanalyzing published weak-lensing catalogs using a common pipeline and unified analysis choices. The authors reanalyzed the cosmic shear studies from four galaxy imaging surveys: the Deep Lens Survey \citep[DLS,][]{Jee2016}, the Canada-France-Hawaii Telescope Lensing Survey \citep[CFHTLenS,][]{Joudaki2017}, the DES Science Verification (SV)  data \citep[DES-SV,][]{Abbott2015}, and the 450 deg$^{2}$ KiDS data \citep[KiDS-450,][]{Hildebrandt2017}. First, they attempted to reproduce the published results, and, through that process discovered subtle issues in each pipeline. Some examples include being overly aggressive in terms of including small scales, an error in defining the angular bins, and an outdated choice of model for the nonlinear power spectrum. They next studied the impact on the cosmological constraints when unifying three specific analysis choices: angular scale cuts, model parameters/priors, and the covariance model. They showed that once these analysis choices were unified, the cosmological constraints appeared much less consistent than the published results and the relative constraining power appear to change significantly too. This work was a sobering reminder of how sensitive cosmological constraints are to the various analysis decisions we make, even after the galaxy catalogs are generated (which is on its own extremely challenging tasks).  It also highlights the importance of transparent, independent cross-checks amongst different experiments, in order to identify issues in the pipelines and analysis choices of previous results.  Additionally, unifying analysis choices and priors allows for a correct comparison between the surveys' results and allows us to quantify their agreement, which cannot be properly computed with the current disparate choices.  

Since \citet{Chang2019}, a longer-term pipeline framework has been established in DESC for a range of analyses using the software package \textsc{Ceci},\footnote{\url{https://github.com/LSSTDESC/ceci}} which is built on the modular workflow engine \textsc{Parsl}\footnote{\url{https://parsl-project.org}} \citep{Babuji2018}. In particular, \textsc{TXPipe}\footnote{\url{https://github.com/LSSTDESC/TXPipe}} (Prat, Zuntz et al. {\em in prep}) is the measurement pipeline designed for measuring various two-point functions for large-scale structure cosmology.  \textsc{TXPipe} is designed to be a single code-base that collects all functionalities under a common structure. The code is also modular, transparent, and well-tested. \textsc{TXPipe} is under constant active development and will be used as the main measurement code in this work. We expect the pipeline to remain the primary ``catalog to two-point measurement'' code up till and continuing into LSST operations. In the coming years, this type of reanalysis work will be extremely valuable in preparation for science with LSST.

The main goal of this paper is to reanalyze the cosmic shear studies carried out by three Stage-III surveys: the first year of DES data \citep[hereafter DES-Y1,][]{Troxel2017}, the first year of HSC-SSP data \citep[hereafter HSC-Y1,][]{Hamana2018,HamanaRevision}, and the 1000 deg$^2$ KiDS data \citep[hereafter KiDS-1000,][]{Asgari2021}. Ultimately we would like to understand if the three datasets are consistent and if so, what is the combined constraint. The shear catalogs used for all three studies are public and will serve as the input to our analysis.  We do not attempt to reevaluate the photometric redshift estimates and redshift distributions but encourage this for future work as a valuable test of DESC photo-z infrastructure as well as a chance to compare photometric redshift approaches across surveys.  Similarly we do not re-measure the shears of galaxies but encourage future analyzes to perform an image-level reanalysis to give valuable insight into shear measurement and calibration.  We adopt the real-space two-point correlation functions as our fiducial statistic, which at the time of this analysis is the statistic that is available to compare to public results for all three surveys.  An assessment of the consistency of the surveys with different statistics would be an interesting exercise for future work.  Following a similar logic as \citet{Chang2019}, we first attempt to reproduce the published results, both in terms of the data vector and the cosmological constraints. Next we test the sensitivity of the cosmological constraints to various analysis choices that are made by each survey. We then present results from the three datasets using a set of unified analysis choices and evaluate the consistency between them. The final step will be to combine the datasets that are consistent with each other and evaluate the consistency of the final result with CMB constraints from {\it Planck}.  This work represents the first use of \textsc{TXPipe} on real data, which poses an opportunity to stress-test the infrastructure.  We aim for this process to provide guidance for LSST and highlight any areas of the ``catalog to cosmology'' software and methodology that needs to be further developed.  

The paper is organized as follows. In Section~\ref{sec:data} we describe the basic characteristics of the three datasets that we use in this work, as well as the basic information of the three cosmic shear analyses we will be investigating. In Section~\ref{sec:analysis} we provide a brief overview of the different elements of the analysis: the theoretical model, the measurement pipeline, and the inference code. In Section~\ref{sec:fiducial} we compare our reanalysis with published results, including comparison of the data vector and the cosmological constraints. 
In Section~\ref{sec:priors} we study the sensitivity of each survey result to the choice of priors on the cosmological parameters. Similarly in Section~\ref{sec:scale_cuts} we study their sensitivity to the treatment of small-scale baryonic model uncertainties. In Section~\ref{sec:unify} we perform a unified analysis with all three datasets assuming the same analysis choices in terms of priors on the cosmological parameters, intrinsic alignment model and small-scale treatment. We evaluate the consistency between the different datasets and combine them to compare with CMB constraints from {\it Planck}. Finally, we conclude in Section~\ref{sec:conclusion}.

\section{Stage-III Cosmic Shear Analyses}
\label{sec:data}

\begin{figure*}
    \centering
    \begin{subfigure}[b]{0.45\textwidth}
        \centering
        \includegraphics[height=3in]{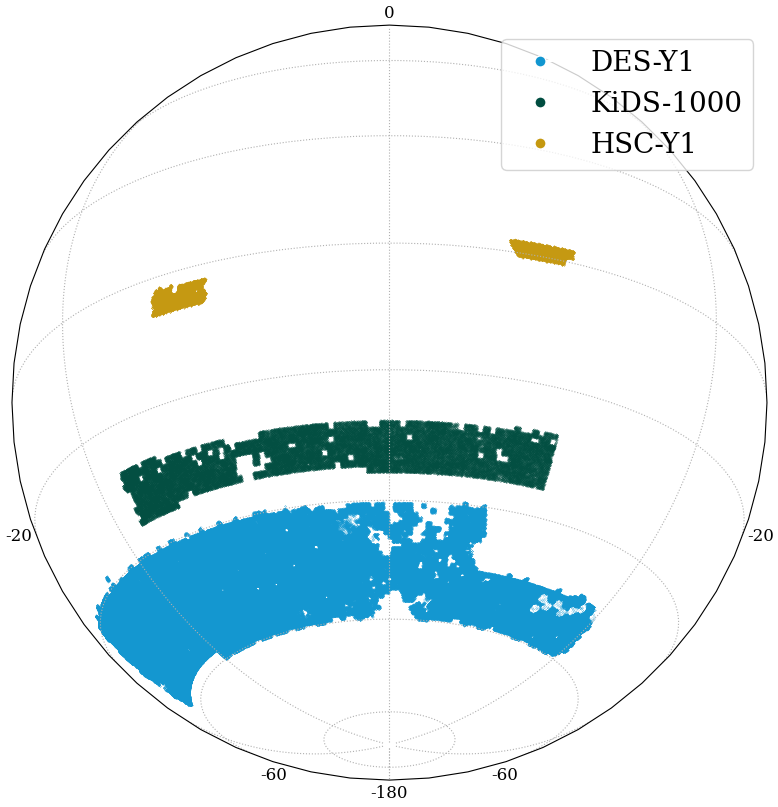}
    \end{subfigure}%
    ~ 
    \begin{subfigure}[b]{0.45\textwidth}
        \centering
        \includegraphics[height=3in]{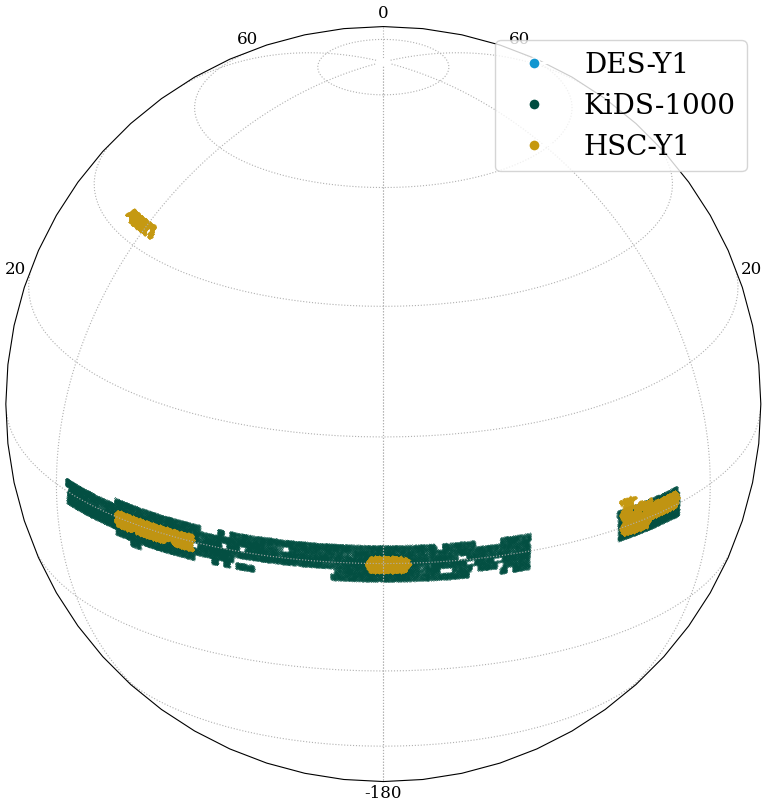}
    \end{subfigure}
    \caption{Sky coverage for the three datasets used in this work.  The left shows a projection of the southern sky and the right shows a projection of the northern sky.  The left shows the DES-Y1 field, KiDS-1000 S and HSC-Y1's XMM and VVDS fields. While there is no overlap between the DES-Y1 data and the other two, three of the HSC-Y1 fields overlap with the north part of KiDS-1000, (left panel), namely GAMA09H, GAMA15H and WIDE12H.  The upper non-overlapping HSC region is the HECTOMAP field.  In order to combine them, the cross-correlations between the datasets would need to be accounted for in the covariance between the overlapped fields. This is discussed further in Section~\ref{sec:unify}.}
\label{fig:footprints}
\end{figure*}

\begin{figure}
	\includegraphics[width=0.45\textwidth]{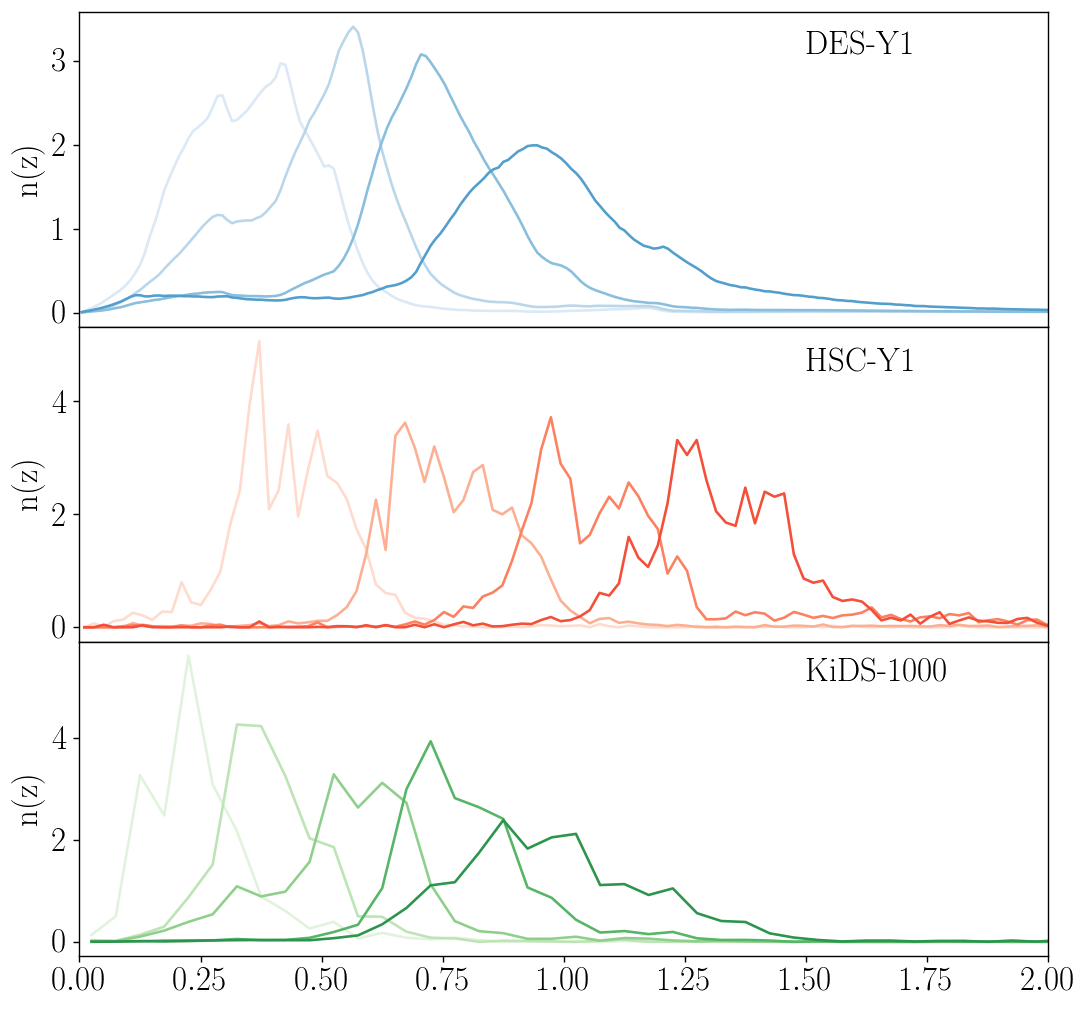}
	\caption{Estimates of the redshift distributions for each of the source samples in each of the three surveys. The $n(z)$'s are derived from photometric redshift estimation codes (\textsc{BPZ} for DES-Y1 and KiDS-1000 and \textsc{Ephor AB} for HSC-Y1). The calibration procedure varies for each survey and is described further in Section~\ref{sec:data}.  The redshift binning is optimized for each survey, and so the distributions for each bin will vary between the surveys based on the underlying physical properties of the samples.}
	\label{fig:nz}
\end{figure}

Since the first detection of cosmic shear \citep{Kaiser2000,Bacon2000,Wittman2000,VanWaerbeke2000}, the field has seen a rapid growth. In particular, a number of large surveys have delivered cosmic shear results with competitive cosmological constraints in the past few years \citep{Heymans2013,Becker2015,Jee2016,Hildebrandt2017,Joudaki2017,Troxel2017,Hikage2019,Hamana2018,Asgari2021,Amon2021,Secco2021}, while recent and future surveys will deliver data in much larger volumes and better quality.

We focus on the reanalysis of three cosmic shear studies using three independent Stage-III surveys: \citet{Troxel2017,Hamana2018,Asgari2021}. These correspond to the latest version of publicly available shear catalogs from DES, HSC-SSP and KiDS at the time when our analysis began. We show in Figures~\ref{fig:footprints}\footnote{This plot was made using the code \textsc{cartosky} (\url{https://github.com/kadrlica/cartosky})
} and \ref{fig:nz} the footprints and redshift distributions of the three datasets. Overall, we note that the footprints of the three datasets are largely non-overlapping except three of the six HSC-Y1 fields; GAMA09H, GAMA15H and WIDE12H with KiDS-1000 North.  The redshift distributions span a similar range in DES-Y1 and KiDS-1000, and the HSC-Y1 survey is about $0.5z$ deeper.  

We now describe in more detail below the characteristics of each dataset as well as an overview of the cosmic shear analyses carried out in \citet{Troxel2017,Hamana2018} and \citet{Asgari2021}. Some of the important information from each survey is summarized in Table~\ref{survey_summary}.

\subsection{DES-Y1}

The Dark Energy Survey (DES) first-year dataset consists of observations from DES between 2013 and 2014 as described in \citet{Drlica-Wagner2017}. The survey used 
the Dark Energy Camera \citep{Flaugher2015} on the Cerro Tololo Blanco 4m telescope with five filter bands ($grizY$). 

The DES-Y1 cosmology analysis from weak lensing was presented in \citet{Troxel2017}. The work shows consistent results between two independent galaxy shape catalogs whose details and validation were recorded in \cite{Zuntz2017}. In this work we use the catalog produced by the \textsc{Metacalibration} \citep{Huff2017,Sheldon2017} shear measurement pipeline run on the $riz$ bands, which contains 26.1 million galaxies and covers an area of 1321 deg$^{2}$. The $riz$ $5\sigma$ limiting magnitude and seeing of this dataset are $\sim24.0$ and $0.96\arcsec$.  The algorithm was run in the fast Bayesian fitting framework \textsc{ngmix} \citep{Sheldon2014} fitting each galaxy to a Gaussian (convolved with a PSF model) to determine an estimate of its ellipticity.  Images are then artificially sheared and the ellipticity is re-measured to calculate a response matrix that is used to calibrate the shear measurement. The self-calibration scheme additionally accounts for selection effects by performing a similar calculation with and without a given selection applied.

We use the same redshift distribution estimates as \citet{Troxel2017} which are described in detail in \cite{Hoyle2017} for tomographic binning and $n_{i} (z)$ distributions.  Galaxies are assigned a tomographic bin based on the mean of the photo-z posterior estimate derived using the Bayesian photometric redshift code \citep[BPZ,][]{Benitez2000} and redshift distributions are estimated from stacking MC draws from the photo-z posteriors.\footnote{In \citet{Malz2018} a more mathematically robust method for treating photo-z posteriors, is introduced to account for the assumptions of stacking photo-z posteriors in the methods used by the surveys considered in this paper.  We do not re-derive the n(z)'s in this work.}  All catalogs are publicly available\footnote{\url{https://des.ncsa.illinois.edu/releases/y1a1}}.

The analysis pipeline used in \citet{Troxel2017} is based on the software package \textsc{CosmoSIS} \citep{Zuntz2014}, which is the same cosmology inference framework we use in this paper.

\begin{table*}
  \caption{Summary of the basic characteristics of the three cosmic shear analyses that we aim to reanalyze in this paper. We quote the surveys' effective number density as $n_{\rm eff} = \frac{1}{A_{\rm eff}}\frac{(\sum{w_{i}})^{2}}{\sum{w_{i}^{2}}}$ for each tomographic bin.  Similarly, we list the standard deviation of the galaxy shapes, $\sigma_{e}$ for each bin, as the quadrature sum of the measurement error and the shape noise \citep[following the definition in][]{Heymans2012}.  We note there are alternative definitions of these quantities, namely \citet{Chang2012} which includes the purely shape noise component and \citet{Joachimi2020} which includes the contribution from the shape calibration.  Redshift ranges are listed for the tomographic bin edges. The last row lists the final data vector length after scale cuts.}
  \label{survey_summary}
  \small\centering
  \begin{tabular}{||l l l l||}
\hline 
      & \multicolumn{1}{c}{DES-Y1} & HSC-Y1 & KiDS-1000 \\ [0.5ex] 
 \hline
Reference  & \citet{Troxel2017} & \citet{Hamana2018} & \citet{Asgari2021} \\
 \hline
Area (deg$^{2}$)  & 1321.0 & 136.9 & 1006.0 \\
 \hline
Tomographic bins & [0.20, 0.43, 0.63, 0.90, 1.30] & [0.30, 0.60, 0.90, 1.20, 1.50] & [0.1, 0.3, 0.5, 0.7, 0.9, 1.2] \\
 \hline
$\sigma_{e}$ & [0.26, 0.29, 0.27, 0.29] & [0.27, 0.27, 0.29, 0.32]\footnote{5} & [0.27, 0.26, 0.27, 0.25, 0.27] \\
\hline
$n_{\rm eff}$ (arcmin$^{-2}$) & [1.52, 1.55, 1.63, 0.83] & [5.78, 5.85, 4.46, 2.61] & [0.62, 1.18, 1.85, 1.26, 1.31] \\
 \hline
$[\theta_{\rm min}, \theta_{\rm max}]$ (arcmin) & [2.5, 250] & [0.316, 316] & [0.5, 300] \\
 \hline 
Number of angular bins & 20 & 31 & 9 \\
 \hline
Data vector length & 227 & 170 & 225 \\
 \hline
 \multicolumn{4}{l}{\footnotesize $^{7}$ We note that we quote $\sigma_{e}$ for HSC using the shear definition $(|e|=(a-b)/(a+b))$, whereas they adopted the distortion} \\
 \multicolumn{4}{l}{in the original analysis $(|e|=(a^2-b^2)/(a^2+b^2))$.  In this definition the values are [0.41, 0.42, 0.43, 0.45].} \\ 
\label{table:survey_properties}
\end{tabular}
\end{table*}

\subsection{HSC-Y1}
The Hyper Suprime-Cam Subaru Strategic Program (HSC SSP) first-year dataset contains observations from between March 2014 and April 2016 taken in 6 disjoint regions (named XMM, GAMA09H, WIDE12H, GAMA15H, VVDS, and HECTOMAP) with the Subaru telescope \citep{Aihara2018b}. The data processing pipeline used by HSC \citep{Bosch2018} is a customized prototype version of 
the Rubin Observatory’s LSST Science Pipelines  
and thus provides valuable insight into the future data products of LSST.

The HSC-Y1 cosmic shear analyses was performed with both a power spectrum analysis described in \citet{Hikage2019} and real-space measurements described in \citet{Hamana2018} and \citet{HamanaRevision}. For both studies, the shape catalog measurements are validated through an extensive set of tests described in \citet{Mandelbaum2018b}. The catalog covers an area of $136.9$ deg$^{2}$ and contains $13.1$ million galaxies. These catalogs are made public through the S16A public data release described in  \citet{Aihara2018a}.\footnote{\hfill\url{https://hsc-release.mtk.nao.ac.jp/doc/index.php/s16a-shape-catalog-pdr2/}}

Shape measurements were estimated using the $i$-band images using the re-Gaussianization PSF correction method \citep{Hirata2004}. The mean $i$-band seeing and $5\sigma$ limiting magnitude are $0.58\arcsec$ and $\sim 26$.  For the weak-lensing catalog a magnitude cut of $i<24.5$ is applied. 
Our catalog corresponds to the same weak-lensing cuts that were applied in \citet{Hikage2019} and \citet{Hamana2018}.\footnote{We note that the number of galaxies listed in Table:~\ref{table:survey_properties} are different than those in \citet{Hamana2018} and \cite{Hikage2019}.  We have clarified with the authors that the numbers listed here are correct and the sample used in this paper is identical to that in their work.} Six photo-z codes were tested for the \cite{Mandelbaum2018a} catalog; their details are described in \cite{Tanaka2018}. Following their recommendation we perform tomographic binning using the \texttt{Ephor AB best} photo-z point estimate.  We use the same redshift distribution estimates as \citet{Hikage2019} and \citet{Hamana2018}, described in \citet{Tanaka2018}, which were estimated as a histogram of COSMOS $30-$band photo-z's, re-weighted  such that the distributions in a self-organizing map constructed from $grizy$ colors reflect that of the source sample.  The shapes are calibrated using the values and procedure described in \citet{Mandelbaum2018b}.   

\subsection{KiDS-1000}

The Kilo-Degree Survey (KiDS) 1000 deg$^{2}$ dataset represents the most recent data of the three used in this work and was made available in the fourth data release by the KiDS collaboration\footnote{\url{http://kids.strw.leidenuniv.nl/DR4/lensing.php}} \citep{Kuijken2019}. The images were taken on the OmegaCAM CCD Mosaic of the VLT Survey Telescope \citep[VST,][]{Kuijken2015,deJong2015,Kuijken2019}. The data release additionally includes nine-band near-infrared photometry ($ugriZYJHK_{s}$) based on imaging from the fully overlapping VISTA Kilo degree INfrared Galaxy Survey \citep[VIKING;][]{Edge2013}.

Cosmic shear constraints from KiDS-1000 are found in \citet{Asgari2021}. The weak lensing sample consists of two fields (North and South) that total an area of 1006 deg$^{2}$. The galaxy shape measurement process and validation tests are described in \citet{Giblin2021}. The self-calibrating $lens$fit software \citep{Miller2013,FenechConti2017} was used for shape measurements from the $r$-band images, which has a mean seeing of $0.7\arcsec$ and $5\sigma$ limiting magnitude of $\sim25.0$. Photo-z point estimates for tomographic binning are determined from the peak of the posterior produced by running BPZ on the nine-band photometry. The redshift distribution estimation is described in detail in \citet{Hildebrandt2021} and is based on a re-weighted spectroscopic reference catalog using a self-organizing map (SOM). The SOM process determines the ``gold'' sample where sources that do not lie in reference color space are cut, resulting in a final sample of 21.2 million galaxies. 

In \citet{Asgari2021} the cosmic shear constraints from three statistics are considered, namely two-point correlation functions (used in this work), complete orthogonal sets of E/B-integrals (COSEBIs), and band power spectra.  Each are linear transformations of the observed cosmic shear two-point correlation function.  They adopt the COSEBIs analysis as their fiducial results, since COSEBIs have the advantage of separating the E and B-modes of the cosmic shear signal.  COSEBIs can be calculated from two-point correlation functions computed in a large number of finely spaced angular bins.  In their work they show the COSEBIs, band powers and two-point correlation function results are consistent. There have also been two analyses of the KiDS-1000 data using the pseudo-Cl approach \citep{Loureiro2021,Troster2022}.

\section{Analysis}
\label{sec:analysis}

\subsection{Theoretical Background}
The two-point correlation function of galaxy shapes, $\xi_{\pm}(\theta)$ \citep{Bartelmann2001}, is a common statistic used to extract weak lensing information.  Assuming the flat-sky approximation, these two-point functions are connected to the lensing power spectrum $C(\ell)$ via
\begin{equation}
\xi^{ij}_{\pm}(\theta) = \frac{1}{2\pi}\int d\ell \, \ell J_{0/4}(\theta \ell) \, C^{ij}(\ell),
\label{eq:xipm}
\end{equation}
where $J_{0/4}$ are the 0th/4th-order Bessel functions of the first kind. The $i$ and $j$ indices specify the two samples of galaxies (or in the case of $i=j$, the same galaxy sample) from which the correlation function is calculated. Usually these samples are defined by a certain redshift selection. Under the Limber approximation \citep{Limber1953,Loverde2008} and in a spatially flat universe,\footnote{For a non-flat universe, one would replace $\chi$ by $f_{K}(\chi)$ in the following equations, where $K$ is the universe's curvature, $f_{K}(\chi)=K^{-1/2}\sin(K^{1/2}\chi)$ for $K>0$ and $f_{K}(\chi)=(-K)^{-1/2}\sinh((-K)^{1/2}\chi)$ for $K<0$.} the lensing power spectrum encodes cosmological information through 
\begin{equation}
C^{ij}(\ell) = \int_{0}^{\chi_{H}} d\chi \frac{q^{i}(\chi)q^{j}(\chi)}{\chi^2} P_{\rm NL}\left( \frac{\ell + 1/2}{\chi}, \chi \right),
\label{eq:Cl}
\end{equation}
where $\chi$ is the radial comoving distance, $\chi_{H}$ is the distance to the horizon, $P_\text{NL}$ is the nonlinear matter power spectrum, and $q(\chi)$ is the lensing efficiency defined via
\begin{equation}
q^{i}(\chi) = \frac{3}{2} \Omega_{\rm m} \left( \frac{H_{0}}{c}\right)^{2} \frac{\chi}{a(\chi)} \int_{\chi}^{\chi_{H}}d\chi ' n_{i}(\chi') \frac{dz}{d\chi'} \frac{\chi' - \chi}{\chi'},
\label{eq:lensing_efficiency}
\end{equation}
where $\Omega_{\rm m}$ is the matter density today, $H_{0}$ is the Hubble parameter today, $a$ is the scale factor, and $n_{i}(\chi)$ is the redshift distribution of the galaxy sample $i$. 

We note that although we focus on the $\xi_{\pm}$ statistics, several alternative cosmic shear statistics have been used in the literature aside from Equation~\eqref{eq:xipm}. These include the Fourier space lensing power spectrum \citep[i.e. Equation~\eqref{eq:Cl}, see][]{Nicola2021,Camacho2021} and the Complete Orthogonal Sets of E/B-Integrals \citep[COSEBIs, see][]{Schneider2010, Asgari2017}. A comprehensive analysis of the different two-point estimators can be found in \citet{Asgari2021}.
 
\subsection{Modeling Systematic Effects}

In addition to the background model, we account for a number of observational and astrophysical systematic effects described below. 

\subsubsection{Intrinsic Alignment}
When galaxies form near the same large-scale structure, their shapes can be coherently distorted by the gravitational field. Additionally, background galaxies lensed by large-scale structure can have correlated shapes with those that formed in the structure's gravitational field. Additionally, the evolutionary processes and galaxy mergers can induce intrinsic alignments between objects.

This Intrinsic Alignment effect (IA) causes their shapes to be correlated as a function of proximity, systematically affecting the weak lensing signal. A commonly used IA model in cosmic shear analyses is the nonlinear alignment model \citep[NLA,][]{Hirata2004,Bridle2007,Joachimi2011}.  The model assumes the IA power spectrum scales with the nonlinear matter power spectrum and includes the contribution from the correlations of intrinsic galaxy shapes that have evolved in the same local field, dubbed the ``intrinsic shear -- intrinsic shear (II) term'', and the ``gravitational shear -- intrinsic shear (GI) term'', which accounts for the correlation between galaxies that are lensed by a structure and galaxies that are intrinsically aligned with the same structure, see, e.g., the review \cite{TroxelIshak2014} and references therein.  In our adopted model, the IA power spectrum scales with the nonlinear matter power spectrum by $F[\chi (z)]$, given by
\begin{equation} 
F[\chi (z)] = A_{\rm IA}C_{1}\rho_{\rm crit} \frac{\Omega_{\rm m}}{D_{+}(z)} \left( \frac{1+z}{1+z_0} \right) ^{\eta},
\end{equation} 
where $A_{\rm IA}$ is the amplitude parameter, $C_{1} = 5 \times 10^{-14} h^{-2}M_{\odot}^{-1} \text{Mpc}^{3}$, $\rho_{\rm crit}$
is the critical density at $z = 0$, and $D_{+}(z)$ is the linear growth factor normalized to unity at $z = 0$ \citep{Bridle2007}. For HSC-Y1 and DES-Y1 the redshift term was allowed to vary. For DES-Y1, the pivot redshift of $z_0 = 0.62$ was adopted, which corresponds to the mean of the redshift sample, the typical choice for NLA.  HSC-Y1 also adopted $z_0 = 0.62$.  We check that changing this to the mean of their redshift distribution (deeper than DES-Y1) does not change the results. 
The KiDS-1000 analysis did not include the redshift-dependent power-law in their fiducial analysis. 

\subsubsection{Photo-z Systematics}
The weak lensing signal is most sensitive to the mean redshift of the galaxy sample, and thus a suitable approximation for our purposes is to adopt a model for the mean value as a nuisance parameter, as was done for previous surveys.  We model the uncertainty on this quantity with the nuisance parameter $\Delta z_{i}$ that shifts the measured $n(z)$ for each bin $i$ such that 
\begin{equation}
    n_{i}(z) = n_{\rm obs,i}\left(z - \Delta z_{i}\right).
\end{equation}
It is possible that accounting for uncertainty in the width of these distributions, at the precision level of e.g. LSST, could have an impact on results, but this is beyond the scope of this paper.  For the KiDS-1000 paper the shift parameters are correlated (due to the SOM formalism of their redshift calibration) so the priors for these quantities are correlated.  We keep this fiducial choice for the analysis for the KiDS-1000 priors, and keep the HSC-Y1 and DES-Y1 priors uncorrelated.\footnote{In theory, the $\Delta z_{i}$ redshift parameter shifts could be correlated between surveys if the redshift calibration was done with overlapping spectroscopic samples.  In our unified analysis we assume this is independent, but Garc\'ia-Garc\'ia ({\em in prep}) suggests that the effect on cosmological constraints is minimal.}
\subsubsection{Shear Calibration Uncertainty}
The calibration procedure for shear measurement has an associated uncertainty.  The residual impact on the observed shear $\gamma_{\textrm{observed}}$ within the weak regime is often modeled with both multiplicative $m_{i}$ and additive components $c_{i}$ \citep{Huterer2006,Heymans2006,Bridle2007}, that scale the true shear $\gamma_{\textrm{true}}$ as
\begin{equation}
    \gamma_{\textrm{observed}} = \gamma_{\textrm{true}} (1+m_{i}) + c_{i},
\end{equation}
where $m_i$ and $c_i$ are constant terms within the $i$th redshift bin.  Physically, systematics such as residual effects from PSF modeling can depend on galaxy properties and therefore not be constant within a tomographic bin, in which case additional terms can be used to model the uncertainty.  However, the dominant terms are often modelled in this convention. In our analysis, we adopt the approaches used by each survey to account for this uncertainty, as the individual surveys have extensively validated their shear calibration scheme and derived these models for their uncertainty:
\begin{itemize}
    \item In DES-Y1 each redshift bin $i$ is assigned a multiplicative shear calibration nuisance variable $m_{i}$ that is independent for each bin and marginalized over in the parameter estimation stage.  The Gaussian priors for each bin are identical but each variable is allowed to vary independently.  Additionally, in their systematics analysis, DES-Y1 found a nonzero residual additive shear component, which they correct for by subtracting the mean shear in each tomographic bin. 
    \item In HSC-Y1,  a single redshift-independent free parameter $m_0$ is used to account for the multiplicative shear calibration uncertainty for all redshift bins (so the multiplicative shear calibration is $100\%$ correlated between the redshift bins).  The parameter is marginalized over in the parameter estimation stage. In an additional chain, they account for the residual additive shear by marginalizing over an additive $c_{0}$ term that is consistent across bins, however this was not found to affect the final constraints.
    \item In KiDS-1000, the multiplicative shear calibration uncertainty is accounted for in the covariance matrix according to \citet{Asgari2021}. Additionally, they subtract the weighted mean ellipticity from each tomographic bin.  A constant term $c_0$ models the residual additive shear uncertainty (which is assumed to be redshift independent).
\end{itemize}
 
In the HSC analysis, two additional nuisance parameters were introduced to account for residual uncertainty in the PSF calibration, that can additively bias the observed shear  \citep[see][for details]{Mandelbaum2018a,Hamana2018}.  The PSF modeling and conditions are individual to each survey, so we maintain this choice and do not unify this modeling. 

\subsubsection{Small-Scale Modeling}
\label{small-scale}
At small-scales, the effects of baryons causes the matter power spectrum to become nonlinear.
Previous studies have generally adopted two approaches in order to mitigate the effects of potential biases that come from inaccuracies in modeling of the nonlinear power spectrum at small scales. The first approach (adopted by DES-Y1 and HSC-Y1) is to use a fixed nonlinear power spectrum (\textsc{HALOFIT}) and remove angular scales that can be contaminated at a certain level by baryon effects from the fit. The particular implementation from DES-Y1 was as follows: First, they calculated theoretical data vectors $\xi_{\pm}$ with baryon contamination by scaling the nonlinear power spectrum as
\begin{equation}
    P_{\rm NL}(k,z) = \frac{P_{\rm DM+Baryon}}{P_{\rm DM}}P_{\rm NL}(k,z),
\end{equation}
where the ``Dark Matter (DM) + Baryon'' power spectrum is from the OverWhelmingly Large Simulations project \citep[OWLS,][]{Schaye2010,vanDaalen2011} AGN simulation and the DM power spectrum is from the OWLS dark-matter only simulations.  The OWLS-AGN case is one of the more extreme in terms of baryonic effects of similar simulations, and thus helps to characterize a conservative cut.
Next, they compared these contaminated data vectors with the uncontaminated ones and determine the scale cut by requiring that the two not differ beyond $2\%$. 
The approach by HSC-Y1 was similar, albeit slightly more lenient in terms of contamination, 
adopting a cut at a roughly $5\%$ level based on the feedback model from \citet{Harnois2015}.
For the first approach, we implement a procedure that is modified from the DES-Y1 approach and used in the more recent DES cosmic shear analysis using the first 3 years of data  \citep[DES-Y3,][]{Krause2021,Amon2021,Secco2021}. The new method takes into account the relative systematic effect of the baryon modeling to the overall survey's uncertainty. We describe the method below.

Similar to that done in DES-Y1, we take a simulated fiducial data vector $\xi_{\pm}$ for each survey and use the $P_{\rm DM+Baryon}$ and $P_{\rm DM}$ ratios from OWLs to contaminate a data vector with baryonic effects. We then look at the $\Delta \chi^{2}$ between the two for each tomographic bin pair, $i;j$, 
\begin{equation}
    (\xi_{\rm \pm \mathrm{Baryon}}^{i,j}-\xi_{\pm \mathrm{Fiducial}}^{i,j})^{t} \textbf{C}_{i,j}^{-1} 
    (\xi_{\pm \mathrm{Baryon}}^{i,j}-\xi_{\pm \mathrm{Fiducial}}^{i,j})
    < \frac{\Delta \chi^{2}}{N}, 
\end{equation}
as a function of angular scales that are included, where $\textbf{C}_{i,j}^{-1}$ is the inverse covariance matrix for the tomographic bin.  Following the DES-Y3 approach, we exclude scales for each survey where the total $\Delta \chi^{2}>0.5$, split between $N$ tomographic bins. We then adopt this cut for an original and contaminated data vector, run an MCMC chain, and confirm  
the difference in the $S_8$ constraint 
is $<0.2 \sigma$. We do this for both the individual surveys, and the combined survey constraint.  The second step is mainly a sanity check -- in all cases it does not actually change the scale cuts. We note that this procedure is not exactly the same as DES-Y3 due to the difference in modelling choices, but provides a good framework to uniformly determine scale cuts across the three surveys.  The scale cuts should lead to approximately the same level of bias assuming a certain baryonic contamination.

The second approach \citep[used by KiDS,][]{Hildebrandt2017,Asgari2021} 
adopts the model implemented in 
\textsc{HMCode}, which is an augmented variant of the Halo Model with physically-motivated parameters fit to N-body simulations. \textsc{HMCode} parameterizes the effect of baryonic feedback with a halo bloating parameter $\eta_{0}$ and the amplitude of the halo mass-concentration relation $A_{\rm baryon}$ \citep{Joachimi2020}.  In their configuration the bloating parameter can be related to the amplitude parameter via
\begin{equation}
    \eta_{0} = 0.98 - 0.12A_{\rm baryon}.
\end{equation}
We adopt this convention when implementing the \textsc{HMCode} model. The uncertainty on this feedback effect on small scales is captured by marginalizing over this parameter with a top-hat prior. A scale cut is still imposed, in particular on $\xi_{-}$ which is more sensitive to nonlinear effects at small scales.   However, more data, in particular for $\xi_{+}$ is used.

\subsection{Covariance Matrix}
\label{sec:cov}
For this analysis we adopt the public covariance matrices used by the three surveys.  DES-Y1 adopt an analytical joint-probe covariance described in \citet{Krause2017}.  KiDS-1000 similarly adopt a joint-probe analytical covariance, as described in \citet{Joachimi2020}. 

HSC-Y1 used 2268 realizations of mock HSC catalogs to directly compute a numerical covariance that accounts for their particularly complicated survey geometry \citep[see][for details]{Shirasaki2019}.  Following the HSC-Y1 survey's decision, 
a calibration factor of ($N_{r}-N_{d}-2)/(N_{r}-1)$ is applied to the inverse covariance, where $N_{r}=2268$ is the number of mock realizations, and $N_{d}=170$.  This calibration accounts for biases that arise from covariances from numerical realizations \citep{Hartlap2007}.  Additionally, the simulations used to compute the covariance can have shears that are underestimated due to the finite thickness effect described in \citet{Shirasaki2019} and Appendix B of \citet{Tanaka2018}.  We account for this in the inverse covariance in the same method as the authors by applying a factor of $(1/0.92)^{2}$ (this factor was discussed with the author in private communication). 

\subsection{Modeling Pipeline: \textsc{CosmoSIS}}
\label{sec:cosmosis}
For our likelihood pipeline we use the cosmology likelihood code \textsc{CosmoSIS}\footnote{\url{https://bitbucket.org/joezuntz/cosmosis/wiki/Home}} package \citep{Zuntz2014}. The code utilizes the Boltzmann and background integrator \textsc{CAMB}\footnote{\url{http://camb.info}} to model the linear matter power spectrum \citep{Lewis2000,Howlett2012}.  The nonlinear matter power spectrum is modeled by either \textsc{HMCode} \citep{Mead2016}, implemented in \textsc{PyCamb} within \textsc{CosmoSIS}, or \textsc{HALOFIT} \citep{Smith2003,Bird2012,Takahashi2012}, implemented in \textsc{CosmoSIS}.  The projection to $C_{l}$ and $\xi_{\pm}$ space uses the Limber approximation.

\subsection{Measurement Pipeline: \textsc{TXPipe}}
\label{sec:txpipe}

Motivated by the enormous stream of data that will be available from the LSST, \textsc{TXPipe} was developed in DESC to produce the necessary data vectors for a cosmology analysis. This pipeline was validated with the mock galaxy catalogs described in Prat et al. ({\em in prep}) where they show that the input cosmology for the simulation can be reproduced. The pipeline is designed to run efficiently on a large set of data and is structured in different stages. 
Each of these stages is wrapped as a Python class, specifying input and output files required for each, and launching them using the \textsc{Ceci} library and executable which automatically interfaces them to workflow management frameworks.\footnote{\url{https://github.com/LSSTDESC/Ceci}} This work is the first to use \textsc{TXPipe} on real data and is an important milestone on DESC's readiness for LSST data. 

We input the public galaxy shape catalogs from each of the surveys.  We do not attempt to reproduce their photometric redshift estimates or redshift distributions, but future studies along these lines would be a useful exercise of the pipeline, and to give insight into different algorithms performances on characteristically different surveys. Additionally, we note that the previous surveys' cosmological analyses were conducted blindly, typically achieved by varying the shear values in the original catalogs by a random, unknown number.  For this work, we use the original unblinded catalogs, to directly compare our results to the public data vectors.

\begin{figure*}
\includegraphics[width=0.8\textwidth]{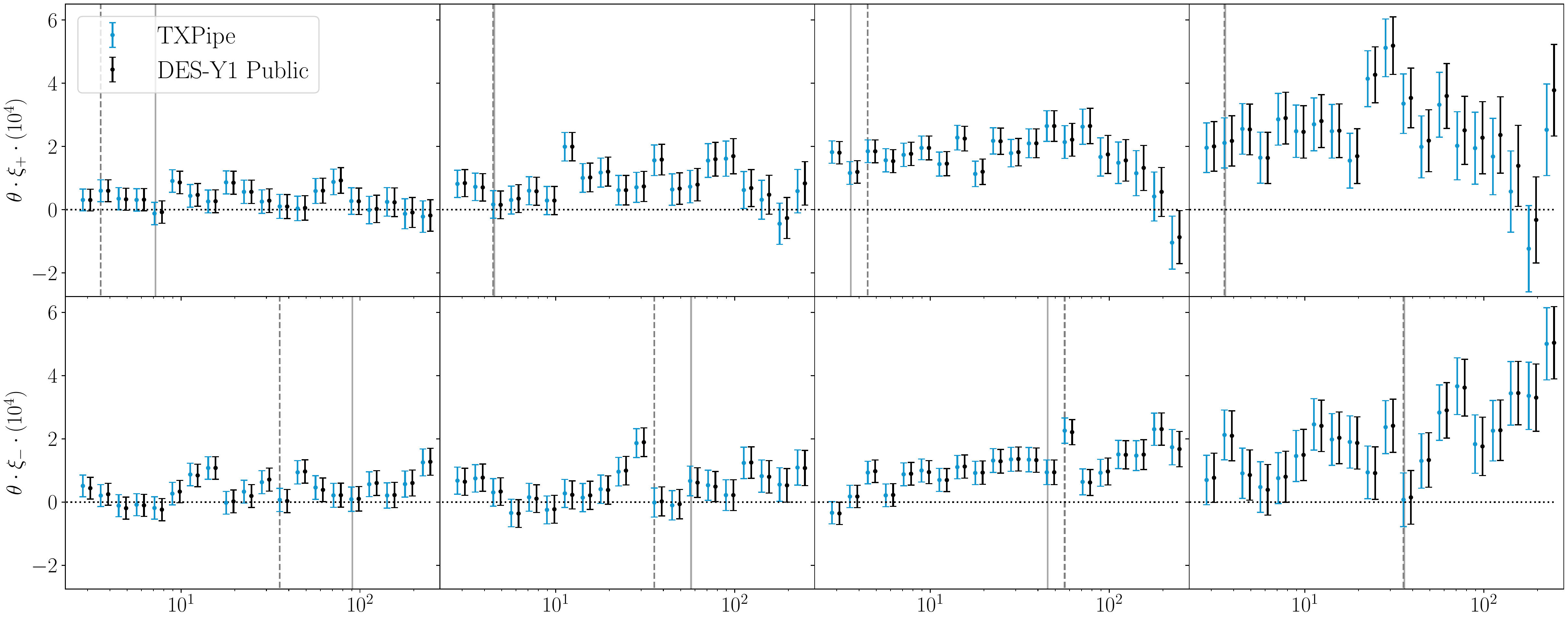}
\includegraphics[width=0.8\textwidth]{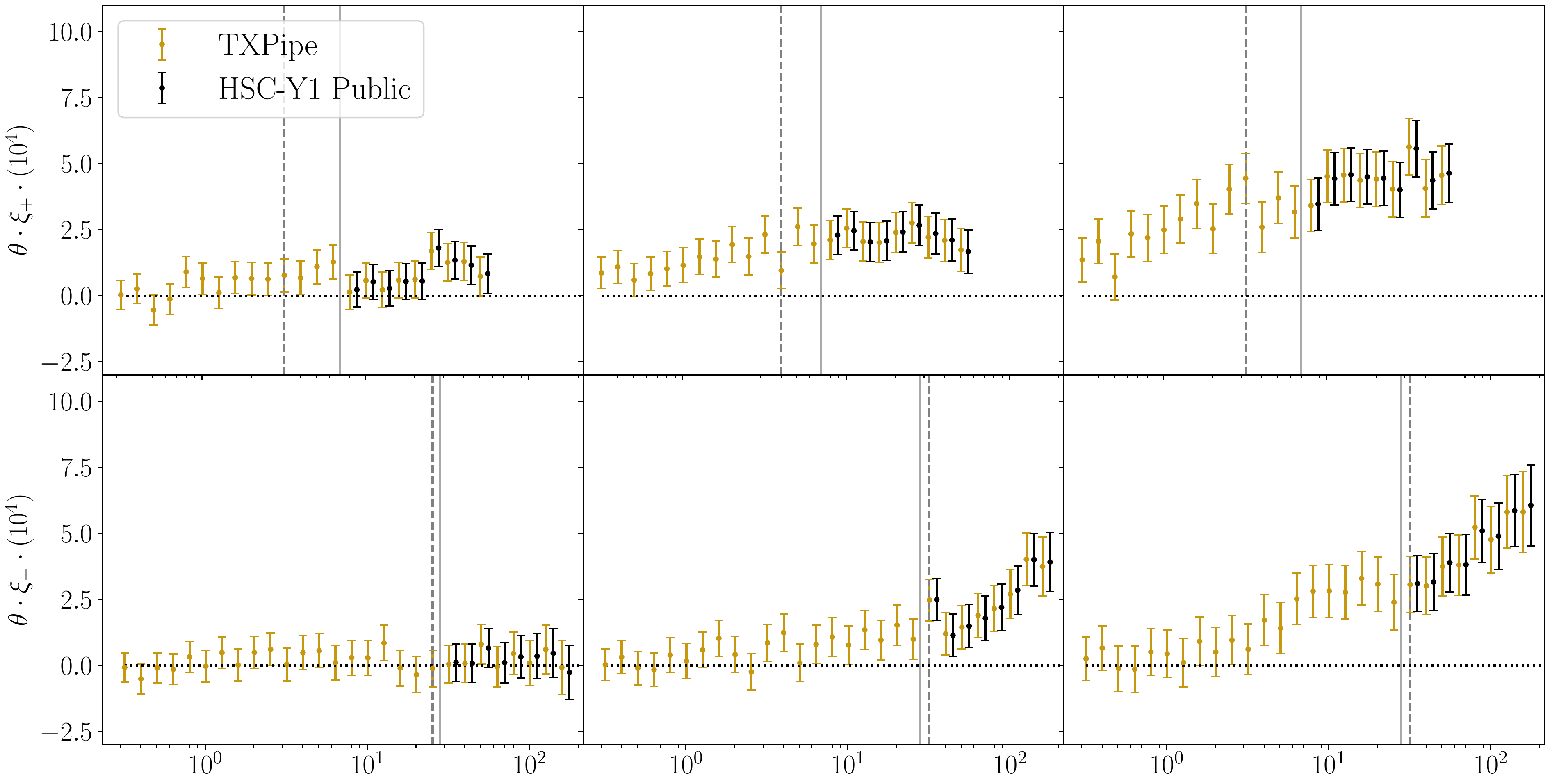} 
\includegraphics[width=0.95\textwidth]{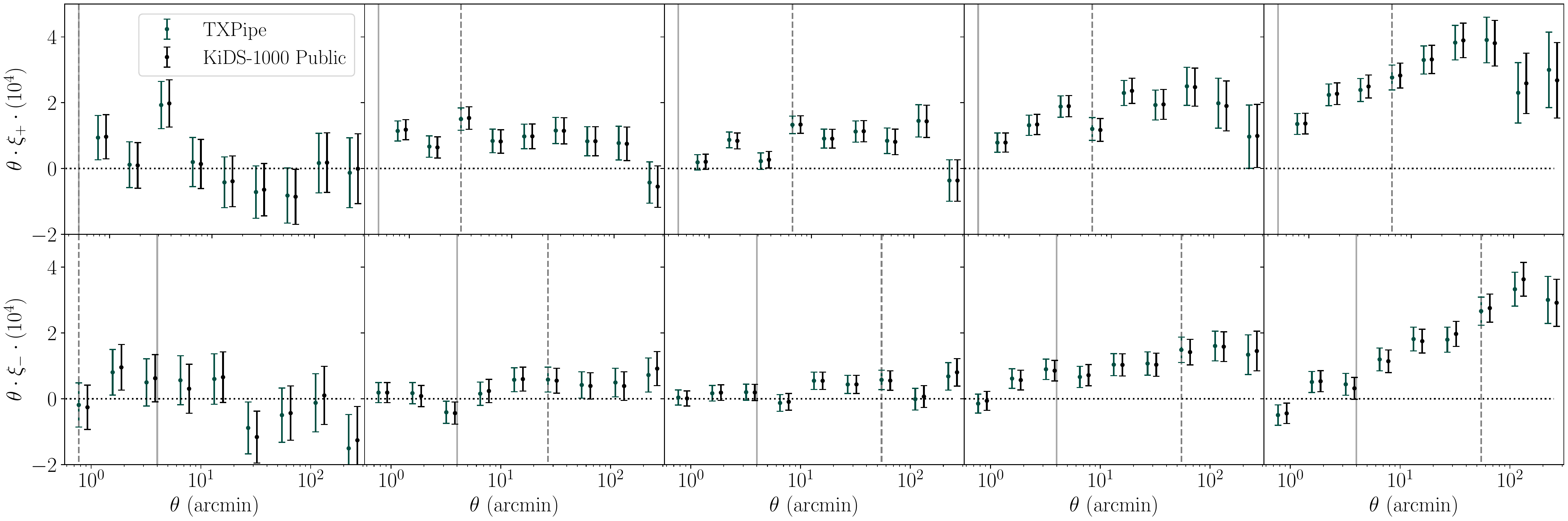} 
\caption{Data vectors and error bars as published (black) compared to those measured from \textsc{TXPipe} for DES-Y1 (top, blue), HSC-Y1 (middle, yellow) and KiDS-1000 (bottom, green). We only show the auto-correlations for clarity. The two sets of vertical lines indicate the scale cuts as published (solid) and as determined from our unified procedure (dashed) described in Section~\ref{sec:scale_cuts}.  Published results are offset horizontally from the \textsc{TXPipe} values for clarity.  To highlight only the differences in signal, the different angular choices for each survey are not shown in this plot but discussed further in Section~\ref{sec:fiducial}.  Errorbars are computed from the published covariance.}
\label{fig:datavec}
\end{figure*}

\subsection{Likelihood and Inference}
\label{sec:likelihood}

We use a Monte Carlo Bayesian likelihood analysis to sample the cosmological and nuisance parameter space.  We assume a Gaussian likelihood $L$, related to the parameters $\textbf{p}$, data $\textbf{D}$, inverse covariance matrix $\textbf{C}^{-1}$ and model $\textbf{M}$ by
\begin{equation}
    -2 \ln L(\textbf{D}|\textbf{p}) = (\textbf{D}-\textbf{M}(\textbf{p}))^{t}\mathcal{\textbf{C}}^{-1}(\textbf{D}-\textbf{M}(\textbf{p})).
\label{eq:likelihood}
\end{equation}

To calculate the cosmology constraints we use \textsc{Cosmo}SIS.  Cosmological constraint plots are shown using the software package C\textsc{hain}C\textsc{onsumer} \citep{Hinton2016}.\footnote{\url{https://samreay.github.io./ChainConsumer/}} We set the \texttt{kde} value in C\textsc{hain}C\textsc{onsumer} to 1.5. The outer and inner contours represent the $68\%$ and $95\%$ confidence levels respectively.  For our fiducial chains we use the sampler \textsc{Multinest} implemented in \textsc{Cosmo}SIS. But we refer the reader to \cite{Lemos2022} for a discussion on how this sampler can underestimate errors in the posterior by up to $\sim10\%$, as well as lead to inaccuracies in the computed evidence.  We do not expect this level of uncertainty to affect the results of this paper. 

\subsection{Consistency Metrics}
\label{sec:metric}

In order to assess both whether the derived model is a good fit to the data, and whether the posteriors for different datasets are consistent, we consider several metrics throughout this work. 

First, to determine whether a model is a good description of the data, we use the goodness-of-fit (GoF) definition following notation in Equation~\eqref{eq:likelihood}, 

\begin{equation}
    {\rm GoF} \equiv \chi^2/\nu = \frac{1}{\nu} (\textbf{D}-\textbf{M}(\textbf{p}))^{t}\mathcal{\textbf{C}}^{-1}(\textbf{D}-\textbf{M}(\textbf{p})),
\end{equation}
where $\nu$ is the degree of freedom, defined as the length of the data vector minus the effectively constrained parameters compared to the priors (calculated using the \textsc{tensiometer} package,\footnote{\url{https://github.com/mraveri/tensiometer}} which implements the definition in \citealt{Raveri2018}). With the $\chi^2$ one could also calculate the corresponding probability-to-exceed (p.t.e) to be 
\begin{equation}
    {\rm p.t.e} = 1 - {\rm CDF} (\chi^2, \nu).
\end{equation}
A low p.t.e. value implies that it is unlikely to have the data and model disagree to this level from pure statistical fluctuation. 

Second, to assess whether two posteriors are consistent under the same model, we consider several metrics. A very simple approach would be to look at the 1D distance in a single parameter. This is in general not a robust method to determine definitive tension between datasets given the high-dimensional nature of most of the cosmological problems.  However, in the case of cosmic shear, the primary parameter that carries the information is $S_{8}$ and part of $\Omega_{\rm m}$, \citep[see e.g. Fig 17 of][]{Secco2021}, and thus comparing the 1D posteriors of $S_8$ and $\Omega_{\rm m}$ gives us intuition into the effects of these changes on the results. We primarily use this as a measure of e.g. how the posterior moves with different analysis choices in Sections~\ref{sec:fiducial}, \ref{sec:priors} and \ref{sec:scale_cuts}. For parameter $\textbf{p}$ constrained by dataset 1 and 2, the distance (in number of $\sigma$'s) between the mean $p$ constraints is calculated via

\begin{equation} 
\Delta \textbf{p}_{12} \equiv \frac{\bar{\textbf{p}}_{1}-\bar{\textbf{p}}_{2}}{\sqrt{\sigma^2(\textbf{p}_{1})+\sigma^2(\textbf{p}_{2})}},
\end{equation}
where $\bar{\textbf{p}}_{1,2}$ and $\sigma(\textbf{p}_{1,2})$ are the mean and standard deviation of the marginalized 1D posterior of parameter $\textbf{p}$ constrained by dataset 1 and 2.

  In \cite{Lemos2021} several metrics are explored in the context of assessing tension metrics between DES-Y3 and Planck.  We adopt two metrics described in \cite{Lemos2021} in the context of assessing tension metrics between DES-Y3 and Planck discussed in this work.
\begin{itemize}
\item \textbf{Bayesian Suspiciousness} ($S$): We select this method because of its robustness to wide uninformative priors, as we adopt for this analysis.  The metric accounts for the mutual prior dependence for the combined surveys by subtracting the dependence on the prior volume. Specifically, 
\begin{equation}
    \log S = \log R - \log I,
\end{equation}
where $I$ is the information ratio, which quantifies the gain in information from the posterior to the prior, and $R$ is the commonly used Bayes ratio.  We use the python package \textsc{anesthetic} \footnote{\url{https://github.com/williamjameshandley/anesthetic}} to implement this metric \citep{anesthetic}.
\item  \textbf{MCMC parameter difference} This method is described in \cite{Raveri2021}, and is one of a number \citep[e.g.][]{Lin2017,Raveri2018,Lin2019} of metrics based on an estimate of the probability of a parameter difference between two experiments.  The metric is computed from a kernel density estimate (KDE) of the parameter difference posterior, and corresponds to the probability of a parameter difference. We compute this statistic over all shared parameters between the two posteriors.  It is particularly advantageous to adopt this metric in addition to the suspiciousness metric, as unlike the suspiciousness, it is robust to posteriors that are highly non-Gaussian.  For this metric we use the \textsc{tensiometer} code.

 Both metrics can be represented in terms of a p-value or $n-\sigma$.  We adopt the threshold $p>0.01$ as indicating that there is no sufficient evidence for disagreement of surveys. 
\end{itemize} 
We note that all the metrics adopted assume statistical independence between the datasets.

\begin{figure*}
\centering
\includegraphics[width=0.33\textwidth]{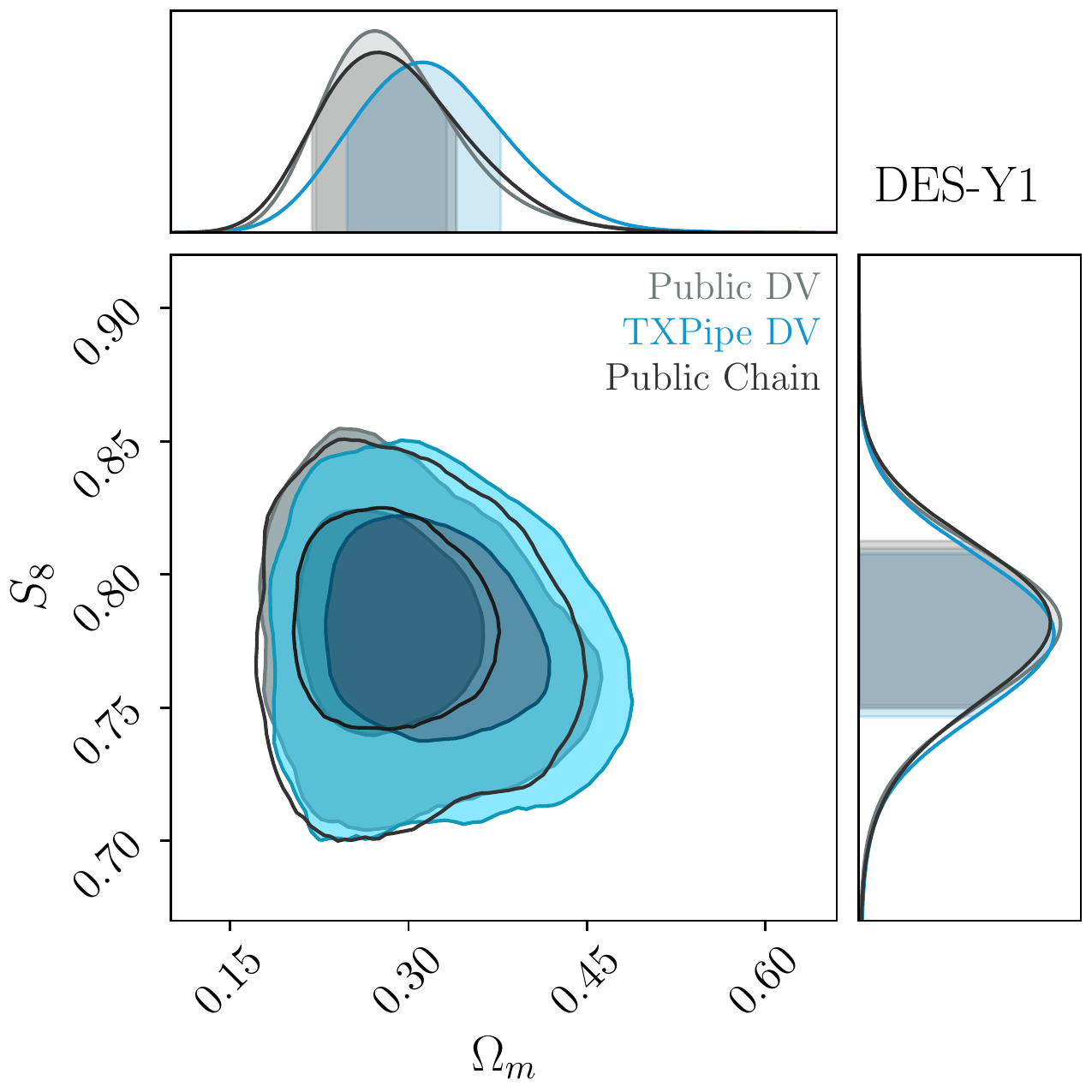}
\includegraphics[width=0.33\textwidth]{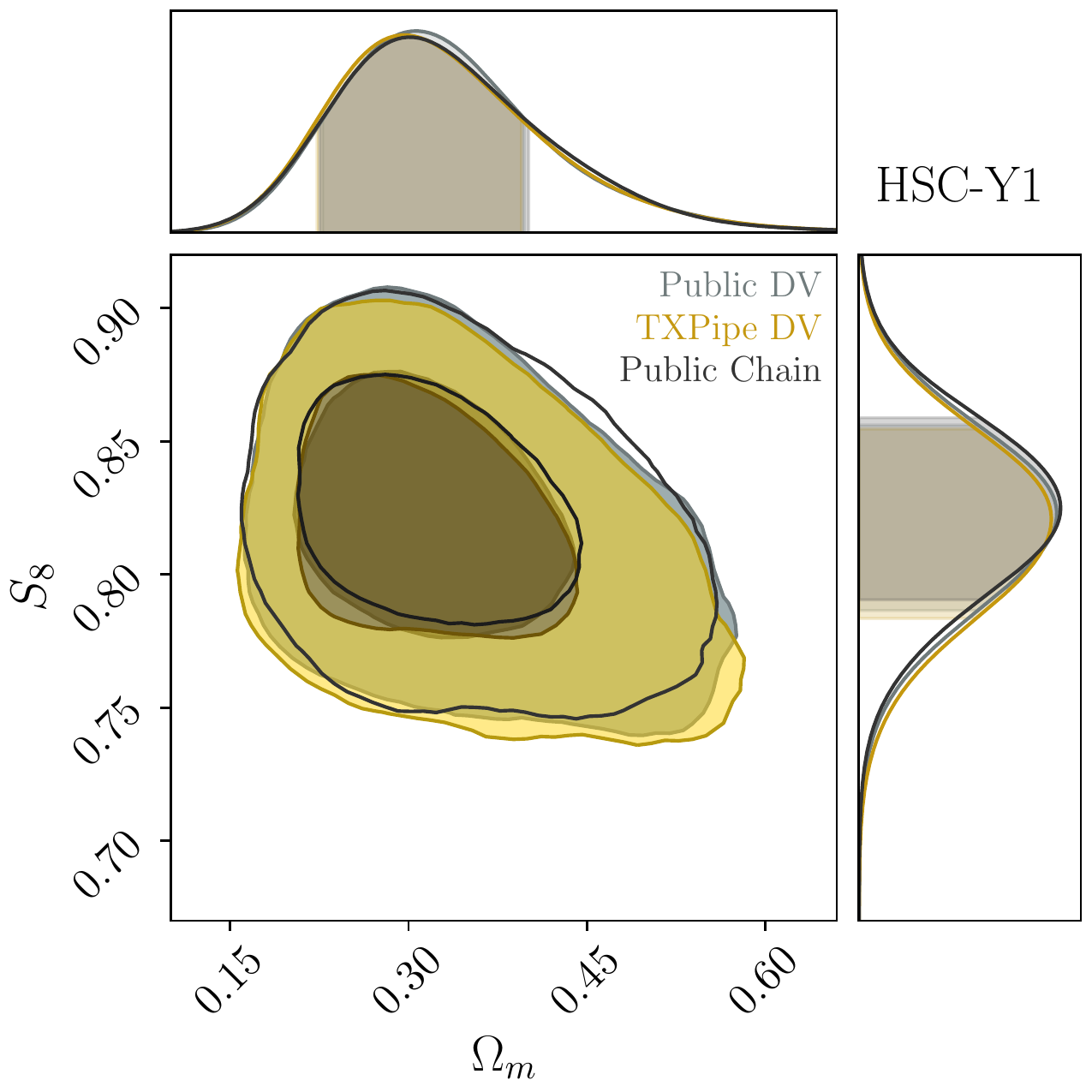}
\includegraphics[width=0.33\textwidth]{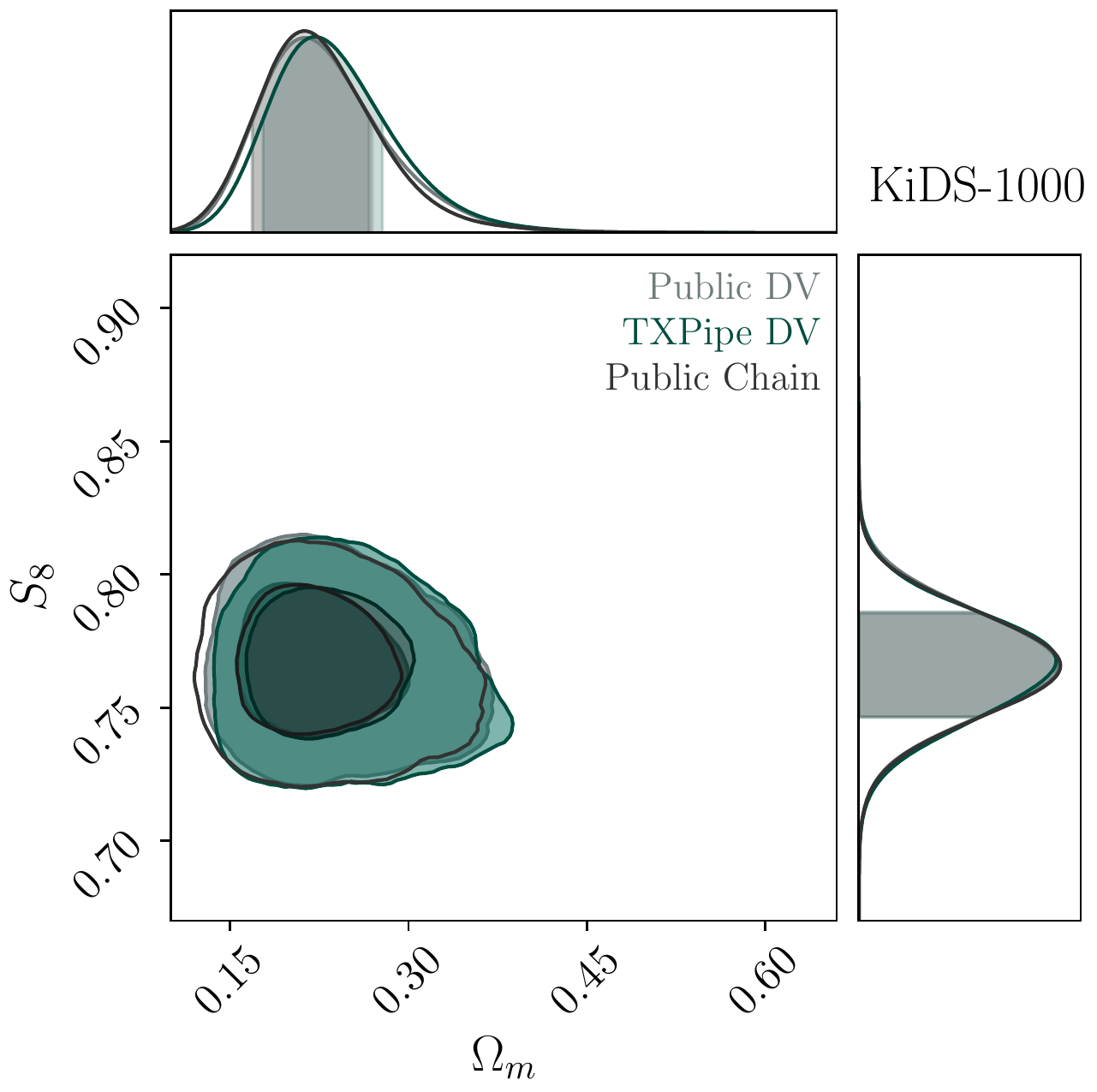}
\caption{Comparison of the $S_8$-$\Omega_{\rm m}$ constraints between 1) the published chain from each survey (grey), 2) our reanalysis of the published data vectors (colors), and 3) our reanalysis of the data vectors as measured by \textsc{TXPipe} (black). The results in all panels (and all subsequent contour plots in this paper) were computed using the published covariances.}
\label{published_comparison}
\end{figure*}

\section{Comparison with Published Analyses}
\label{sec:fiducial}

In this section we compare our reanalysis with those published in \citet{Troxel2017,Hamana2018,Asgari2021,HamanaRevision}. The intent here is to match as closely as possible the analysis process in the publications, but using independently developed software whenever possible. This provides a very strong test of the robustness of the published results. We first examine the measurement of the data vector in Section~\ref{data_vector} as well as the covariance from \textsc{TXPipe} compared to the published results. Next in Section~\ref{sec:txpipe_comparison_cosmo} we compare the posterior of our cosmological inference using \textsc{CosmoSIS} with the published chains. 

\subsection{Data Vector}
\label{data_vector}

In Figure~\ref{fig:datavec} we show the comparison between the two-point shear correlation function measurements from \textsc{TXPipe} compared to the published results. Overall we are able to reproduce the mean results for the two-point measurements. We find small differences which can be attributed to several identified features in the different pipelines.

First, we note that the two-point measurement codes used in the previous surveys have settings that can lead to small approximations. Namely this is the \texttt{binslop} parameter in the \textsc{TreeCorr} package (used by KiDS-1000 and DES-Y1) and the \texttt{OATH} parameter in the \textsc{Athena} package \citep[used in HSC,][]{athena}.  This is an approximation to speed up computation that allows for a small threshold of angular bin precision when placing the galaxy pairs into bins of angular separation. For our analysis we set this to zero which results in galaxy pairs being placed in the exact angular bins specified. We note that the effect of this is most apparent in the KiDS-1000 comparison. This is because the two-point correlation functions were computed in a large number of angular bins with a high \texttt{binslop}$=1.5$ \citep{Joachimi2020}\footnote{\url{https://github.com/KiDS-WL/Cat_to_Obs_K1000_P1}} setting and then binned into the nine coarser bins. This was done so that the same data could be used to compute the alternative COSEBIS measurements. These differences are small and random (they do not introduce a systematic offset) so we do not expect this to affect the resulting cosmology. To confirm that we are able to reproduce the results of previous surveys with \textsc{TXPipe}, we compare the results of the published data vector and \textsc{TXPipe} data vector with our likelihood pipeline.  The results of this are discussed in Section~\ref{sec:txpipe_comparison_cosmo}.

In our two-point measurement comparison one small systematic difference was found between the DES-Y1 published results and our pipeline. This was determined to come from an error in the original analysis in \citet{Troxel2017} that caused the mean shear subtraction value to be less than intended. This is fixed in our pipeline as well as the current DES pipeline. The effect of this change is small (and most apparent on the larger scales where the uncertainties are higher) so we do not expect this to affect the constraints.

\subsection{Cosmological Constraints}
\label{sec:txpipe_comparison_cosmo}

In Table~\ref{table:survey_priors} we show the priors adopted for the surveys for each of the cosmological and nuisance parameters that we adopt in this analysis.  For each survey we match the modeling choices made by the published results (nonlinear modeling choice, priors, IA modeling, and scale cuts).  One exception is our sampling in ln$A_{s}$ rather than $S_{8}$ for KiDS-1000 (discussed in more detail in this section).  The goal of this comparison is to assess the consistency of both \textsc{TXPipe} and our likelihood pipeline to the published analyses.  

 \begin{table*}
\caption{Priors for cosmological parameters and nuisance parameters used for the cosmological inference. Brackets indicate top-hat priors with the given bounds, while parenthesis indicate Gaussian priors with given $(\mu,\sigma)$. The first three columns show the survey choices for DES-Y1, HSC-Y1 and KiDS-1000 respectively.  We note that for KiDS-1000, instead of sampling in $S_8$, we instead sample in $A_s$ and use the priors in \citet{Hildebrandt2017}, with $\ln(As \times 10^{10}):$ [1.7, 5.0].}
\label{table:survey_priors}
\centering
\begin{tabular}{||l l l||}
\hline 
DES-Y1 & HSC-Y1 & KiDS-1000 \\ 
\hline
\multicolumn{3}{c}{\textsc{Cosmological Parameters}} \\
\hline
$A_{s} \times 10^{9}$: [0.5,5.0] & $\log_{10}\left(A_{s} \times 10^{9}\right)$: [-1.5,2.0]
& $S_{8}=\sigma_{8}\left(\Omega_{\rm m}/0.3\right)^{0.5}:[0.1,1.3]$\\
$\Omega_{\rm m}: [0.1,0.9]$ 
& $\Omega_{\rm c}: [0.01,0.9]$ 
& $\Omega_{\rm c}h^{2}: [0.051,0.255]$ \\
$\Omega_{\rm b}: [0.03,0.07]$ 
& $\Omega_{\rm b}: [0.038,0.053]$ 
& $\Omega_{\rm b}h^{2}: [0.019,0.026]$ \\
$h: [0.55,0.9]$
& $h: [0.64, 0.82]$ 
& $h: [0.64, 0.82]$\\
$n_{s}: [0.87,1.07]$ 
& $n_{s}: [0.87,1.07]$ 
& $n_{s}: [0.84,1.1]$\\
$\Omega_{k}: 0.0$ 
& $\Omega_{k}: 0.0$ 
& $\Omega_{k}: 0.0$ \\
$\Omega_{\nu}h^{2}: [0.0006,0.01]$ 
& $\sum{m_{\nu}}: 0.06 eV $ & $\sum{m_{\nu}}: 0.06 eV $\\
 \hline
\multicolumn{3}{c}{\textsc{Astrophysical Nuisance Parameters}} \\
\hline
$A_{\rm IA}: [-5.0, 5.0]$  
& $A_{\rm IA}: [-5.0, 5.0]$ 
& $A_{\rm IA}: [-6.0, 6.0]$ \\
$\eta_{\rm IA}: [-5.0, 5.0]$ 
& $\eta_{\rm IA}: [-5.0, 5.0]$ 
& $A_{\rm baryon}: [2.0, 3.13]$\\
$z_{0}: 0.62$
& $z_{0}: 0.62$ \\
& &$z_{0}: 0.62$ \\
\hline
\multicolumn{3}{c}{\textsc{Observational Nuisance Parameters}} \\
\hline
$\Delta z_{1}: (0.1,1.6)$  
& $\Delta z_{1}: (0.0,0.0374)$  & $N(\mu,C)$ \\
$\Delta z_{2}: (-1.9,1.3)$  
& $\Delta z_{2}: (0.0,0.0124)$  \\
$\Delta z_{3}: (0.9,1.1)$  
& $\Delta z_{3}: (0.0,0.0326)$ \\
$\Delta z_{4}: (-1.8,2.2)$  
& $\Delta z_{4}: (0.0,0.0343)$ \\
$m_{1..4}: (0.012,0.023)$ 
& $m_{0}: (0.0,0.01)$ 
& $c_{0}: (0,2.3\times 10^{-4})$\\
& $\alpha: (0.029,0.01)$\\ 
& $\beta: (-1.42,1.11)$ \\
\hline
\end{tabular}
\end{table*}

We now compare the cosmology results of our likelihood pipeline to the published results. In Figure~\ref{published_comparison} we show for the three datasets, the results of constraints in the $S_{8}$-$\Omega_{\rm m}$ plane from 1) our data vector + our likelihood pipeline (\textsc{TXPipe} DV); 2) the public data vector + our likelihood pipeline (Public DV); and 3) the publication (Public Chain). The comparison of 1) and 2) is a test that the differences described in Section~\ref{data_vector} do not affect significantly the cosmological constraints, and the comparison of 2) and 3) is a test for any difference between our likelihood pipeline and that used in the published results.  

Overall, we find that the data vectors produced from \textsc{TXPipe} give consistent values for the well constrained $S_8$ parameter with the published data vectors for all three data sets. However, we also found small shifts in less well constrained parameters including $\Omega_{\rm m}$. In the following, we will mainly focus on the comparison in $S_8$.  We will also point out changes in $\Omega_{\rm m}$, keeping in mind that it is intrinsically less well-constrained and more sensitive to noise. 
For the comparison of 2) and 3), or the comparison of the likelihood pipelines, we detail below for each of the datasets separately.

\begin{itemize}
\item For DES-Y1, we find that our likelihood code gives a constraint in $S_8$ very consistent with the published results, with the public data vector and \textsc{TXPipe} data vector constraints differing by $<0.1\sigma$ with that from the public chain. 
There is a shift in $\Omega_{\rm m}$ at $\sim$0.4$\sigma$.  This parameter is less well-constrained, and is more sensitive to variation due to noise between runs of the sampler, in particular for the \textsc{Multinest} settings that were used in DES-Y1.  By running multiple chains with the DES-Y1 \textsc{Multinest} settings, we confirm that there can be up to $\sim$0.5$\sigma$ shifts in $\Omega_{\rm m}$ solely from sampler noise. In our analysis, we run with a lower tolerance for this value and the variation in chains is less than when run with less stringent parameters (see \citealt{Lemos2022} for more details, and the end of this section for further discussion).

This test provides confirmation that the difference caused by the mean-shear subtraction issue does not change the cosmology results as demonstrated by the agreement between the \textsc{TXPipe} and published data vectors.

\item For HSC-Y1, we find very good agreement for $S_8$, with our results differing from the public chain by $<0.1\sigma$.  The $\Omega_{\rm m}$ constraints are also very consistent, differing by $<0.1\sigma$.  This indicates both \textsc{TXPipe} and our likelihood pipeline are consistent with the HSC pipelines.  One subtlety that is present is the choice of approximation in the method of comparison between the theory and data.  For our implementation of \textsc{CosmoSIS}, we assume that when comparing the data points with a theoretical model, we evaluate the theory at angular positions that correspond to a fixed angular bin definition for the data vector. \textsc{TXPipe} implements the pair-weighted mean of the angular bins whereas the HSC analysis used the area-weighted mean of the angular bin.  Our comparison finds consistent results between the public data vector and the \textsc{TXPipe} data vector, and we have further checked that the results do not change when the \textsc{TXPipe} data vector is evaluated with area-weighted means. The lack of bias between these comparisons is likely further explained by the fact that the angular binning for HSC is relatively narrow, and the effect is larger for coarser bin sizes.  

\item For KiDS-1000, we are reproducing the published $S_{8}$ (within $<0.05\sigma$) and $\Omega_{\rm m}$ (within $<0.1\sigma$).  We note for the public data vector in this analysis we adopt the method of \citet{Asgari2021} which integrates the theoretical prediction over the width of the angular bin.  As mentioned previously, \textsc{TXPipe} implements the pair-weighted mean of the angular bins instead of integrating the theory over the bin \citep{Krause2017}.  We find this adoption to be unbiased compared to the published results.  Because of this, we continue to adopt the pair-weighted mean for all subsequent analysis.  An additional subtlety comes from the different parameterization of the cosmological parameters used for KiDS-1000 -- they sampled in $S_8$ instead of the $\Omega_{\rm m}$ and $A_{s}$ parameterization in our pipeline. To accommodate this we have used the $A_{s}$ priors from an earlier KiDS analysis \citep{Hildebrandt2017}, setting the priors on $\ln(A_{s} \times 10^{10})$ to be a flat tophat at [1.7, 5.0]. As described in \citet{Joachimi2020}, we do not expect this to cause a bias in the mean $S_{8}$ posterior because it is well constrained compared to the prior.
\end{itemize}
Performing this comparison, we found a few issues with the \textsc{Multinest} sampler.  Primarily, we find that the less well-constrained parameters, e.g. $\Omega_{\rm m}$ could have varying posteriors at up to the $\sim0.5\sigma$ level when adopting certain values for the tolerance settings.  For this analysis, we adopt the settings used by \citet{Asgari2021}, which gives good results in terms of stability between chains.  However, we note the results of \cite{Lemos2021} that show that even with tighter tolerance settings this sampler can underestimate the uncertainty by $\sim10\%$.

Having validated the results from each of the publications using our independent pipeline, in the following sections we investigate the sensitivity of these constraints to different analysis choices. The two most significant class of analysis choices that could be made at this stage are: 1) the choice of priors on cosmological parameters and the model for astrophysical nuisance parameters (intrinsic alignment in the case of cosmic shear), and 2) the choice of how one treats the uncertainty in the small scale modeling. We discuss them in Sections~\ref{sec:priors} and \ref{sec:scale_cuts} respectively.

\begin{table}
\center
\caption{Priors for cosmological parameters and nuisance parameters used for the unified cosmological inference. Brackets indicate top-hat priors with the given bounds.  Note that the $A_{\textrm{baryon}}$ parameter is only varied in the \textsc{HMCode} scenario.}
\label{table:unified_priors}
\begin{tabular}{|| c||}
\hline 
Unified Analysis\\ 
\hline
\multicolumn{1}{c}{\textsc{Cosmological Parameters}} \\
\hline
 $\log_{10}\left(A_{s} \times 10^{9}\right)$: [-1.5,2.0]\\
 $\Omega_{\rm m}: [0.05,0.95]$ \\
 $\Omega_{\rm b}: [0.03,0.07]$ \\
$h: [0.55,0.9]$\\
$n_{s}: [0.84,1.1]$\\ $\Omega_{k}: 0.0$ \\ $\Omega_{\nu}h^{2}: [0.0006,0.01]$\\
 \hline
\multicolumn{1}{c}{\textsc{Astrophysical Nuisance Parameters}} \\
\hline
$A_{\rm IA}: [-6.0, 6.0]$ \\
$\eta_{\rm IA}: [-5.0, 5.0]$ \\
$A_{\rm baryon}: [2.0, 3.13]$ \\
$z_{0}: 0.62$
\end{tabular}
\end{table}

\begin{figure*}
\centering
\includegraphics[width=0.33\textwidth]{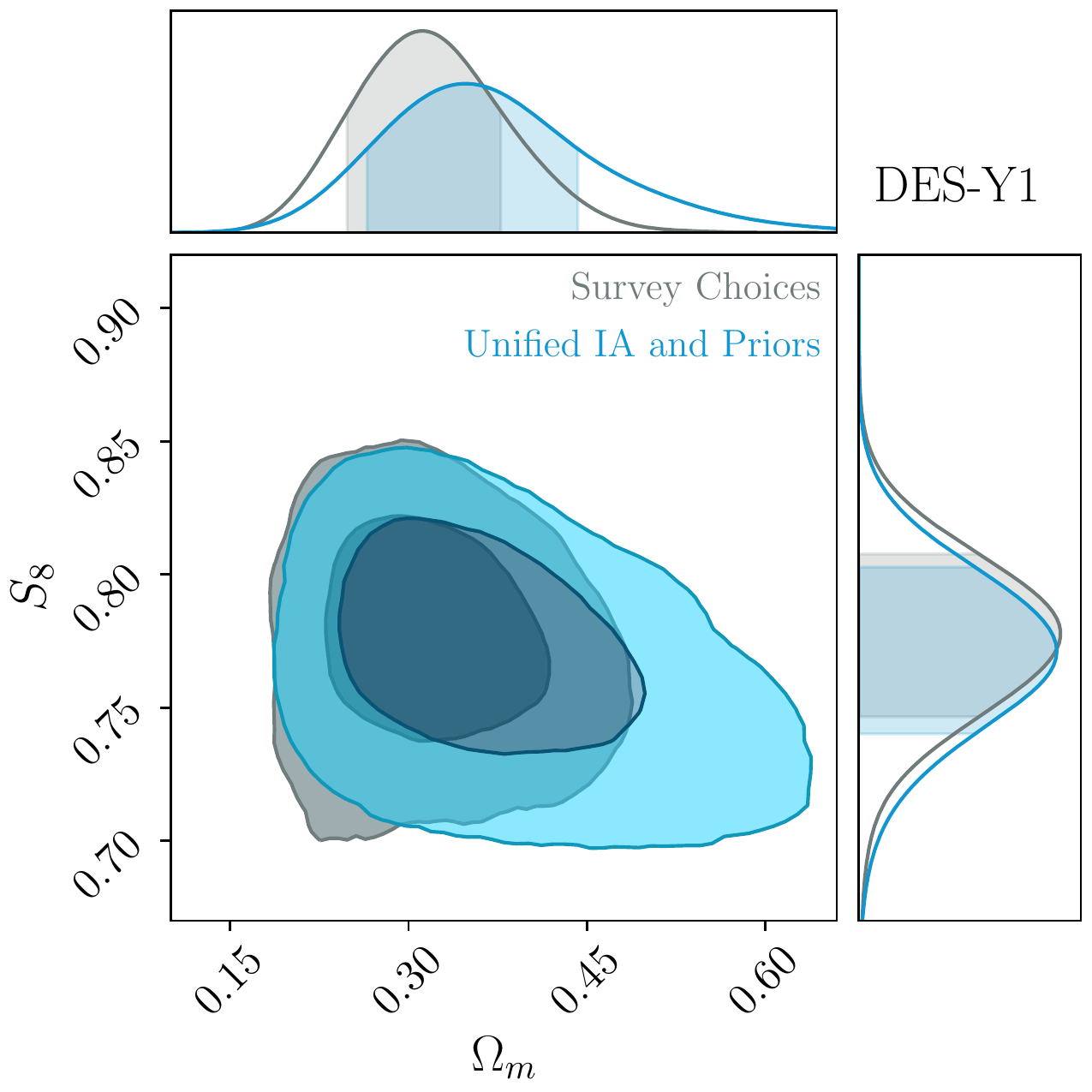}
\includegraphics[width=0.33\textwidth]{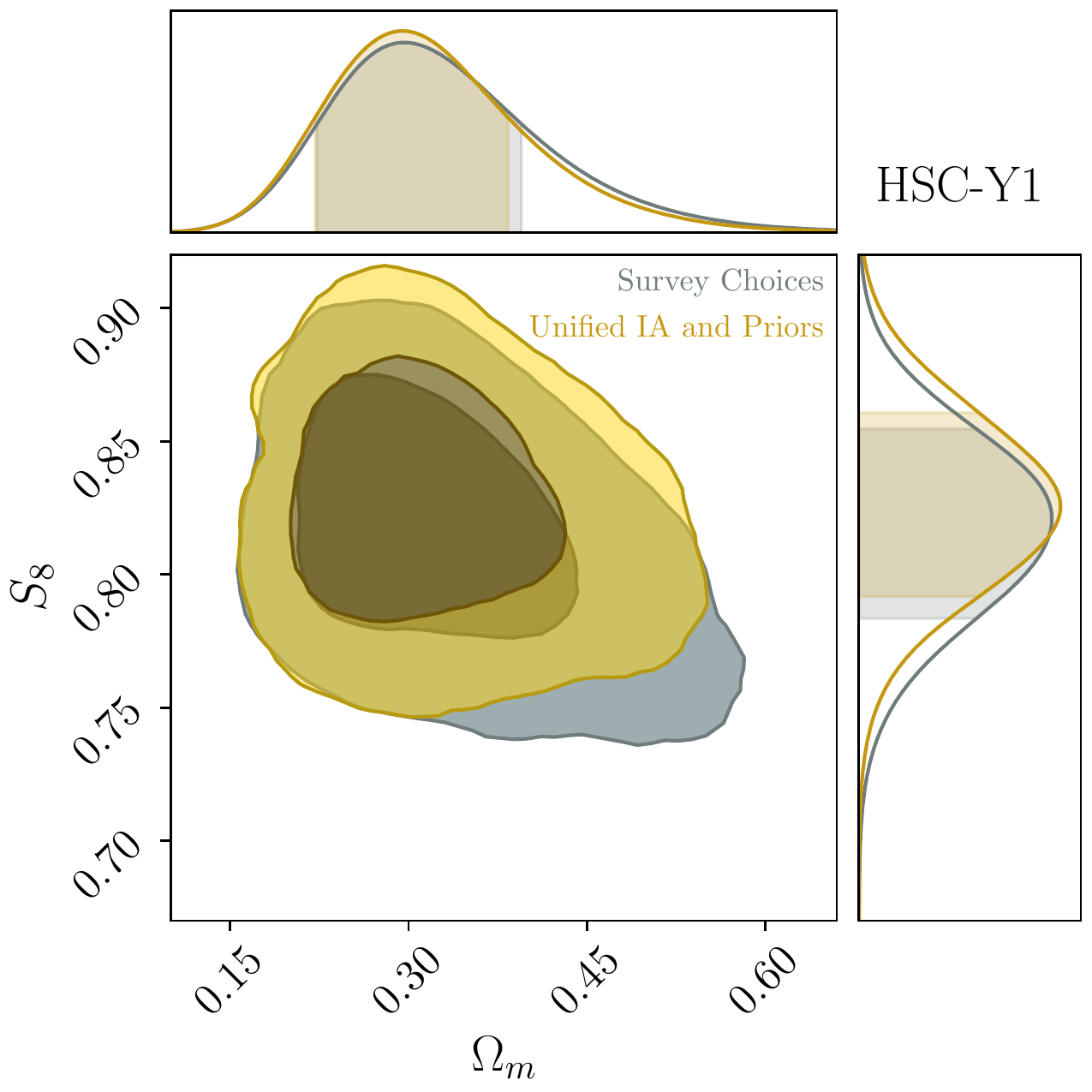}
\includegraphics[width=0.33\textwidth]{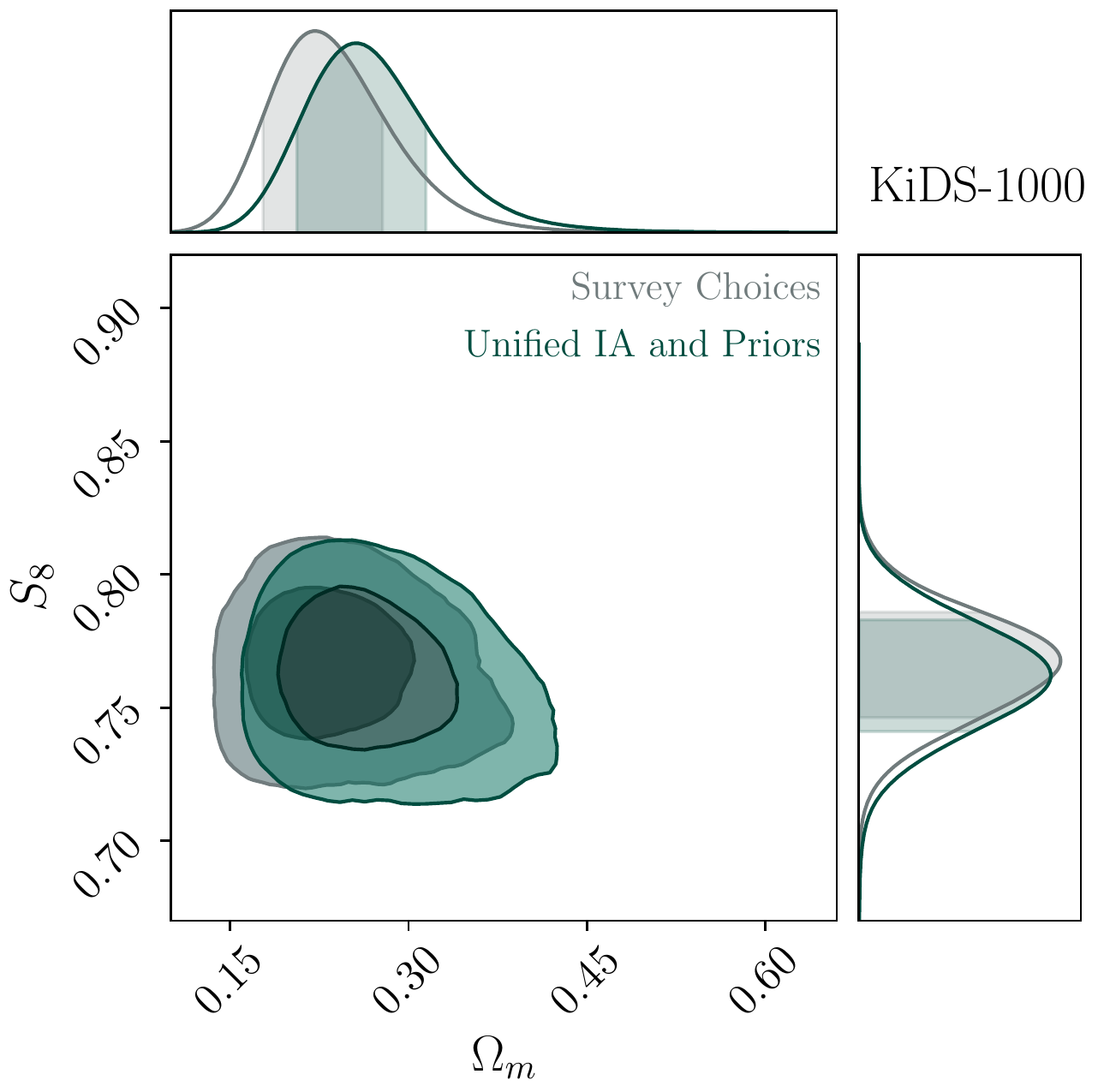}
\caption{Comparison of the $S_8$-$\Omega_{\rm m}$ constraints between 1) our reanalysis using the survey analysis choices (grey) and 2) our reanalysis unifying only the priors for cosmological parameters and intrinsic alignment (colors). The grey contours are identical to the ``\textsc{TXPipe} DV + \textsc{CosmoSIS}'' contours in Figure~\ref{published_comparison}.} 
\label{fig:unified_priors}
\end{figure*}

\section{Priors on Model Parameters and Intrinsic Alignment}
\label{sec:priors}

As described in Section~\ref{sec:likelihood}, priors for all the model parameters are incorporated when running the inference pipeline. This includes priors for the cosmological parameters as well as astrophysical and observational nuisance parameters. The priors for the model parameters adopted by each survey is shown in Table~\ref{table:survey_priors}. 

For the cosmological parameters we wish to constrain (for cosmic shear, this is primarily $S_8$ but also there is some sensitivity to $\Omega_{\rm m}$), it is important that the priors are wide enough so they do not inform the constraints. However, in the parameter space we work in, there could be implicit priors on derived parameters that impact the constraints indirectly. One example shown in \citet{Chang2019} is that the priors on $h$ could indirectly affect $\Omega_{\rm m}$, which would then propagate into $S_8$. As this is unavoidable, it is important to at least unify the priors when making comparisons between two datasets, which is what we will ultimately do in Section~\ref{sec:unify}.

In addition, different analyses choose different models for astrophysical and observational systematic effects. For observational systematic effects (e.g. photometric redshift uncertainties, shear calibration uncertainties), it does not make sense to unify the modeling between the different experiments when comparing since the datasets are different and the different teams have individually characterized them. On the other hand, for astrophysical systematic effects (e.g. IA, baryonic effects), it is reasonable to unify the modeling approach if the basic properties of the galaxy samples are not drastically different between the surveys. In the unified analysis in Section~\ref{sec:unify} we will use a consistent IA model for all three datasets. For modeling of baryonic effects, we separately discuss the treatment of small scales in Section~\ref{sec:scale_cuts}.

As can be seen in Table~\ref{table:survey_priors}, the model parameter priors adopted by DES-Y1 and HSC-Y1 primarily differ in $A_s$, $h$ and the neutrinos. The HSC-Y1 analysis samples $A_{s}$ in log space and the range is much larger than that of DES-Y1 (it translates to 0.03$\times 10^{-9}$ -- 100$\times 10^{-9}$)\footnote{In the DES-Y1 multiprobe analysis \citep{Abbott2019} a range of $[0.5, 10.0]$ is used for $10^{9}A_{s}$.}. The $h$ prior is slightly narrower for HSC-Y1.  In addition, both KiDS-1000 and HSC-Y1 fix the neutrino mass in their fiducial analysis, whereas DES-Y1 allows it to vary. Aside from these differences, the approaches are similar, adopting wide priors for the cosmological parameters with the same IA treatment ($z$-dependent NLA) and identical priors for these model parameters. The KiDS-1000 approach differs in their choice to sample in $S_{8}$ instead of $A_s$ and they adopt the $\Omega_{\rm c}h^{2}$ and $\Omega_{\rm b}h^{2}$ parameterization. Additionally, they do not adopt the redshift-dependent power-law in their fiducial IA modeling\footnote{In \citet{Asgari2021} they test adding this to the modeling and did not find evidence for a redshift evolution in the KiDS sample.}.  To test how changing the cosmological parameter priors and IA treatment affects each survey, we now use a set of common choices listed in the last column of Table~\ref{table:unified_priors}. 

For our unified cosmological analysis we choose to adopt top-hat priors in the $\Omega_{\rm m}$, $\log_{10}(A_{s})$, and $h$ parameterization.  Our primary goal is to unify the choices between the surveys, and therefore the choice of prior and parameterization is somewhat arbitrary.  However, we generally aim to err on the side of caution in terms of informing the posterior for the parameters of interest.  For this reason we choose the bounds for our unified choice to correspond to the widest adopted for each of the previous surveys. 

In Appendix~\ref{sec:sample_as} we show the effect of sampling $A_s$ in logarithmic and linear space, and show that the prior range we adopt for $\log_{10}(A_s)$ is flat in the parameter range of interest for $S_{8}$. For this reason we choose to sample in $\log_{10}(A_s)$ instead of $A_s$.  As mentioned previously, because this prior is wide compared to the constraint, we do not expect this choice of $\log_{10}(A_s)$ vs. $S_{8}$ to affect the results.  The constraint on $\Omega_{\rm m}$ on the other hand is more comparable to the width of the priors, and therefore we sample in this parameter of interest to be the least informative in the posterior.  
Also as explained above, we keep any additional systematic nuisance parameters (i.e. shifts in mean redshift, multiplicative bias) the same as each survey's original settings. In terms of intrinsic alignment modeling, we adopt a $z$-dependent NLA IA model for all three surveys.

The $S_8$-$\Omega_{\rm m}$ constraints produced from the unified prior+IA treatment scheme are shown in Figure~\ref{fig:unified_priors}. For reference, the ``Survey Choice'' constraints are shown in grey on the same plots, which correspond to the ``\textsc{TXPipe} DV + \textsc{CosmoSIS}'' contours in Figure~\ref{published_comparison}. That is, the only difference between the two contours in the same panel of Figure~\ref{fig:unified_priors} are coming from the change in cosmological parameter priors and IA treatment listed in Table~\ref{table:survey_priors}. 

We find that for DES-Y1 and HSC-Y1; there is a small shift in the $S_8$ constraint when we unified the cosmological priors and IA treatment of $0.15\sigma$ and $0.21\sigma$ respectively. 
The change in $\Omega_{\rm m}$ for DES-Y1 is noticeable, there is a $0.5\sigma$ shift and the constraint is $\sim$40$\%$ wider.  This change is primarily driven by the change in the prior on $A_s$.  Though the survey and unified $\Omega_{\rm m}$ priors are flat, the degeneracy between $A_{s}$ and $\Omega_{\rm m}$ means the prior space in the original DES-Y1 bounds are not completely flat.  The new $A_{s}$ bounds are flat in $\Omega_{\rm m}$ (see Appendix~\ref{sec:sample_as}) and the new bounds widen the constraint on $\Omega_{\rm m}$.

For KiDS-1000 we find unifying the IA and prior choices gives an $S_{8}$ constraint that is lower than the survey choices by $\sim$0.13$\sigma$.  In Appendix~\ref{appendix:IA} we isolate the effect of IA and prior choice individually.
We see a shift in $\Omega_{\rm m}$
from the combined effect of the redshift dependent IA model and priors analysis. We attribute this to the degeneracy between the IA amplitude, $\Omega_{\rm m}$ and the unified $A_s$ prior.
The constraint on $\Omega_{\rm m}$ is $\sim$0.5$\sigma$ higher.  Additionally the $\Omega_{m}$ constraint is wider by $10\%$.  Overall, the relatively small shifts in the $S_{8}$ constraints are encouraging for the robustness of the results to the analysis choices.  The $\Omega_{\rm m}$ constraint is less well constrained compared to the priors, making it more sensitive to the choices.  We encourage caution when quoting this parameter because of this, but note that this effect is likely less relevant for more tightly constraining datasets.

\section{Effect of Unifying Small-Scale Treatment}
\label{sec:scale_cuts}

\begin{figure*}
\centering
\includegraphics[width=0.33\textwidth]{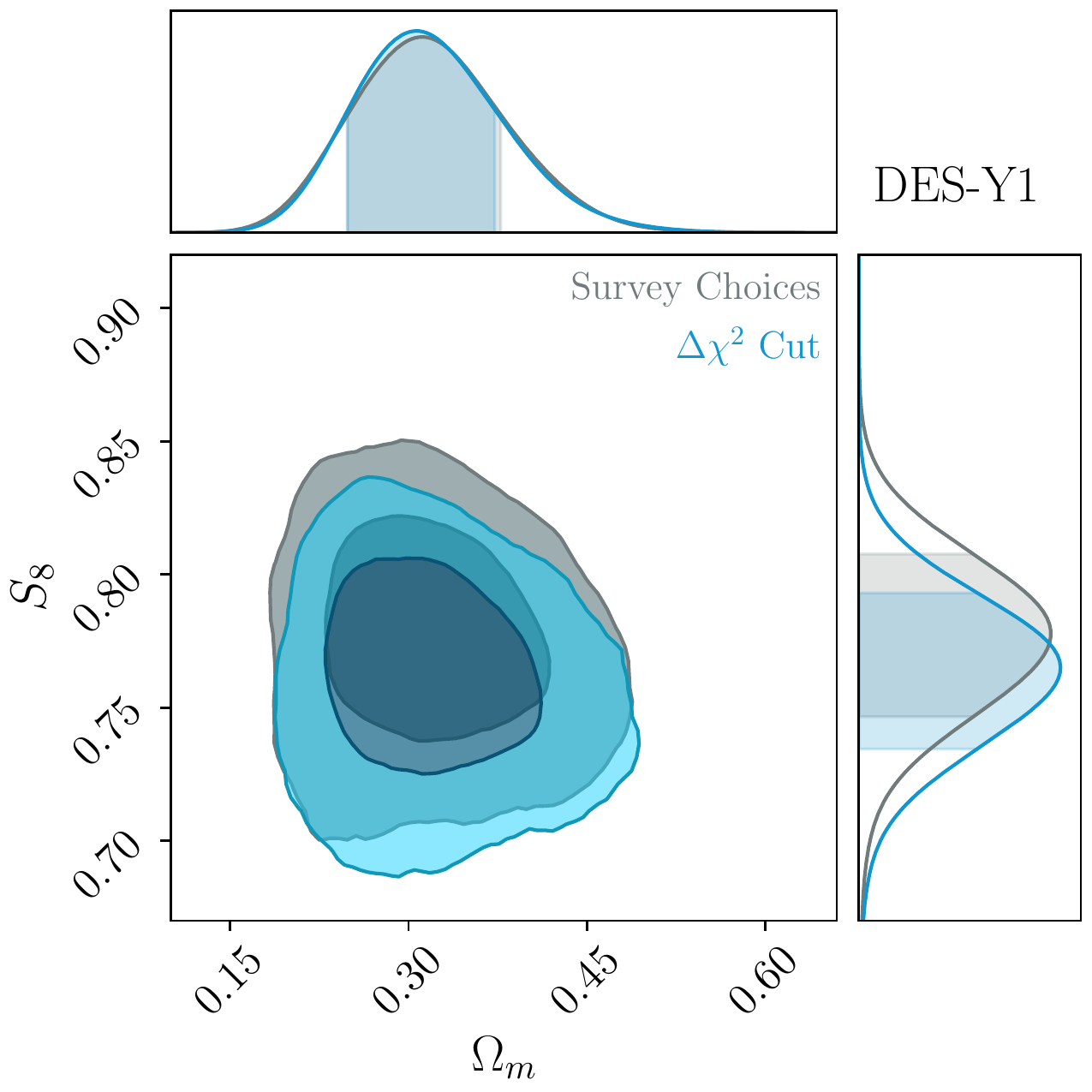}
\includegraphics[width=0.33\textwidth]{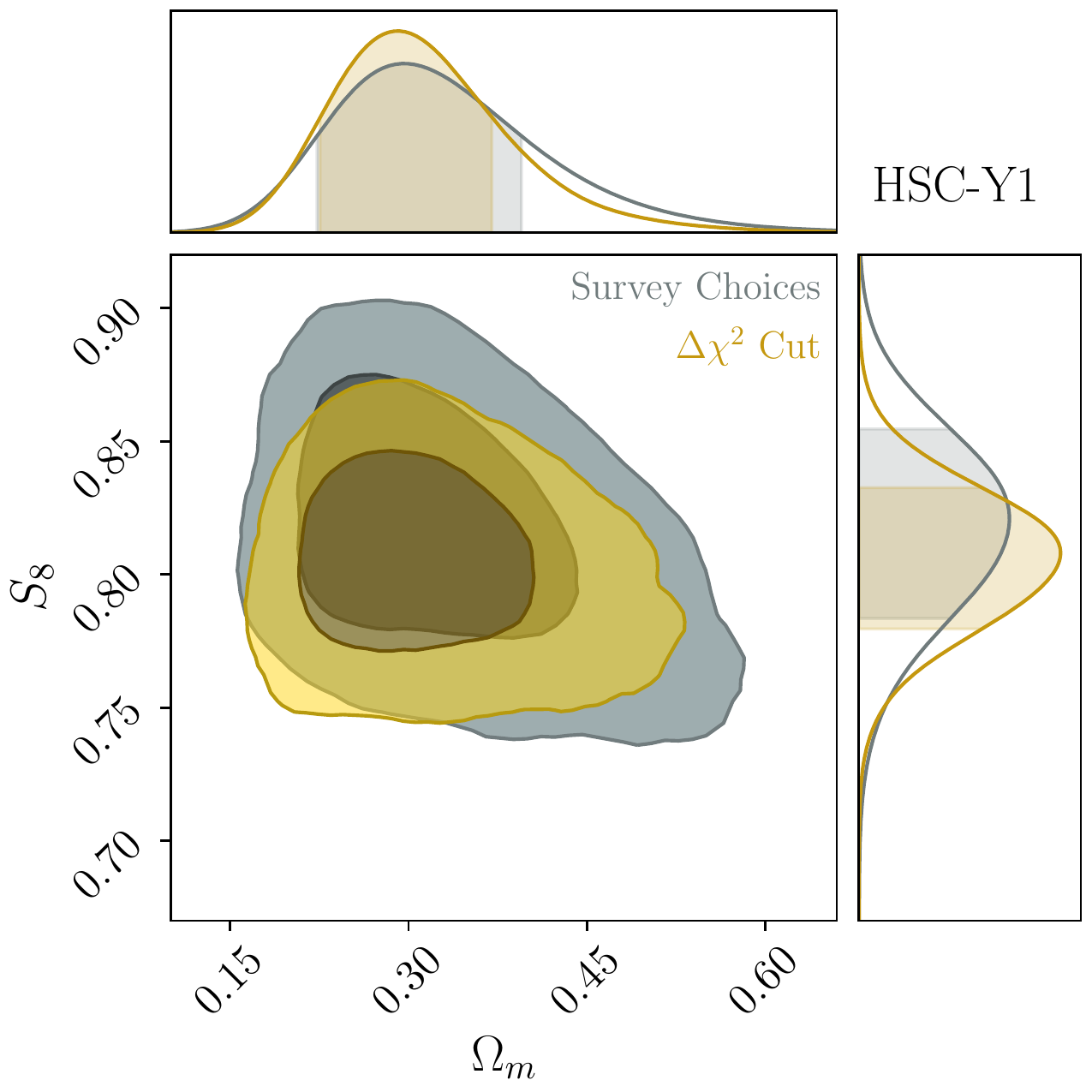}
\includegraphics[width=0.33\textwidth]{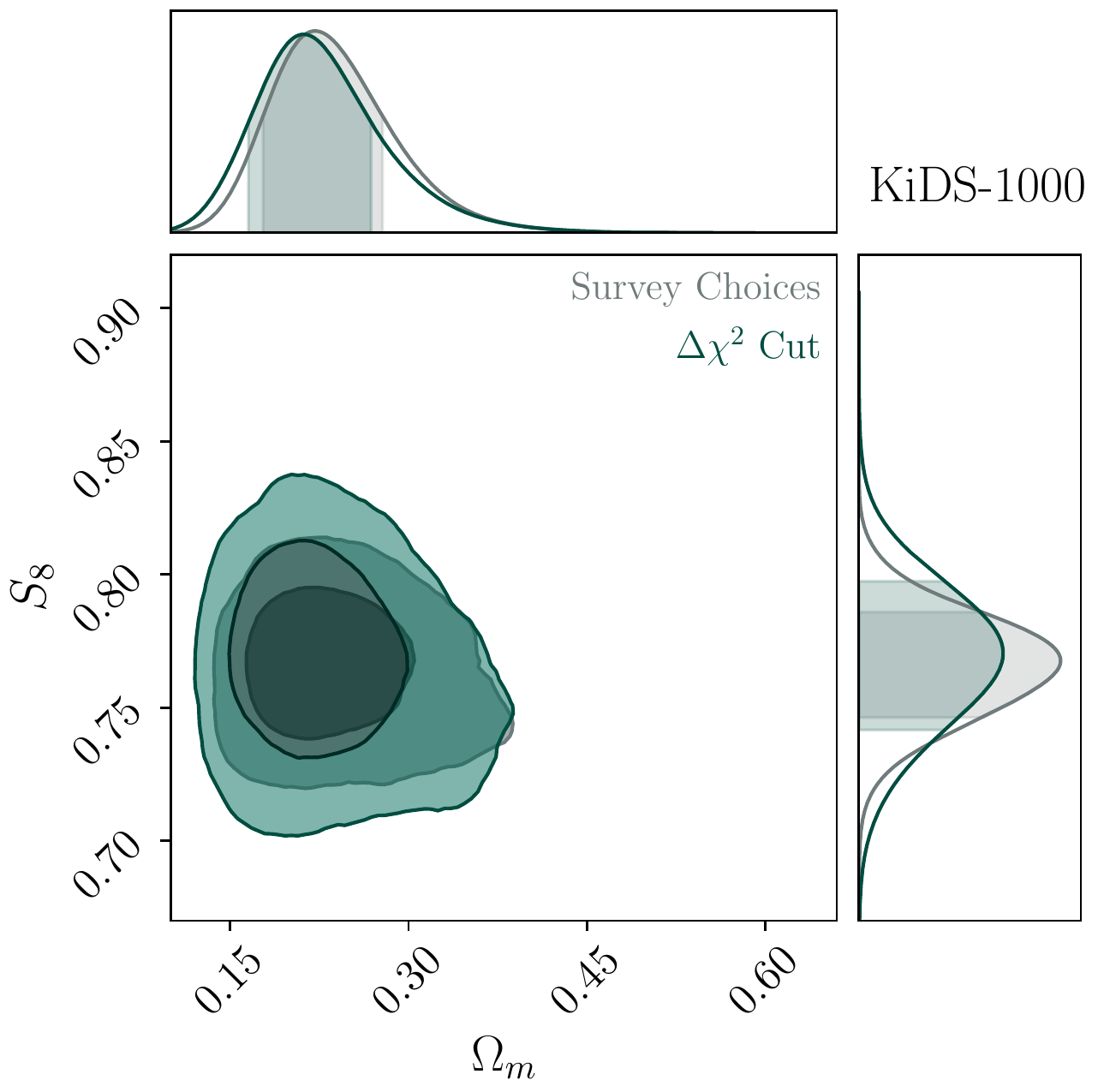} \\

\includegraphics[width=0.33\textwidth]{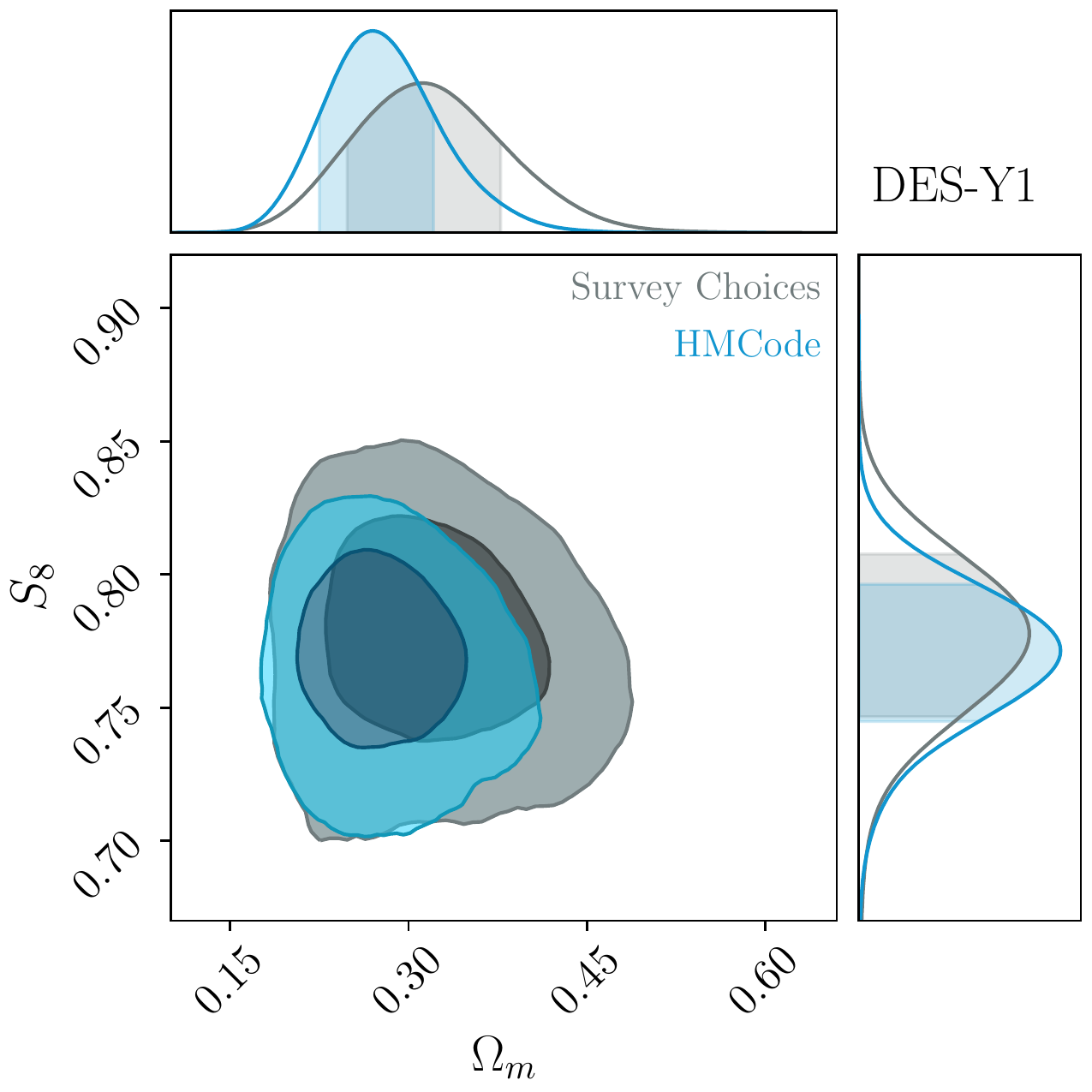}
\includegraphics[width=0.33\textwidth]{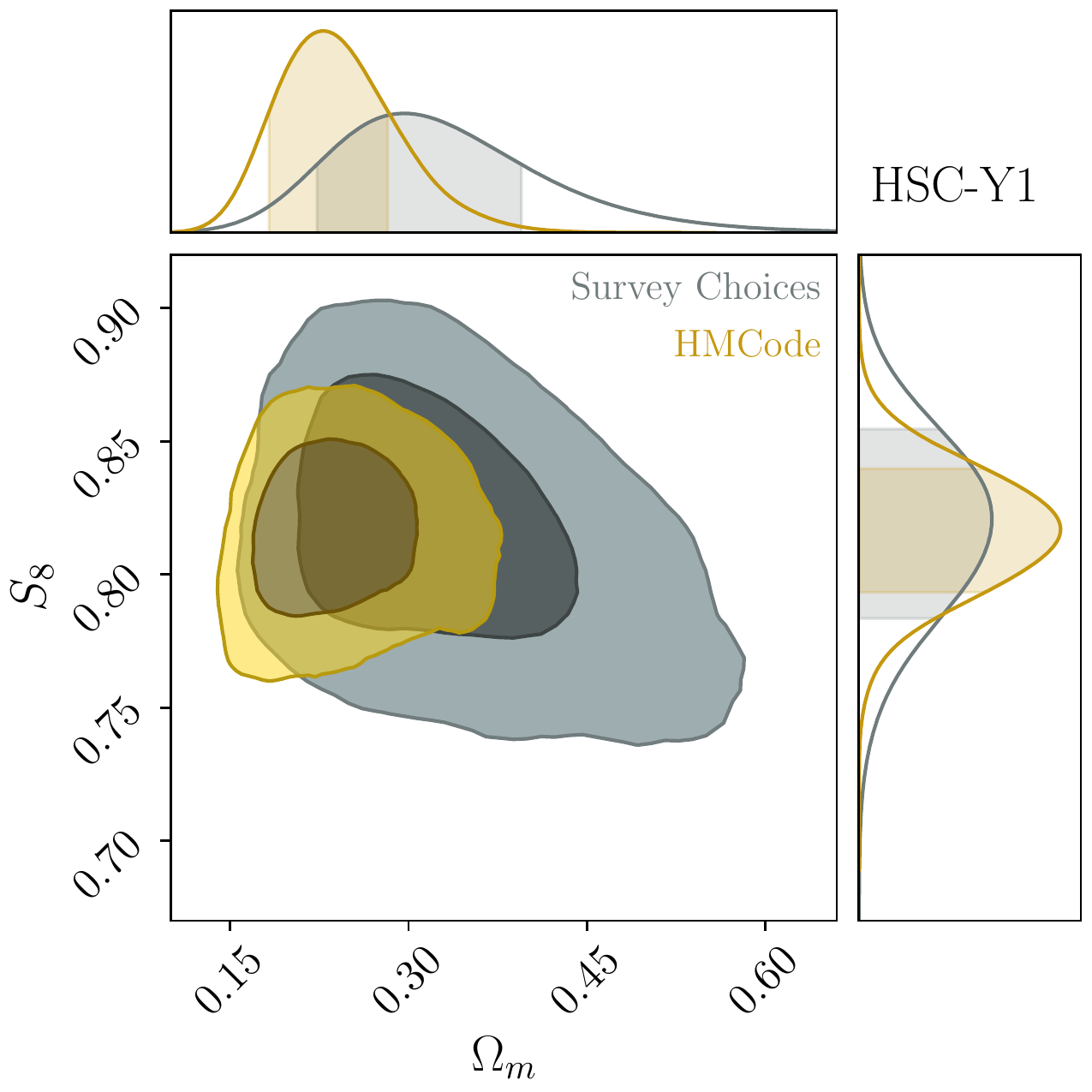}
\includegraphics[width=0.33\textwidth]{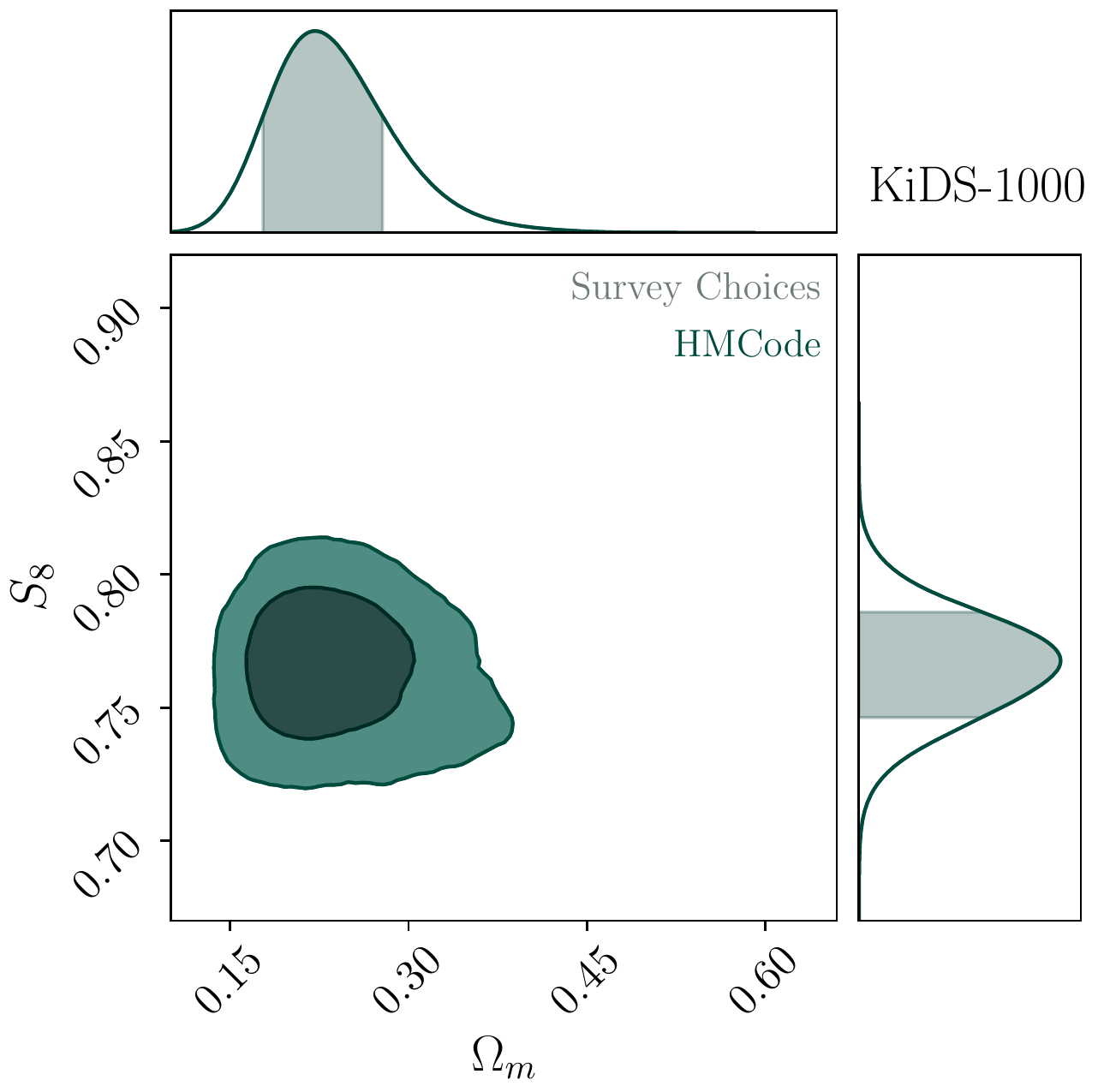}
\caption{Comparison of the $S_8$-$\Omega_{\rm m}$ constraints between 1) our reanalysis using the survey analysis choices (colored) and 2) our reanalysis unifying only the small-scale treatment (grey). In the top row we show the result when we use the same procedure of defining a scale cuts to remove the small scales where the model is inaccurate. In the bottom row we show the results when we use all the small scale information but marginalize over our ignorance of the modelling using \textsc{HMCode}. The colored contours are identical to the ``\textsc{TXPipe} DV + \textsc{CosmoSIS}'' contours in Figure~\ref{published_comparison}.}
\label{fig:small_scale_treatment}
\end{figure*}

Figure~\ref{fig:small_scale_treatment} shows the resulting constraints for the three surveys when unifying the small scale treatment to the two approaches. For comparison, we again overlay the corresponding ``\textsc{TXPipe} DV + \textsc{CosmoSIS}'' contours.  For both the $\Delta\chi^{2}$ cut approach and \textsc{HMCode} approach, we include smaller scales in the data vectors for DES-Y1 and HSC-Y1 than was used in their fiducial analyses.  The chosen cuts for the $\Delta\chi^{2}$ compared to the published choices are shown in Appendix~\ref{appendix:scale_cuts}.  For the \textsc{HMCode} cut we adopt the KiDS-1000 fiducial choice of $0.5\arcsec$ for $\xi_{+}$ and $4.0\arcsec$ for $\xi_{-}$.   For KiDS-1000 we are excluding more scales in the $\Delta\chi^{2}$ cut approach, than their fiducial choice, which matches the \textsc{HMCode} cut.  In particular, we note that the systematic tests that were performed by the previous analyses were not used to validate the data vector at these set of scales. 

Overall, as expected, we find that for the $\Delta\chi^{2}$ cut approach, the KiDS-1000 results shows the largest change; while for the \textsc{HMCode} approach, DES-Y1 and HSC-Y1 change more significantly. Also, in general, the \textsc{HMCode} allows us to use data on smaller scales, making the overall constraints in $S_8$ about $15\%$ tighter for DES-Y1, $30\%$ tighter for HSC-Y1, compared to the public analyses. Adopting the $\Delta\chi^{2}$ cut approach for KiDS-1000 widens the $S_{8}$ constraint by $\sim$40$\%$ compared to the \textsc{HMCode} cut. Finally, for KiDS-1000, the unified treatment with \textsc{HMCode} is by definition the same as the survey choice, which is shown by the identical contours in the bottom right panel of Figure~\ref{fig:small_scale_treatment}.  In Appendix~\ref{appendix:chi2_hmcode} we explore the effects of adopting the $\Delta\chi^{2}$ cut and \textsc{HMCode}.

We next examine each set of contours more carefully. We find that in general, the shift in the $S_8$ constraints when changing the small-scale treatments is very significant. For DES-Y1, HSC-Y1 and KiDS-1000, there is a +0.35, +0.30, -0.10 $\sigma$ shift in $S_8$ when we change from the survey choice to the unified $\Delta\chi^2$ cut.  For the HMCode small-scale treatment, we find shifts of -0.20, -0.03, and 0 $\sigma$. This finding is interesting as it implies that including the small scales with the \textsc{HMCode} approach is consistent with the $S_{8}$ values when adopting the more conservative approach. We note that this statement is difficult to make using simulations as \textsc{HMCode} is based on fits to a certain set of simulations. With three independent set of data here, however, it gives empirical support to the robustness of the small-scale treatment with \textsc{HMCode}, at least at the statistical power of these three datasets. There is a noticeable change in the $\Omega_{\rm m}$ constraints for DES-Y1 and HSC-Y1 when switching to the \textsc{HMCode} approach -- the constraints become significantly tighter (26\%, 40\%) and lower in their absolute value by 0.53$\sigma$ and 0.95$\sigma$ respectively. This change is however much less pronounced in the KiDS-1000 contours. We note that one possibility is that the covariance matrices for DES-Y1 and HSC-Y1 are not well-validated on the small scales since they were not used in the published results, and $\Omega_{\rm m}$ is more sensitive to these subtleties as it is not well constrained. We do not have sufficient information here to come to a definite conclusion.

\section{Unified Analysis}
\label{sec:unify}

With the understanding of the impact of the individual analysis choices, we now look at the combined effect when we unify all analysis choices considered here across the three surveys -- priors on cosmological parameters and IA model (unified treatment as listed in Table~\ref{table:unified_priors}), and small-scale treatment (either the $\Delta\chi^{2}$ cut or \textsc{HMCode}). 
  
We emphasize again that it is expected the shear calibration parameters, photo-z bias and PSF systematic parameters are survey-specific and as such we adopt the survey's choice for these components.  Additionally, we caution the reader that the systematic tests that were performed individually for the surveys were adopted with a target precision in mind, and are not necessarily valid for the additional data that we include in this analysis.  We do not attempt to reevaluate these analyses for the data.

\subsection{Individual Unified Constraints}

\begin{figure*}
    \centering
    \includegraphics[width=0.32\textwidth]{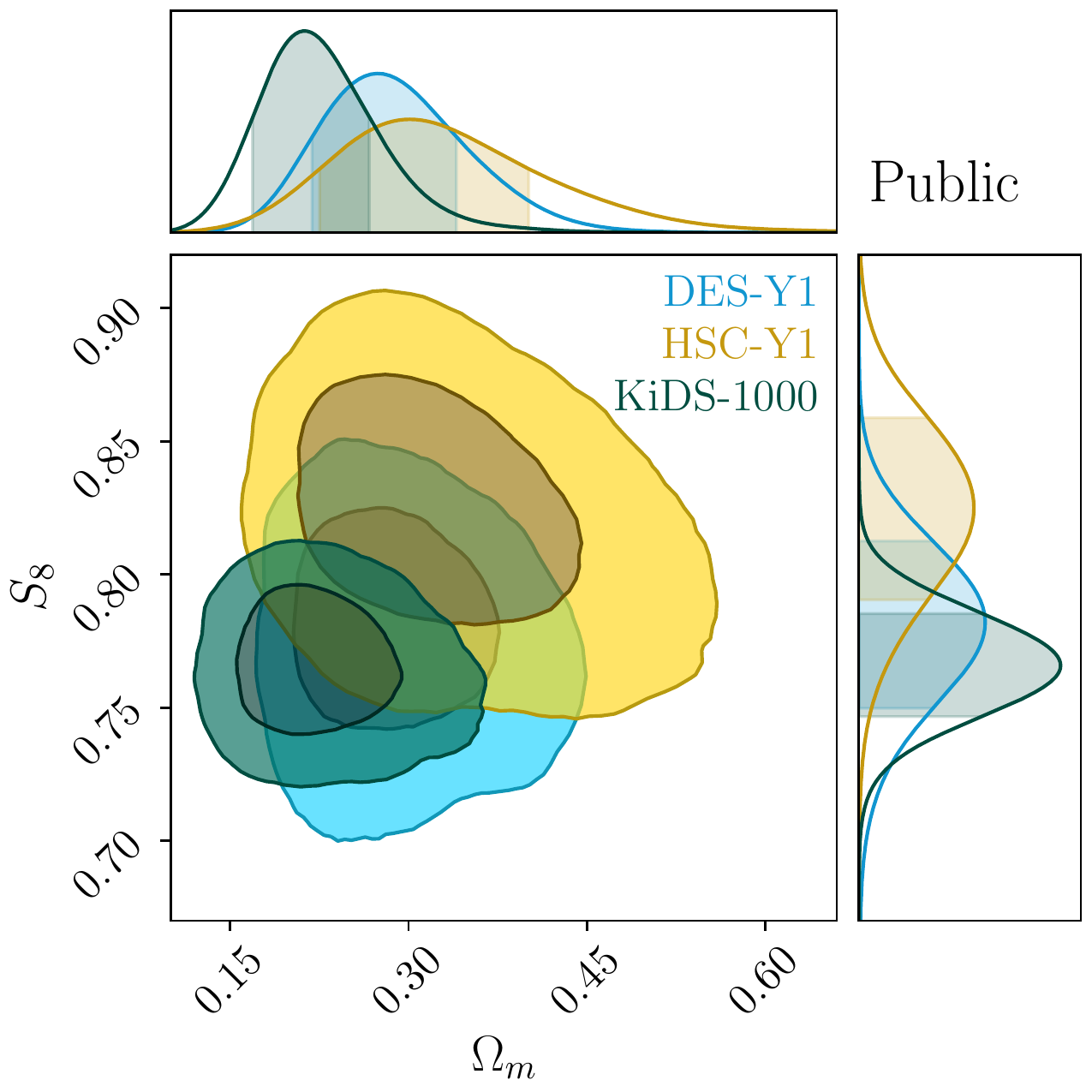}
    \label{fig:combined_published}
    \includegraphics[width=0.32\textwidth]{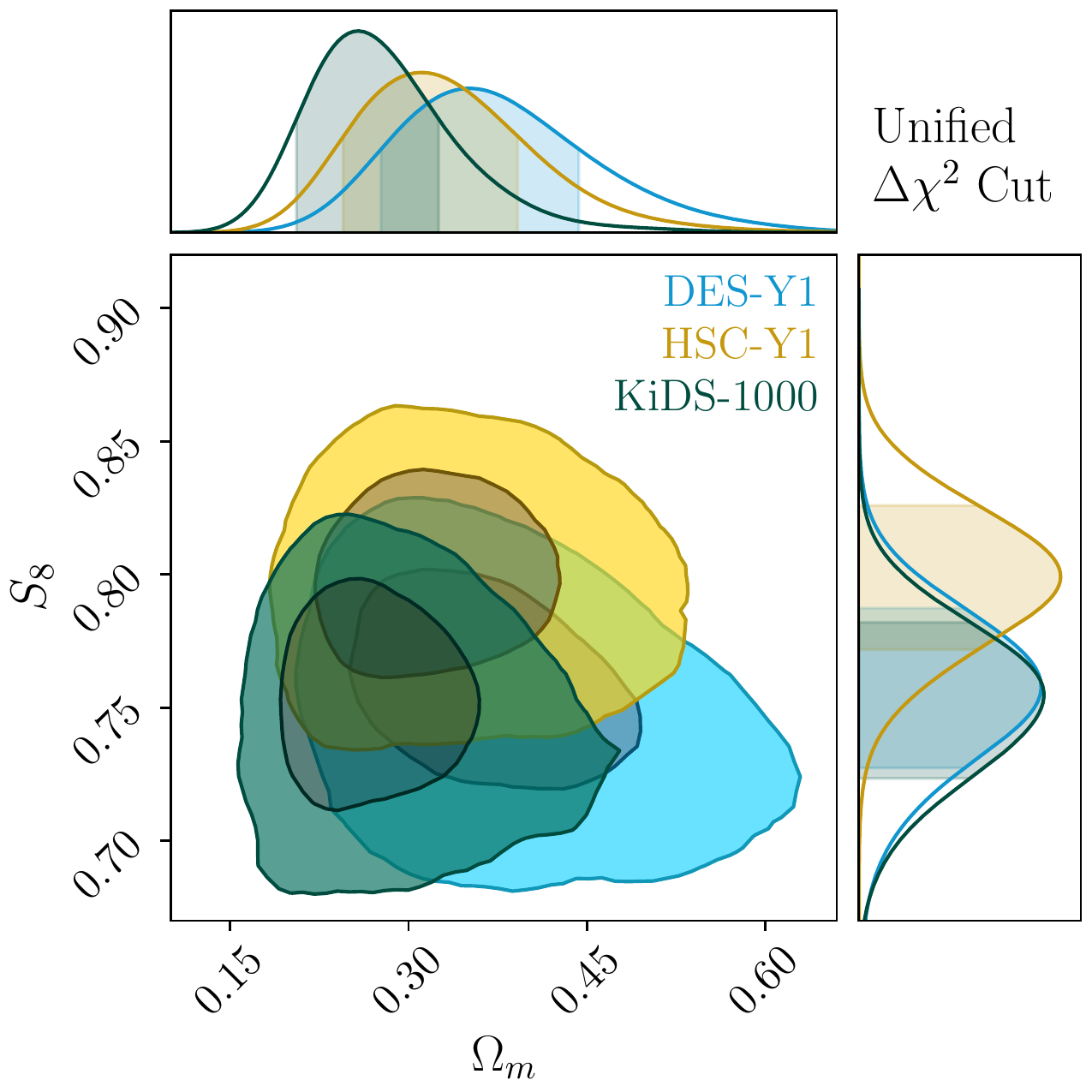} 
    \label{fig:combined_unified}
    \includegraphics[width=0.32\textwidth]{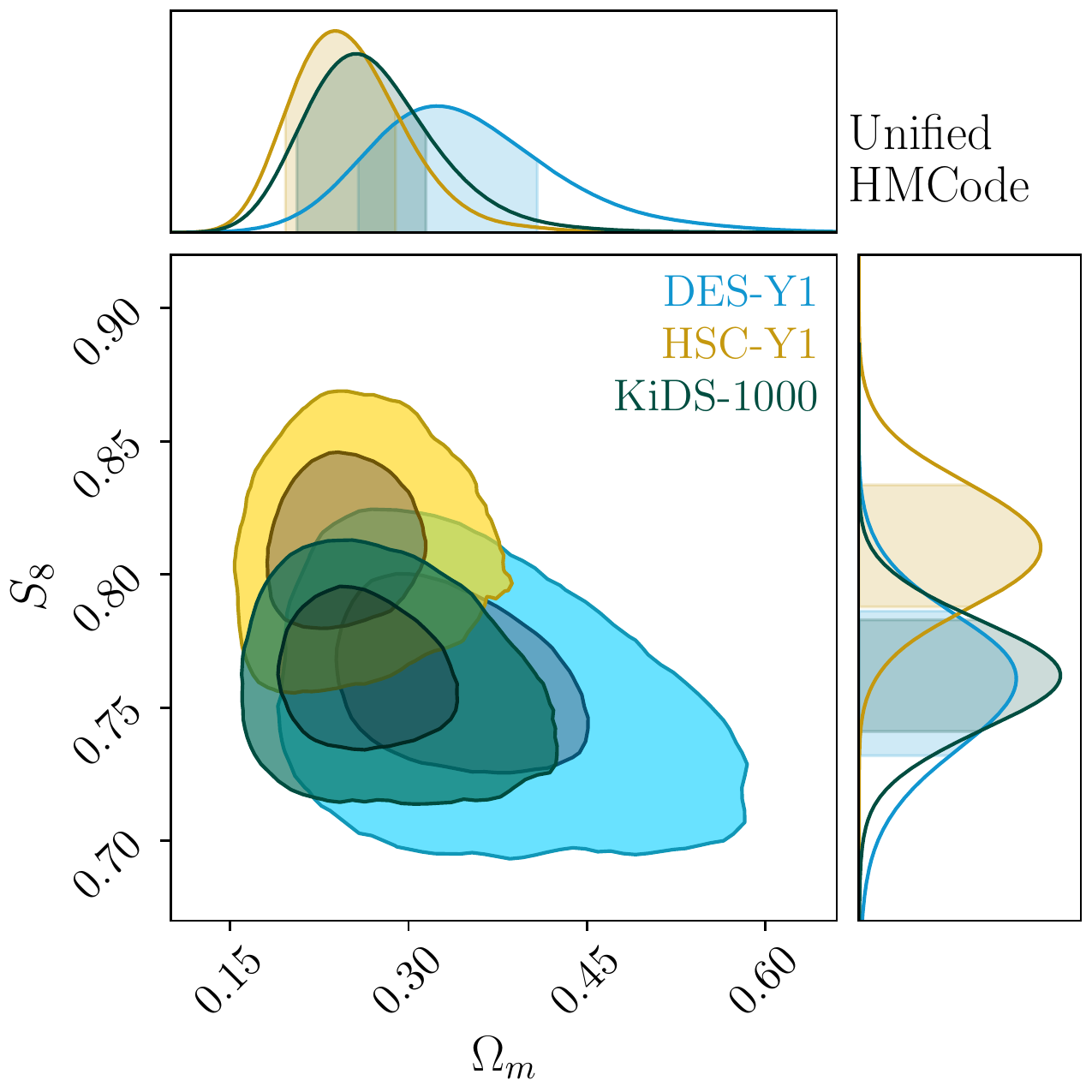}
    \label{fig:combined_unified_small_scales}
    \caption{Comparison of the $S_8$-$\Omega_{\rm m}$ constraints from the published results from \citet{Troxel2017,Hamana2018,Asgari2021,HamanaRevision} (left) and our unified reanalysis of the same data (middle and right). In addition to unifying the measurement and modeling pipeline, the priors on cosmological models and IA, we present two different treatments of the small-scale information either using a scale cut determined by the uncertainty in the modeling (middle) or the marginalization over baryonic effects on those scales using \textsc{HMCode} (right). 
    }
    \label{fig:unified_choices}
\end{figure*}
\begin{figure*}
\centering
\includegraphics[width=0.4\textwidth]{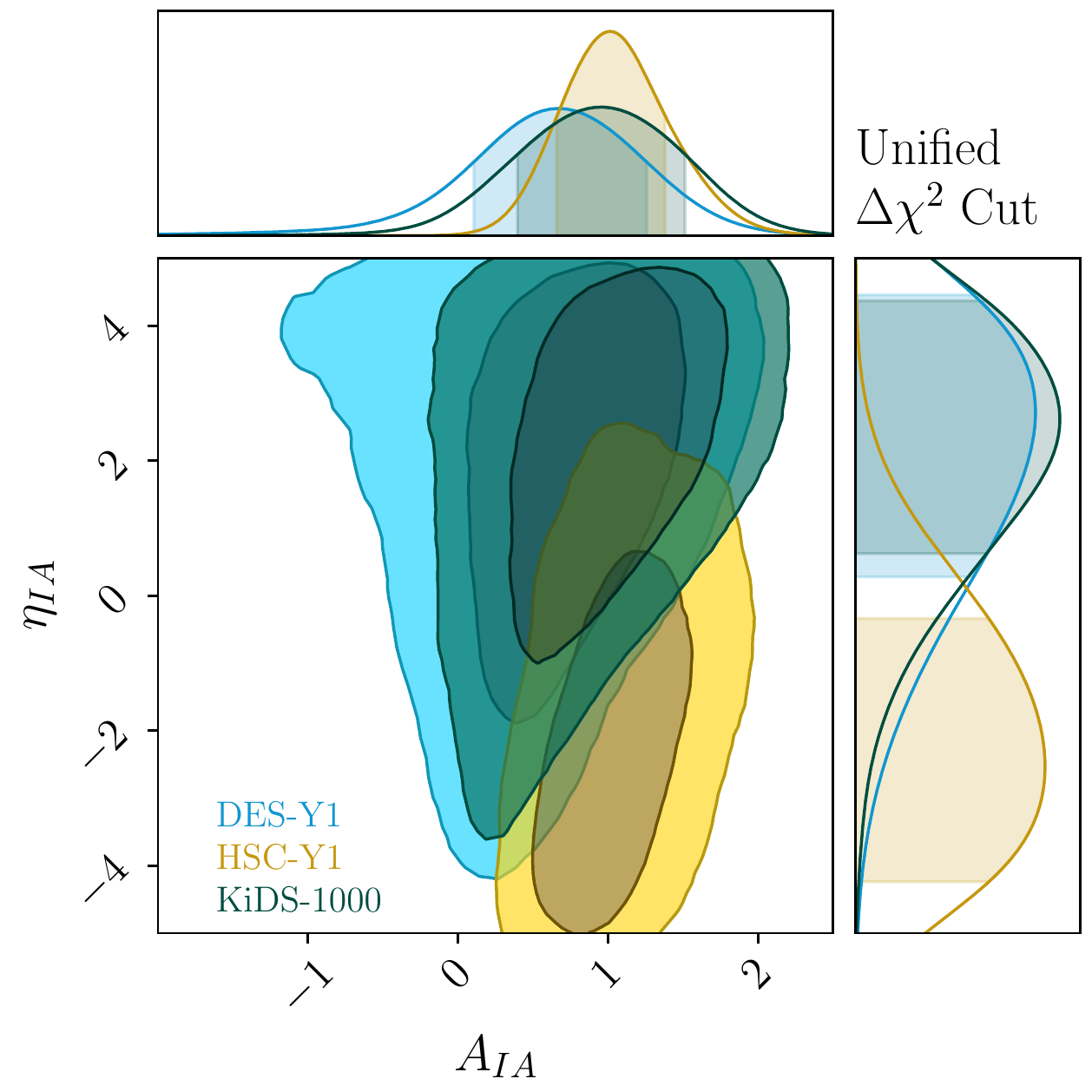}
\includegraphics[width=0.4\textwidth]{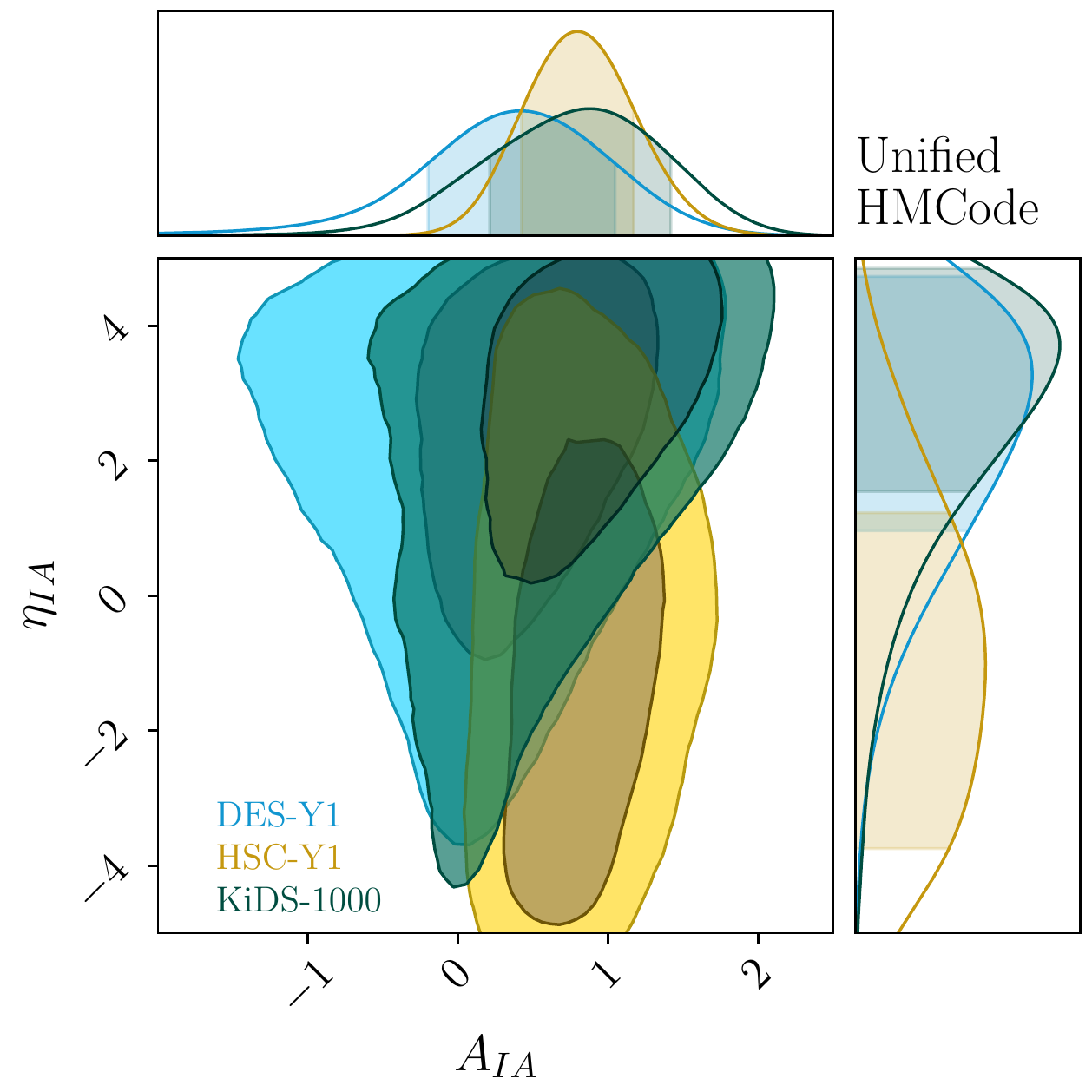}
\caption{IA parameter constraints for each of the datasets when we unify the analysis choices and use the $\Delta\chi^2$ cut (right), and HMCode (left) approach for small-scale treatment.}
\label{fig:IA_constraints}
\end{figure*}

Figure~\ref{fig:unified_choices} shows the $\Omega_{\rm m}$-$S_8$ constraints for the unified choices (middle and right panels) compared to the published results (left panel) using the three datasets. We present two unified choices using two different approaches for small-scale treatments.  We note that for DES-Y1 and HSC-Y1, we are including smaller scales than were originally tested in the previous analyses.  Overall, we find the relative relation between the three contours does not change significantly compared to the published results.  Below we discuss the small differences we do observe, in terms of shifts compared to the published contours.

First we compare the {constraints from the $\Delta\chi^2$ scale cut approach, shown in the middle panel, to the constraints of these parameters for the fiducial published analyses which are shown in the left panel} of Figure~\ref{fig:unified_choices}. From our previous results, we find that $\Omega_{\rm m}$ is sensitive to the choice of priors.  This is most noticeable in the change for DES-Y1, where the posterior extends to large $\Omega_{\rm m}$ values.  Overall, this parameter is not well constrained compared to the prior at the level of the individual surveys.  Overall, for KiDS-1000, the contours became less constraining while HSC-Y1 contours became more constraining — this primarily comes from the fact that the unified $\Delta\chi^{2}$ cut effectively resulted in using less small scales for KiDS-1000 and more scales for HSC-Y1. The mean of the $S_8$ constraint shifts by -0.62$\sigma$, -0.64$\sigma$, and 0.46$\sigma$ going from the public chain to the unified analysis with $\Delta\chi^2$ cut for DES-Y1, HSC-Y1 and KiDS-1000, respectively. The constraining power increased by $5\%$ and $20\%$ for DES-Y1 and HSC-Y1, and decreased by $50\%$ for KiDS-1000.

It is also interesting to look at the IA constraints in the unified analysis. Unlike the cosmological constraints, we expect these could differ between the surveys, as differences in the selection of the galaxies could result in samples with different properties.  We show in Figure~\ref{fig:IA_constraints}, the IA parameter constraints for the $\Delta\chi^2$ cut and HMCode case respectively. We find that the DES-Y1 and KiDS-1000 IA constraints are fairly consistent, with weak positive detection on the amplitude of IA, or $A_{\rm IA}$ (with $\sim$60$\%$, $\sim$80$\%$ and $\sim$70$\%$ of the marginalized posterior greater than zero for DES-Y1, HSC-Y1 and KiDS-1000 respectively).  Out of the datasets, HSC-Y1 has a slightly stronger detection of $A_{\rm IA}$. We find that the parameters are consistent between two unified modeling cases.  None of the three datasets constrain $\eta$ very well. We note that the HSC-Y1 source sample is looking at fainter and higher redshift galaxies, so the qualitative difference in the IA constraints could be a result of this. In Appendix~\ref{appendix:IA_combined} we examine the effect on the combined constraint when we adopt separate IA parameters for each survey.

Next we compare the resulting constraints for the \textsc{HMCode} approach, shown in the right, with the published constraints in the left panels of Figure~\ref{fig:unified_choices}. We find tighter and lower constraints on $\Omega_{\rm m}$ for HSC-Y1 coming from change in the small-scale treatment as shown in Figures~\ref{fig:small_scale_treatment}. For DES-Y1 the change in priors resulted in a higher and less constraining $\Omega_{\rm m}$. For $S_8$, the relative relation between the mean of the constraints remain largely unchanged from the published results, with DES-Y1 and HSC-Y1 gaining some constraining power due to the use of smaller scales. The mean of the $S_8$ constraint shifts by $0.59\sigma$, $0.33\sigma$, and $0.17\sigma$ for DES-Y1, HSC-Y1 and KiDS-1000, respectively, going from the public chain to the unified analysis with \textsc{HMCode}. The constraining power increased for DES-Y1 and HSC-Y1 by $12\%$ and $29\%$ respectively, primarily driven by the inclusion of more scales, and KiDS-1000 decreased by $15\%$ with unified priors.  In general the relatively small shifts in $S_{8}$ from the choice of small scale modeling is encouraging, and suggests that using more data can lead to tighter constraints without a large bias in the results.  We note this conclusion would assume that there are not additional systematics that are present at the smaller scales.

\begin{table*}
  \caption{Below we show, for the unified analyses across the three surveys, the $S_8$ constraints and the goodness-of-fit. We quote in terms of $\chi^{2}/$(d.o.f.-constrained parameters) and the resulting reduced $\chi^2$ (p-value).  The top and bottom halves of the table show two different cases of treatment of the small scales in the data vectors. The top half removes the small scales based on a $\Delta\chi^2$ criteria while the bottom half uses \textsc{HMCode} to marginalize over uncertainties in baryonic physics that affect small scales.}
  \label{tab:constraints_unified}
  \centering
  \begin{tabular}{l c c c}
\hline 
Dataset & DES-Y1 & HSC-Y1 & KiDS-1000 \\
\hline
$\Delta\chi^2$ cut & & &  \\
$S_{8}$& $0.754 ^{+0.031}_{-0.024}$ & $0.798^{+0.026}_{-0.024}$  & $0.751^{+0.029}_{-0.024}$ \\
$\chi^{2}$/D.O.F. & 257.08/(258-6.79) & 253.16/(213-7.59) & 197.26/(164-6.60)\\
Reduced $\chi^{2}$ (p-value) & 1.02 (0.386) & 1.23 (0.012) & 1.25 (0.0171)\\
 \hline
\textsc{HMCode} & & &  \\
$S_{8}$& $0.757^{+0.027}_{-0.021}$ & $0.812^{+0.021}_{-0.021}$ & $0.761^{+0.021}_{-0.019}$ \\
$\chi^{2}$/D.O.F. & 381.12/(380-7.28) & 431.60/(380-7.53) & 257.17/(225-7.28) \\
Reduced $\chi^{2}$ (p-value) & 1.02 (0.371) & 1.16 (0.019)  & 1.17 (0.034) \\
 \hline
\end{tabular}
\end{table*}

\begin{table}
  \caption{Metrics for consistency between the ``Dataset 1'' and ``Dataset 2'' -- we show our metrics, ($\Delta S_{8}$, MCMC parameter shift, or Par Diff, and Suspiciousness), as described in Section~\ref{sec:metric}. The top and bottom halves of the table show two different cases of treatment of the small scales in the data vectors. The top half removes the small scales based on a $\Delta\chi^2$ criterion while the bottom half uses \textsc{HMCode} to marginalize over uncertainties in baryonic physics that affect small scales.  We quote MCMC and Suspiciousness in terms of n$\sigma$ (p-value).}  Our threshold for agreement between surveys is a p-value $>0.01$.  Note that the tension metrics here assume independence between the datasets, and do not account for the cross-correlation due to the survey footprint overlap.
  \label{tab:tension_pairwise}
  \centering
  \begin{tabular}{l c c c}
\hline 
Dataset 1 & DES-Y1 & DES-Y1 & KiDS-1000 \\
Dataset 2 & HSC-Y1 & KiDS-1000 & HSC-Y1 \\
\hline
$\Delta\chi^2$ cut & & &  \\
$\Delta S_{8}$ & 1.15$\sigma$  & 0.09$\sigma$  & 1.26$\sigma$  \\
Par Diff & 1.01$\sigma$ (0.31) & 0.04$\sigma$ (0.97) & 1.36$\sigma$ (0.17) \\
Suspiciousness & 0.31$\sigma$ (0.77) & 0.55$\sigma$ (0.58) & 1.25$\sigma$ (0.22) \\
\hline 
\textsc{HMCode} & & &  \\
$\Delta S_{8}$ & 1.67$\sigma$ & 0.15$\sigma$ & 1.71$\sigma$ \\
Par Diff & 1.24$\sigma$ (0.21) & 0.43$\sigma$ (0.67) & 0.88$\sigma$ (0.38) \\
Suspiciousness & 1.52$\sigma$ (0.15) & 0.56$\sigma$ (0.57) & 1.22$\sigma$ (0.22) \\
 \hline
\end{tabular}
\end{table}

\begin{figure*}
	\centering
 	\includegraphics[width=0.95\columnwidth]{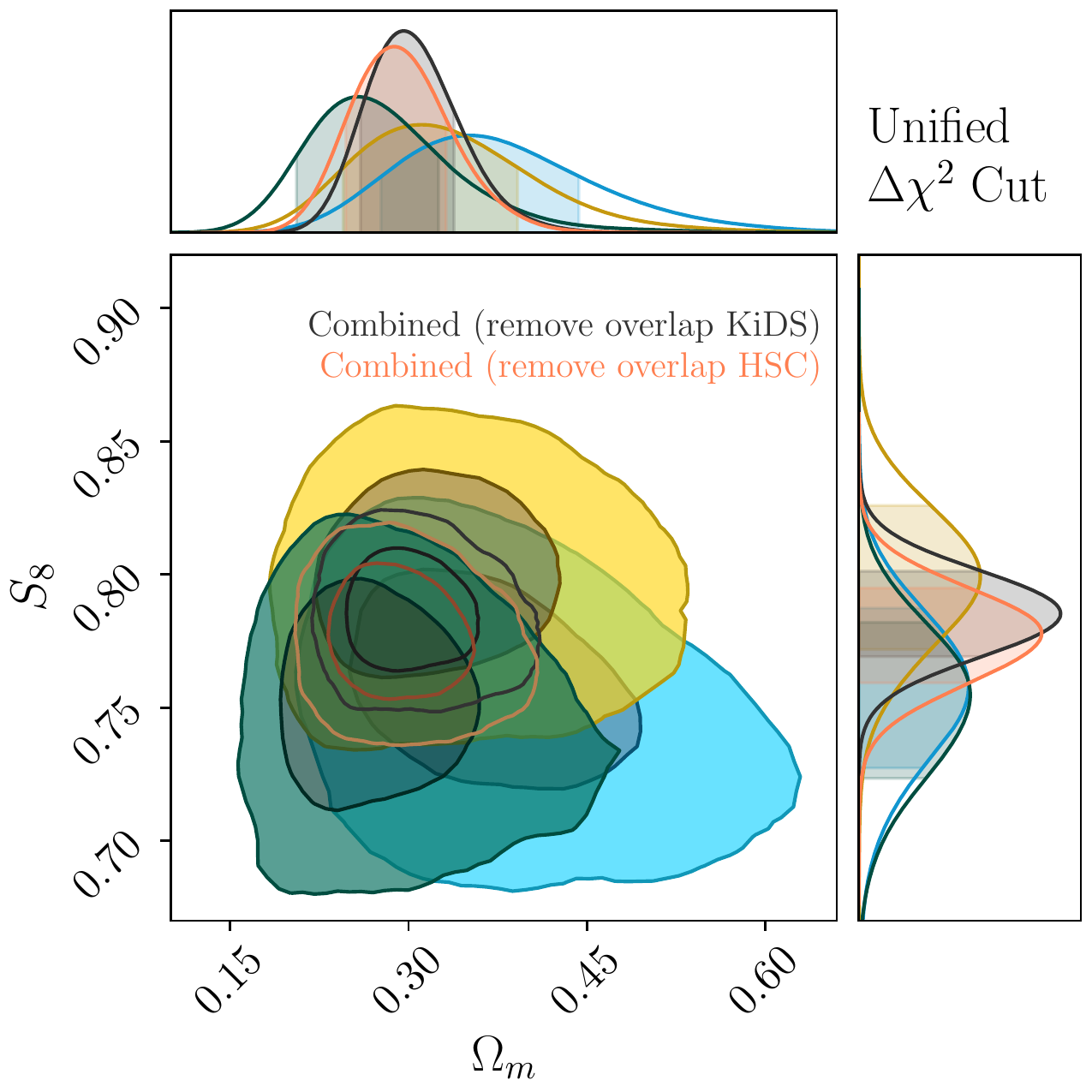}
	\includegraphics[width=0.95\columnwidth]{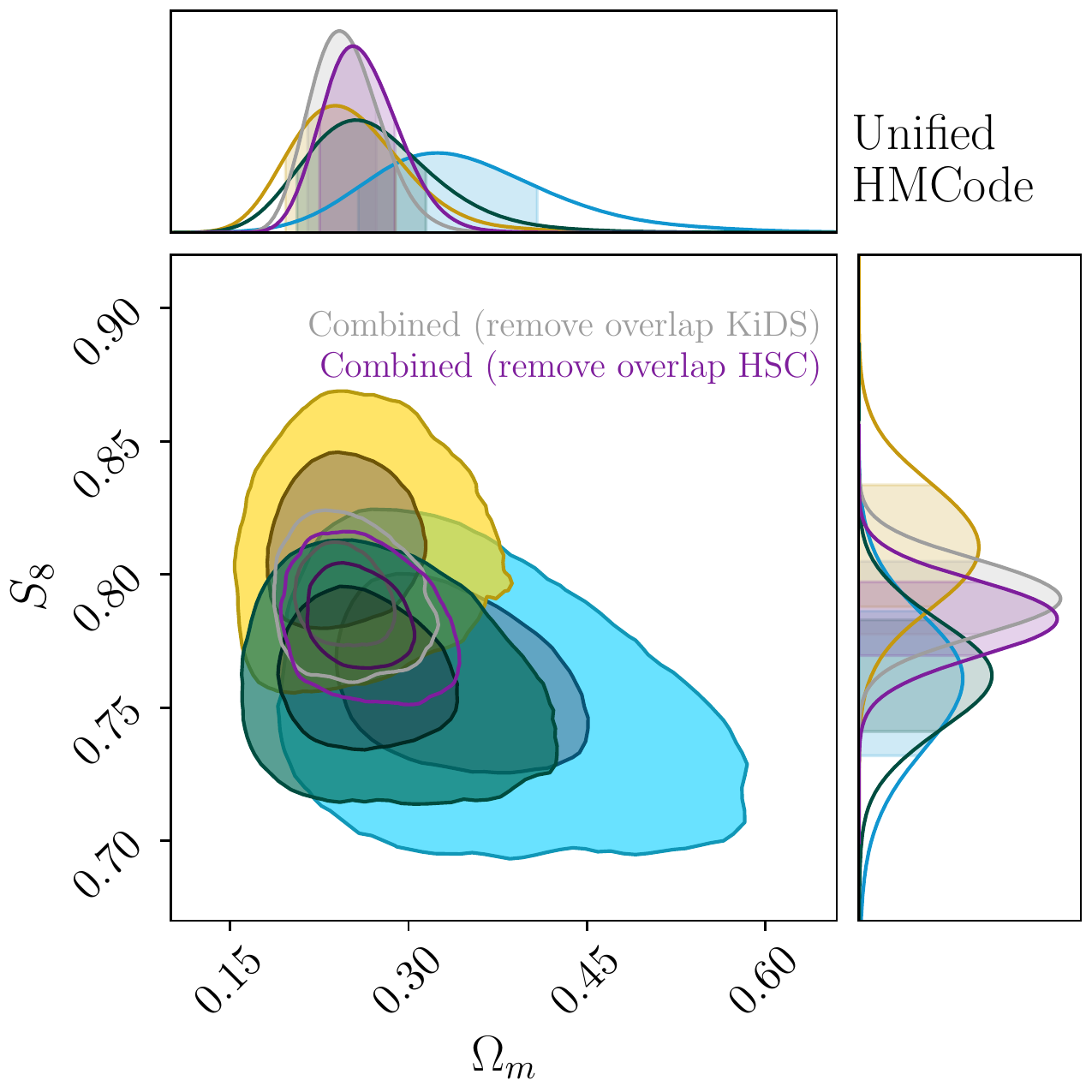}
	\caption{$S_8$-$\Omega_{\rm m}$ constraints from the unified analysis of the three individual surveys (DES-Y1, HSC-Y1 and KiDS-1000) and the combined constraints. Two sets of combined constraints are shown assuming the North part of KiDS-1000 and HSC-Y1 are fully correlated (black) and fully independent (grey).  
	The two panels show two different treatments of the small-scale information either using a scale cut determined by the uncertainty in the modeling (left) or the marginalization of those scales using \textsc{HMCode} (right).}
	\label{fig:planck_combined}
\end{figure*}

A summary of the final marginalized posterior mean and 68$\%$ confidence interval on $S_{8}$ for each of the unified chains described is listed in Table~\ref{tab:constraints_unified}, together with the goodness-of-fit for each of the constraints -- we find that all the results show acceptable goodness-of-fit values. In Table~\ref{tab:tension_pairwise} we list the tension metrics between each pair of the surveys. In general, we find that all metrics pass our criteria in that we do not deem any of them indicating inconsistency between each pair of datasets.

In all the combinations, the results of the tension metrics are highest between KiDS-1000 and HSC-Y1.  Using the $\Delta\chi^{2}$ cut, we find a $1.36\sigma$ (0.17) and $1.25\sigma$ (0.22) based on the parameter difference and suspiciousness metric respectively.  For the HMCode approach, we find $0.88\sigma$ difference (p-value 0.38) and $1.22\sigma$ (p-value 0.22) for the parameter difference and suspiciousness metric respectively.   This means if the two datasets come from the same underlying cosmology, there is still at the least, a 17\% ($\Delta\chi^2$ cut), 38\% (HMCode) chance that a discrepancy at this level could appear due to statistical fluctuation, and in our pre-set threshold we view them as not inconsistent (see Section~\ref{sec:metric}). While the experiments are formally consistent under our criteria, we note that the  $S_{8}$ value from HSC-Y1 remains relatively high compared to the other surveys.  With this in mind, we proceed to combine the datasets.  Additionally, we note that because these metrics were computed from chains that use the original published covariance, they assume independence between the datasets and do not account for the cross-correlation from the area overlap.

\begin{figure}
	\includegraphics[width=0.95\columnwidth]{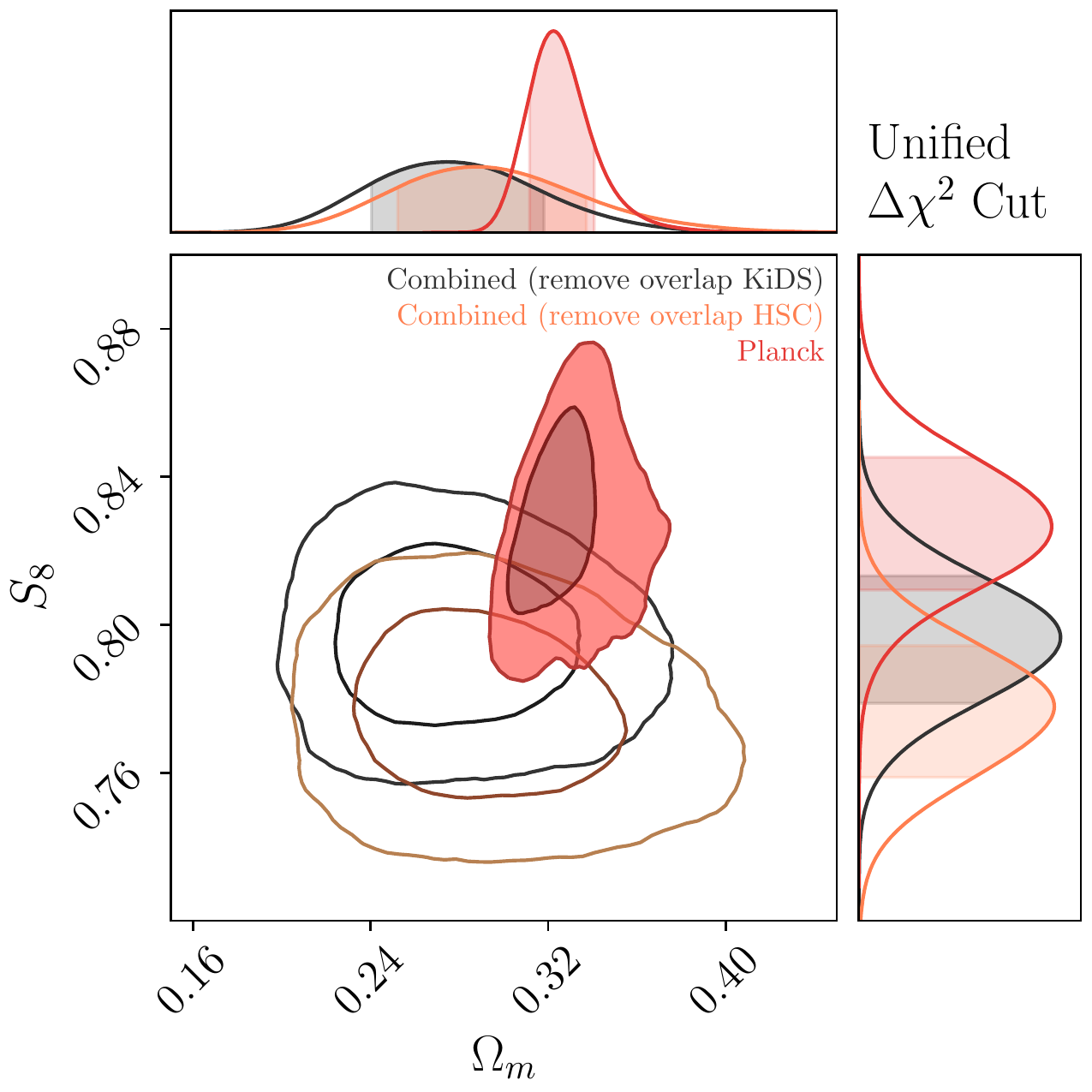}
	\caption{$S_8$-$\Omega_{\rm m}$ constraints from the combined constraints from DES-Y1, HSC-Y1 and KiDS-1000 assuming two different small-scale treatments and two scenarios for accounting for the covariance between the surveys. These constraints are compared with the primary CMB probes of {\it Planck} \citep{Planck2018}.} 
	\label{fig:planck_compare}
\end{figure}

\begin{figure}
	\includegraphics[width=0.95\columnwidth]{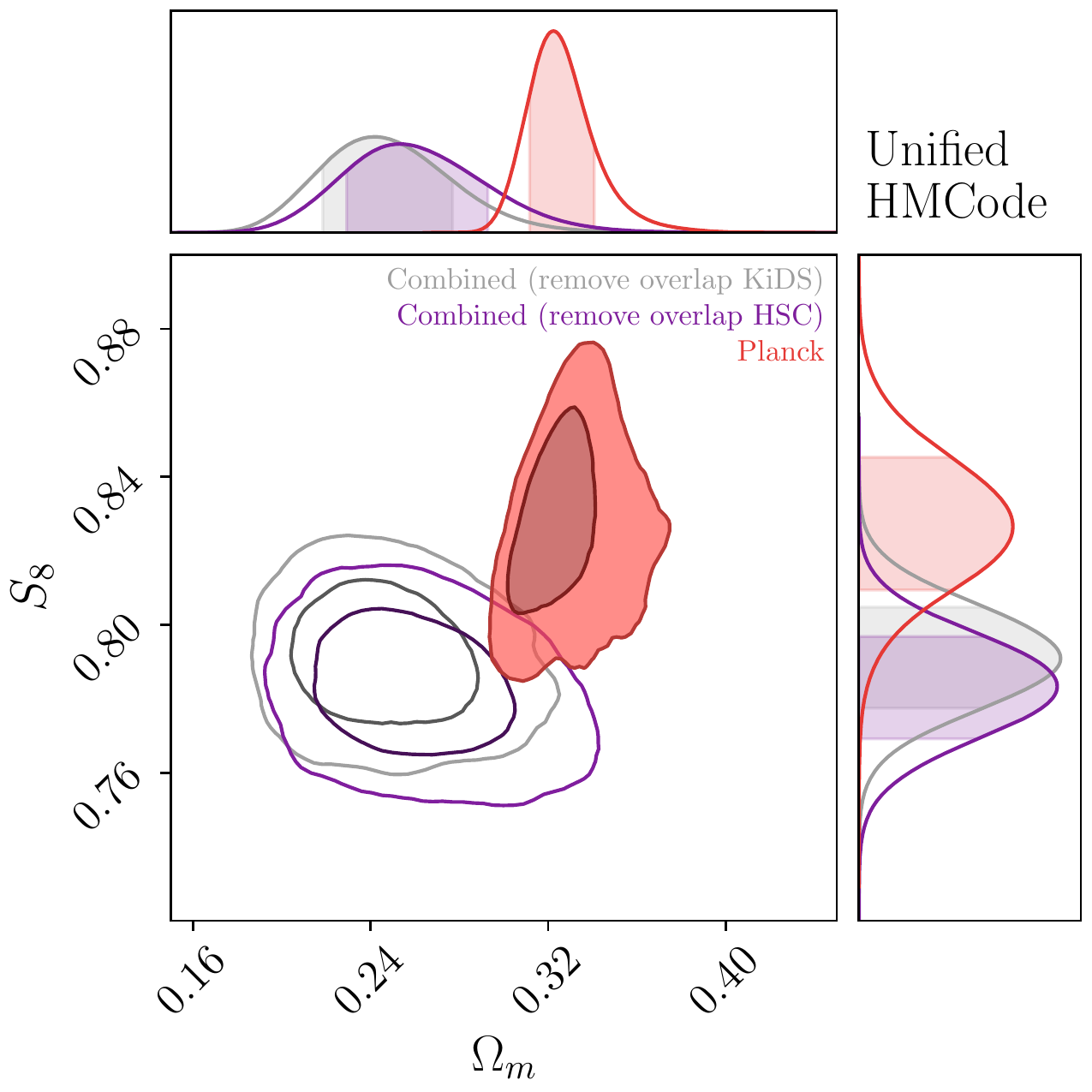}
	\caption{$S_8$-$\Omega_{\rm m}$ constraints from the combined constraints from DES-Y1, HSC-Y1 and KiDS-1000 assuming two different small-scale treatments and two scenarios for accounting for the covariance between the surveys. These constraints are compared with the primary CMB probes of {\it Planck} \citep{Planck2018}.} 
	\label{fig:planck_compare_small}
\end{figure}

\subsection{Combined Constraints and Comparison with {\it Planck}}

To combine the three datasets, we first need to consider the fact that the datasets are not fully independent. In particular, as shown in Figure~\ref{fig:footprints}, half of the HSC-Y1 footprint overlaps with the northern footprint of KiDS-1000. The actual covariance is also complicated by the fact that the surveys do not fully overlap in redshift -- HSC-Y1 is somewhat deeper than KiDS-1000. A full quantification of the covariance between HSC-Y1 and KiDS-1000 is beyond the scope of this paper. Rather, we take an approximate treatment to bracket the constraining power when combining the three datasets. We assume that the northern footprint of KiDS-1000 is fully covariant with $\sim$half of the HSC-Y1 footprint. We could treat this as effectively removing the statistical constraining power of either the northern part of the KiDS-1000 dataset or half of the HSC-Y1 dataset. In practice, we model this by enlarging either the KiDS-1000 or the HSC-Y1 covariance by the corresponding ratio of sky area between the full footprint and the partial footprint, then combining the three datasets assuming the two dataset are independent.  We take this approach with the aim of a simple method that should approximate the largest effect of the cross-covariance. To approximate the sky fraction we use \textsc{HEALPix}\footnote{\url{http://healpix.sourceforge.net}} \citep{Gorski2005, Zonca2019} with an $N_{\text{side}} = 4096$ to find a factor of $A_{\text{full}}/A_{\text{partial}} = 1.82$ for KiDS-1000 and $A_{\text{full}}/A_{\text{partial}} = 2.36$ for HSC. 

We note that this does not account for the actual covariance in the data vector and therefore is an approximation. This approximation could be mitigated by, for example, cutting out the overlapping sources between the two survey samples.  However, the resulting shear correlation function could be incorrectly calibrated by shear bias parameters that were computed for the entire survey sample.  The $n(z)$'s are also computed on the full sample, and would therefore need to be evaluated on the subsample. In addition, the priors on the shear uncertainty, redshift uncertainty and any additional survey-specific systematics have been computed for the entire sample.  Therefore they would need to be reevaluated if the cut sample was not fully representative of the original catalogs (a condition that might be met, but is not a priori expected due to potential variations in depth, seeing, etc. across the fields).  Nevertheless, it should give a reasonable estimate of the constraining power of the three datasets combined.

\begin{table*}
  \caption{When combining all three surveys under a unified analysis framework, the $S_{8}$ constraints and metrics for internal consistency. We quote $S_{8}$ in terms of the marginalized one-dimensional constraints mean and 68$\%$ confidence interval.} Same as Table~\ref{tab:constraints_unified}, the top and bottom half of the table shows two different cases of treatment of the small scales in the data vectors. The two columns assume two scenarios of the footprint overlap: first (second) column enlarges the HSC-Y1 (KiDS-1000) covariance when combining, approximating the scenario where we remove the overlapped HSC-Y1 (KiDS-1000) footprint.
  \label{tab:constraints2}
  \centering
  \begin{tabular}{l cc}
\hline 
Overlapped footprint &  & \\
 removed & HSC-Y1 & KiDS-1000 \\
\hline
$\Delta\chi^2$ cut & &\\
$S_{8}$&  $0.777^{+0.016}_{-0.017}$& $0.785^{+0.015}_{-0.015}$ \\
$\chi^{2}$ / D.O.F. & 667.87/(635-10.89) & 620.02/(635-11.46)\\
Reduced $\chi^{2}$ (p-value) & 0.91 (0.95) & 1.00 (0.53)\\
 \hline
\textsc{HMCode} &  & \\
$S_{8}$& $0.783^{+0.012}_{-0.012}$ & $0.791^{+0.013}_{-0.013}$ \\
$\chi^{2}$ / D.O.F. & 830.78/(985-11.73) &  962.34/(985-12.14)\\
Reduced $\chi^{2}$ (p-value) & 0.85 (0.99)  & 0.99 (0.59)\\
 \hline
\end{tabular}
\end{table*}

Figure~\ref{fig:planck_combined} shows, for both the $\Delta\chi^2$ cut (left) and the \textsc{HMCode} (right) approach, the combined $\Omega_{\rm m}$-$S_8$ constraints for the two cases of the covariance treatment. We show the results for each combined case in the full parameter space in Appendix~\ref{appendix:combined_full}.  We find fairly intuitive behaviors of the contours — the combined constraints are overall tighter and located where the three individual constraints overlap.  When enlarging the KiDS-1000 covariance, the relative contribution from KiDS-1000 becomes lower and the combined $S_8$ value is moved higher due the relatively higher contribution of HSC-Y1. On the contrary, when enlarging the HSC-Y1 covariance we get a lower $S_8$ value due to the higher relative contribution of DES-Y1 and KiDS-1000.  

Table~\ref{tab:constraints2} lists the constraint on $S_8$ for these different scenarios where we combine all three datasets, together with the goodness-of-fit for each case -- we find good goodness-of-fit values. Noticeably, we find that the combined constraints on $S_8$ are between 1.6 and 1.9\%. This is at a similar constraining level as the most constraining cosmic shear results today \citep{Amon2021,Secco2021}. This is expected as it combines the constraining power of three datasets that are already very tight individually. We caution the readers on a number of factors to consider when comparing these numbers with other cosmic shear analyses. First, as discussed above, there is an approximation in our treatment in the cross-covariance between two of three datasets. We have taken a rather conservative approach in assuming the overlapping footprint to be fully covariant. However, a full treatment of this cross-covariance could in principle yield different results. We refer the reader to Appendix~\ref{appendix:pairwise} which shows the pair-wise constraints from each survey combination.  Note, the pairwise constraints, KiDS-1000+DES-Y1 and HSC-Y1+DES-Y1, do not include the area overlap estimation as these datasets are independent. Second, the analyses from \citet{Amon2021,Secco2021} have both adopted more complex IA models \citep{blazek2019} and included additional priors on the IA parameters through a technique called lensing ratio \citep{y3-shearratio}, making it hard to directly compare. Finally, our scale cuts were evaluated with the constraining power from individual surveys instead of the combined, suggesting that they could be insufficient in the combined case.

Nonetheless, this analysis has given some intriguing insights.  In particular, we find that although at the individual survey level,  $\Omega_{\rm m}$ was not generally well constrained compared to the prior,  
the combined survey results in a 
fairly well-constrained $\Omega_{\rm m}$ relative to the prior.  The constraint generally trends to a lower value than the previous survey results, ranging at the lowest $\Omega_{\rm m}=0.248^{+0.023}_{-0.031}$ for the combined (remove KiDS-1000 overlap) \textsc{HMCode} case to highest with $\Omega_{\rm m}=0.304^{+0.030}_{-0.042}$ for the combined (remove KiDS-1000 overlap) $\Delta\chi^{2}$ cut case.

Taking the relatively conservative approach of the approximated constraints as brackets of the $S_{8}$ constraint from all three surveys, the $S_{8}$ means range $~0.777-0.791$.  This is slightly higher, albiet consistent, with the DES-Y3 cosmic shear results, which found for their fiducial analysis $S_{8} = 0.759^{+0.025}_{-0.023}$ and for optimized scales $S_{8} = 0.772^{+0.018}_{-0.017}$ \citep{Amon2021,Secco2021}.  Interestingly, our unified analysis still results in slightly lower $S_{8}$ values than Planck.    

Given the potential concerns with the combined constraints, we refrain from making quantitative statements of the consistency with {\it Planck}. We do show, however, the 2D contours of the different scenarios of our combined results compared with the primary CMB constraints from the \textit{Planck} 2018 TT, TE, EE+lowE likelihood \citep{Planck2018}. We note that with our choice of priors (Table~\ref{table:survey_priors}), and in particular the choice to allow the neutrino mass parameter to vary, results in wider posteriors in the $\Omega_{\rm m}$-$S_{8}$ plane than the published \textit{Planck} 2018 results. Visually, we can see the expected behaviour of the four cosmic shear contours, with the ``$\Delta\chi^2$ cut/overlap KiDS'' closest to {\it Planck} and the ``\textsc{HMCode}/overlap HSC'' case furthest from {\it Planck}. Taken at face value, the apparent difference compared to {\it Planck} is roughly similar (in terms of a by-eye comparison in the $\Omega_{\rm m}$ vs. $S_{8}$ plane) to the levels of tension seen in current datasets.  This relatively similar picture remains even though we have combined 3 independent datasets and the statistical uncertainties have decreased (approximately $2\%$ constraint on $S_{8}$).  That is, the results are slightly lower than the Planck values, but there is no evidence of a significant obvious tension any greater than the $\sim$2$\sigma$ difference previously observed for cosmic shear. Therefore, our results do not dramatically change our view of the potential tension between cosmic shear and the primary CMB.

\section{Discussion and Conclusion}
\label{sec:conclusion}

We perform a systematic reanalysis of three published cosmic shear analyses \citep{Troxel2017,Hamana2018,Asgari2021} -- we attempt to reproduce the published results, compare them assuming different analysis choices, and eventually combine them. The reanalysis and examination of consistency between the datasets are important as the community considers potential tensions in the $\Lambda$CDM model where the constraints on $S_8\equiv \sigma_8 \sqrt{\Omega_{\rm m}/0.3}$ in recent lensing surveys appear systematically low compared to that inferred from measurements of the primary CMB anisotropies.  Additionally, testing a unified framework to perform the cosmic shear measurements from the catalog level is a highly useful exercise in preparation for DESC's analysis with LSST. 

In this work we start with the weak lensing shear catalogs provided by the three surveys (DES, HSC, KiDS) and perform the measurement of the shear two-point function using tools developed by the LSST Dark Energy Science Collaboration (DESC). We then perform cosmological inference while systematically unifying priors on the cosmological parameters, the intrinsic alignment model, and the treatment of the small scales. Overall we are able to explain the changes in the cosmological constraints coming from these analysis choices, and demonstrate the importance of being transparent in these analysis choices and cautious when comparing across surveys. Our final unified analysis finds no evidence for tension between the three datasets. We highlight some interesting findings in different parts of this unified analysis:
\begin{itemize}
    \item Different choices in priors at the level of the differences between the various surveys’ choices can result in shifts in less well constrained parameters such as $\Omega_{\rm m}$, while $S_8$ remains very robust.  This information is practical to consider, when interpreting for example the future constraints from LSST.  The change in $\Omega_{\rm m}$ we see is primarily coming from the priors on $A_s$ which is degenerate with $\Omega_{\rm m}$. We also find that our results are sensitive to whether we sample $A_s$ in linear or logarithmic space. 
    \item Changing from a more conservative treatment of the small scales (removing small scales based on a $\Delta\chi^2$ cut) to a more aggressive one (modeling the small scales using \textsc{HMCode} and marginalizing over the model parameters), the constraint on $S_8$ becomes up to $30\%$ tighter. There is the most statistical gain for KiDS-1000 compared to DES-Y1 and HSC-Y1. This could be related to the small-scale covariance matrix in DES-Y1 and HSC-Y1, which were not validated on the small scales.
    \item When unifying all analysis choices, we find for the $\Delta\chi^2$ cut approach, $S_8 = 0.754^{+0.031}_{-0.024}$ for DES-Y1, $S_8 = 0.798^{+0.026}_{-0.024}$ for HSC-Y1, and $S_8 = 0.751^{+0.029}_{-0.024}$ for KiDS-1000. HSC-Y1 both has the largest best-fit $S_8$ value and the tightest constraint.
    \item When unifying all analysis choices for the \textsc{HMCode} case, we find $S_8 = 0.757^{+0.027}_{-0.021}$ for DES-Y1, $S_8 = 0.812^{+0.021}_{-0.021}$ for HSC-Y1, and $S_8 = 0.761^{+0.021}_{-0.019}$ for KiDS-1000. HSC-Y1 both has the largest best-fit $S_8$ value, and KiDS-1000 has the tightest constraint.
    \item We examine the consistency between all pairs out of the three datasets with three consistency metrics and find no evidence for disagreement. The largest inconsistency is between KiDS-1000 and HSC-Y1, when using the \textsc{HMCode}, at $\sim$1.7$\sigma$.
\end{itemize}

Due to the complication of overlapping footprints, we take an approximate approach to combine the three datasets. We examine two scenarios for the small-scale treatment and two scenarios for accounting for the cross-survey covariance. With the ``$\Delta\chi^{2}$ cut'' scenario, we find $S_8=0.777^{+0.016}_{-0.017}$ ($S_8=0.785^{+0.015}_{-0.015}$) if we effectively remove part of the HSC-Y1 (KiDS-1000) footprint that overlaps with KiDS-1000 (HSC-Y1). With the \textsc{HMCode} scenario, we find $S_8=0.783^{+0.012}_{-0.012}$ ($S_8=0.791^{+0.013}_{-0.013}$) if we effectively remove part of the HSC-Y1 (KiDS-1000) footprint that overlaps with KiDS-1000 (HSC-Y1).  The combined constraints shift by $\sim$0.3$\sigma$ each depending on the covariance method.  Interestingly, we also do not find a large shift in the constraints between the two small-scale treatments, suggesting the small-scale model used in \textsc{HMCode} is not significantly different from what is in the data.    

We caution the reader that these results contain several simplifications.  Given the uncertainty in the combined results due to the overlapping footprint, we only perform a qualitative comparison of the combined results with the constraints from the primary CMB as measured by {\it Planck}. Roughly, the combined constraint result is fairly consistent with the current picture of the ``$S_8$ tension'' with individual datasets, though now the combined power of three datasets.

In addition to the comparison between the three datasets, this work also demonstrated that the DESC software package \textsc{TXPipe} can now be applied to Stage-III data. This is a crucial milestone in preparation for the LSST with Rubin Observatory. The statistical power from the LSST dataset will be unprecedented, and will allow us to test, to an extremely high precision, the validity of the $\Lambda$CDM model. As we prepare for the arrival of LSST data, a thorough understanding of Stage-III results will pave the way for a successful Stage-IV dark energy program. This work represents a continued effort in re-examining, digesting, and understanding the results from Stage-III cosmic shear surveys. 

\section{Data Availability}
The \textsc{TXPipe} data products dervied and used in this project are available at \url{https://zenodo.org/record/6983861#.YvV2KS1h1pQ}.  

\section{Acknowledgements} 
This paper has undergone internal review by the LSST Dark Energy Science Collaboration, and we kindly thank the reviewers Rachel Mandelbaum, Tilman Tr{\"o}ster and Michael Troxel. We thank Catherine Heymans and Ben Giblin for assistance with the KiDS-1000 weak-lensing catalogs and reanalysis.  We thank Takashi Hamana for assistance with the HSC-Y1 analysis and providing the full-scale HSC-Y1 covariance.

EPL and CW were supported by Department of Energy, grant DE-SC0010007. CC was supported by DOE grant DE-SC0021949. 

RM acknowledges the support of the Department of Energy grant DE-SC0010118.  HM was supported in part by MEXT/JSPS KAKENHI Grant Number JP20H01932.  MESP is funded by the Deutsche Forschungsgemeinschaft (DFG, German Research Foundation) under Germany’s Excellence Strategy – EXC 2121 „Quantum Universe“ – 390833306.  TT acknowledges support from the Leverhulme Trust.  AHW is supported by an European Research Council Consolidator Grant (No. 770935).

The DESC acknowledges ongoing support from the Institut National de 
Physique Nucl\'eaire et de Physique des Particules in France; the 
Science \& Technology Facilities Council in the United Kingdom; and the
Department of Energy, the National Science Foundation, and the LSST 
Corporation in the United States.  DESC uses resources of the IN2P3 
Computing Center (CC-IN2P3--Lyon/Villeurbanne - France) funded by the 
Centre National de la Recherche Scientifique; the National Energy 
Research Scientific Computing Center, a DOE Office of Science User 
Facility supported by the Office of Science of the U.S.\ Department of
Energy under Contract No.\ DE-AC02-05CH11231; STFC DiRAC HPC Facilities, 
funded by UK BEIS National E-infrastructure capital grants; and the UK 
particle physics grid, supported by the GridPP Collaboration.  This 
work was performed in part under DOE Contract DE-AC02-76SF00515.

We finally wish to acknowledge the data sources for each of the surveys used in this paper:
\\
\textbf{DES-Y1:}
This project used public archival data from the Dark Energy Survey (DES). Funding for the DES Projects has been provided by the U.S. Department of Energy, the U.S. National Science Foundation, the Ministry of Science and Education of Spain, the Science and Technology FacilitiesCouncil of the United Kingdom, the Higher Education Funding Council for England, the National Center for Supercomputing Applications at the University of Illinois at Urbana-Champaign, the Kavli Institute of Cosmological Physics at the University of Chicago, the Center for Cosmology and Astro-Particle Physics at the Ohio State University, the Mitchell Institute for Fundamental Physics and Astronomy at Texas A\&M University, Financiadora de Estudos e Projetos, Funda{\c c}{\~a}o Carlos Chagas Filho de Amparo {\`a} Pesquisa do Estado do Rio de Janeiro, Conselho Nacional de Desenvolvimento Cient{\'i}fico e Tecnol{\'o}gico and the Minist{\'e}rio da Ci{\^e}ncia, Tecnologia e Inova{\c c}{\~a}o, the Deutsche Forschungsgemeinschaft, and the Collaborating Institutions in the Dark Energy Survey.
The Collaborating Institutions are Argonne National Laboratory, the University of California at Santa Cruz, the University of Cambridge, Centro de Investigaciones Energ{\'e}ticas, Medioambientales y Tecnol{\'o}gicas-Madrid, the University of Chicago, University College London, the DES-Brazil Consortium, the University of Edinburgh, the Eidgen{\"o}ssische Technische Hochschule (ETH) Z{\"u}rich,  Fermi National Accelerator Laboratory, the University of Illinois at Urbana-Champaign, the Institut de Ci{\`e}ncies de l'Espai (IEEC/CSIC), the Institut de F{\'i}sica d'Altes Energies, Lawrence Berkeley National Laboratory, the Ludwig-Maximilians Universit{\"a}t M{\"u}nchen and the associated Excellence Cluster Universe, the University of Michigan, the National Optical Astronomy Observatory, the University of Nottingham, The Ohio State University, the OzDES Membership Consortium, the University of Pennsylvania, the University of Portsmouth, SLAC National Accelerator Laboratory, Stanford University, the University of Sussex, and Texas A\&M University.
Based in part on observations at Cerro Tololo Inter-American Observatory, National Optical Astronomy Observatory, which is operated by the Association of Universities for Research in Astronomy (AURA) under a cooperative agreement with the National Science Foundation.
\\
\textbf{HSC-Y1:}
The Hyper Suprime-Cam (HSC) collaboration includes the astronomical communities of Japan and Taiwan, and Princeton University. The HSC instrumentation and software were developed by the National Astronomical Observatory of Japan (NAOJ), the Kavli Institute for the Physics and Mathematics of the Universe (Kavli IPMU), the University of Tokyo, the High Energy Accelerator Research Organization (KEK), the Academia Sinica Institute for Astronomy and Astrophysics in Taiwan (ASIAA), and Princeton University. Funding was contributed by the FIRST program from the Japanese Cabinet Office, the Ministry of Education, Culture, Sports, Science and Technology (MEXT), the Japan Society for the Promotion of Science (JSPS), Japan Science and Technology Agency (JST), the Toray Science Foundation, NAOJ, Kavli IPMU, KEK, ASIAA, and Princeton University. 

This paper makes use of software developed for Vera C. Rubin Observatory. We thank the Rubin Observatory for making their code available as free software at http://pipelines.lsst.io/.

This paper is based on data collected at the Subaru Telescope and retrieved from the HSC data archive system, which is operated by the Subaru Telescope and Astronomy Data Center (ADC) at NAOJ. Data analysis was in part carried out with the cooperation of Center for Computational Astrophysics (CfCA), NAOJ. We are honored and grateful for the opportunity of observing the Universe from Maunakea, which has the cultural, historical and natural significance in Hawaii. 
\\
\textbf{KiDS-1000}
Based on observations made with ESO Telescopes at the La Silla Paranal Observatory under programme IDs 177.A-3016, 177.A-3017, 177.A-3018 and 179.A-2004, and on data products produced by the KiDS consortium. The KiDS production team acknowledges support from: Deutsche Forschungsgemeinschaft, ERC, NOVA and NWO-M grants; Target; the University of Padova, and the University Federico II (Naples).

The contributions from the primary authors are as follows. E.P.L. worked out the reproduction of the published results and implemented the analysis choice testing and combined results.  C.C. conceived of the project, guided the reanalysis and combined analysis, and contributed to the technical implementation. E.P.L. and C.C. wrote the paper with input from all authors.  C.W. guided the reanalysis efforts and provided ideas for the unified and combined analysis.  J.Z. is the developer of \textsc{TXPipe} and \textsc{CosmoSIS} which is the primary software used in this analysis.

\bibliographystyle{mnras}
\bibliography{sample.bib}

\appendix

\section{Angular Scale Cuts}
In Tables~\ref{table:published_scale_cuts} and \ref{table:chi2_scale_cuts} we compare the angular scale cuts for the survey choices and the $\Delta\chi^{2}$ scale cuts.  The $\Delta\chi^{2}$ scale cuts results in the usage of more small scales than the survey choices for DES-Y1 and HSC-Y1, and less scales for KiDS-1000.  
\label{appendix:scale_cuts}

 \begin{table*}
 \caption{Published Angular Scale Cuts: The angular scale cuts defined by the previous surveys.  DES-Y1 used a threshold cut of $2\%$ baryon contamination based on the OWLS simulation, HSC-Y1 used a similar approach but adopted a $5\%$ cut and fixed the threshold for $\xi_{+}$ and $\xi_{-}$.  They additionally cut out higher scales that are impacted by PSF systematics.  KiDS-1000 modeled the nonlinear power spectrum with HMCode and went to smaller scales in their analysis.  The resulting data vector lengths are 227, 225 and 170 for DES-Y1, HSC-Y1 and KiDS-1000 respectively.}
 \label{table:published_scale_cuts}
 \centering
 \begin{tabular}{||l l l l||}
 \hline 
 zbin & DES-Y1 & HSC-Y1 & KiDS-1000 \\ 
 \hline
  \\ $\xi_{+}$ / $\xi_{-}$ &  &   & \\
 (1, 1) \\ & [7.2, 250.0] / [90.6, 250.0]  & [7.0, 56.0] / [28.0, 178.0]  & [0.5, 300.0] / [4.0, 300.0]\\
 (1, 2) \\ & [7.2, 250.0] / [71.9, 250.0] & [7.0, 56.0] / [28.0, 178.0]   & [0.5, 300.0] / [4.0, 300.0]\\
 (1, 3) \\ & [5.7, 250.0] / [71.9, 250.0] & [7.0, 56.0] / [28.0, 178.0]   & [0.5, 300.0] / [4.0, 300.0]\\
 (1, 4) \\ & [5.7, 250.0] / [71.9, 250.0] & [7.0, 56.0] / [28.0, 178.0]   & [0.5, 300.0] / [4.0, 300.0]\\
 (1, 5) \\ & -- / -- & -- / --  & [0.5, 300.0] / [4.0, 300.0]\\
 (2, 2) \\ & [4.5, 250.0] /  [57.2, 250.0] & [7.0, 56.0] / [28.0, 178.0]    & [0.5, 300.0] / [4.0, 300.0]\\
 (2, 3) \\ & [4.5, 250.0] /  [57.2, 250.0] & [7.0, 56.0] / [28.0, 178.0]   & [0.5, 300.0] / [4.0, 300.0]\\
 (2, 4) \\ & [4.5, 250.0] / [45.4, 250.0] & [7.0, 56.0] / [28.0, 178.0]   & [0.5, 300.0] / [4.0, 300.0]\\
 (2, 5) \\ & -- / -- & -- / --  & [0.5, 300.0] / [4.0, 300.0]\\
 (3, 3) \\ & [3.6, 250.0] / [45.4, 250.0] & [7.0, 56.0] / [28.0, 178.0]   & [0.5, 300.0] / [4.0, 300.0]\\
 (3, 4) \\ & [3.6, 250.0] / [45.4, 250.0] & [7.0, 56.0] / [28.0, 178.0]   & [0.5, 300.0] / [4.0, 300.0]\\
 (3, 5) \\ & -- / -- & -- / --  & [0.5, 300.0] / [4.0, 300.0]\\
 (4, 4) \\ & [3.6, 250.0] / [36.1, 250.0] & [7.0, 56.0] / [28.0, 178.0]   & [0.5, 300.0] / [4.0, 300.0]\\
 (4, 5) \\ & -- / -- &  -- / -- & [0.5, 300.0] / [4.0, 300.0]\\
 (5, 5) \\ & -- / -- & -- / --  & [0.5, 300.0] / [4.0, 300.0]\\
 \hline
 \end{tabular}
 \end{table*}

 \begin{table*}
 \caption{$\Delta\chi^{2}$ Unified Angular Scale Cuts: The angular scale cuts for the $\Delta\chi^{2}$ case, in which we cut out scales with a $\Delta\chi^{2} < 0.5$ between theoretical data vectors with and without baryon contamination, using the OWLS simulation.  The published covariance matrices were used for calculating the $\Delta\chi^{2}$. The $\Delta\chi^{2}$ scale cuts results in the usage of more small scales than the survey choices for DES-Y1 and HSC-Y1, and less scales for KiDS-1000.  The resulting data vector lengths are 258, 205 and 164 for DES-Y1, HSC-Y1 and KiDS-1000 respectively.  Units are in arcmin.  DES-Y1, HSC-Y1 used four tomographic bins, and KiDS-1000 used five tomographic bins.  Both KiDS-1000 and HSC adopt cuts that are uniform for each $\xi_{+}$ and $\xi_{-}$, whereas DES-Y1 varies per bin based on a $2\%$ cut from baryon contamination.  For the \textsc{HMCode} cut we adopt the KiDS-1000 fiducial choice of $0.5\arcsec$ for $\xi_{+}$ and $4.0\arcsec$ for $\xi_{-}$.}
 \label{table:chi2_scale_cuts}
 \centering
 \begin{tabular}{||l l l l||}
 \hline 
 zbin & DES-Y1 & HSC-Y1 & KiDS-1000 \\ 
 \hline
  z-bin \\ $\xi_{+}$ / $\xi_{-}$ &  &   & \\
 (1, 1) \\ & [3.5, 250.0] / [35.6,  250.0]  & [3.1, 56.0] / [20.11, 178.0] & [0.7,  300.0] / [0.7,  300.0] \\
 (1, 2) \\ & [5.6, 250.0] / [44.8,  250.0]  & [4.0, 56.0] / [31.8, 178.0]  & [0.7, 300.0] / [0.7, 300.0] \\
 (1, 3) \\ & [5.6, 250.0] / [56.4,  250.0] & [4.0, 56.0] / [31.8, 178.0]   & [3.1,  300.0] / [13.2,  300.0] \\
 (1, 4) \\ & [4.4, 250.0] / [44.8, 250.0]  & [3.1, 56.0] / [25.3, 178.0]  &  [3.1, 300.0] / [13.2,  300.0]  \\
 (1, 5) \\ & -- / -- & -- / --  & [3.1, 300.0] / [13.2,  300.0]\\
 (2, 2) \\ & [4.4, 250.0] / [35.6,  250.0]   & [4.0, 56.0] / [31.8, 178.0]    & [3.1,  300.0] / [26.9, 300.0]\\
 (2, 3) \\ & [5.6, 250.0] / [56.4,  250.0]  & [4.0, 56.0] / [31.8, 178.0]   & [6.5, 300.0] / [54.7, 300.0] \\
 (2, 4) \\ & [4.4, 250.0] / [44.8,  250.0] & [3.1, 56.0] / [31.8, 178.0]   & [6.5, 300.0] / [54.7,  300.0]  \\
 (2, 5) \\ & -- / -- & -- / --  & [6.5,  300.0] / [54.7,  300.0] \\
 (3, 3) \\ & [4.4, 250.0] / [56.4,  250.0] & [3.1, 56.0] /  [31.8, 178.0] & [6.5,  300.0] / [54.7, 300.0]\\
 (3, 4) \\ & [4.4, 250.0] / [56.4,  250.0]  & [3.1, 56.0] / [25.3, 178.0]   & [6.5, 300.0] / [54.7,  300.0] \\
 (3, 5) \\ & -- / -- & -- / --  & [6.5,  300.0] / [54.7,  300.0 \\
 (4, 4) \\ & [3.5, 250.0] / [35.6,  250.0] & [2.5, 56.0] / [20.11, 178.0]   & [6.5,  300.0] / [54.7,  300.0] \\
 (4, 5) \\ & -- / -- &  -- / -- & [6.5, 300.0] / [54.7,  300.0] \\
 (5, 5) \\ & -- / -- & -- / --  & [6.5,  300.0] / [54.7,  300.0] \\
 \hline
 \end{tabular}
 \end{table*}

\section{Full parameter space for combined constraints} 
\label{appendix:combined_full}

\begin{figure*}
	\includegraphics[width=0.6\textwidth]{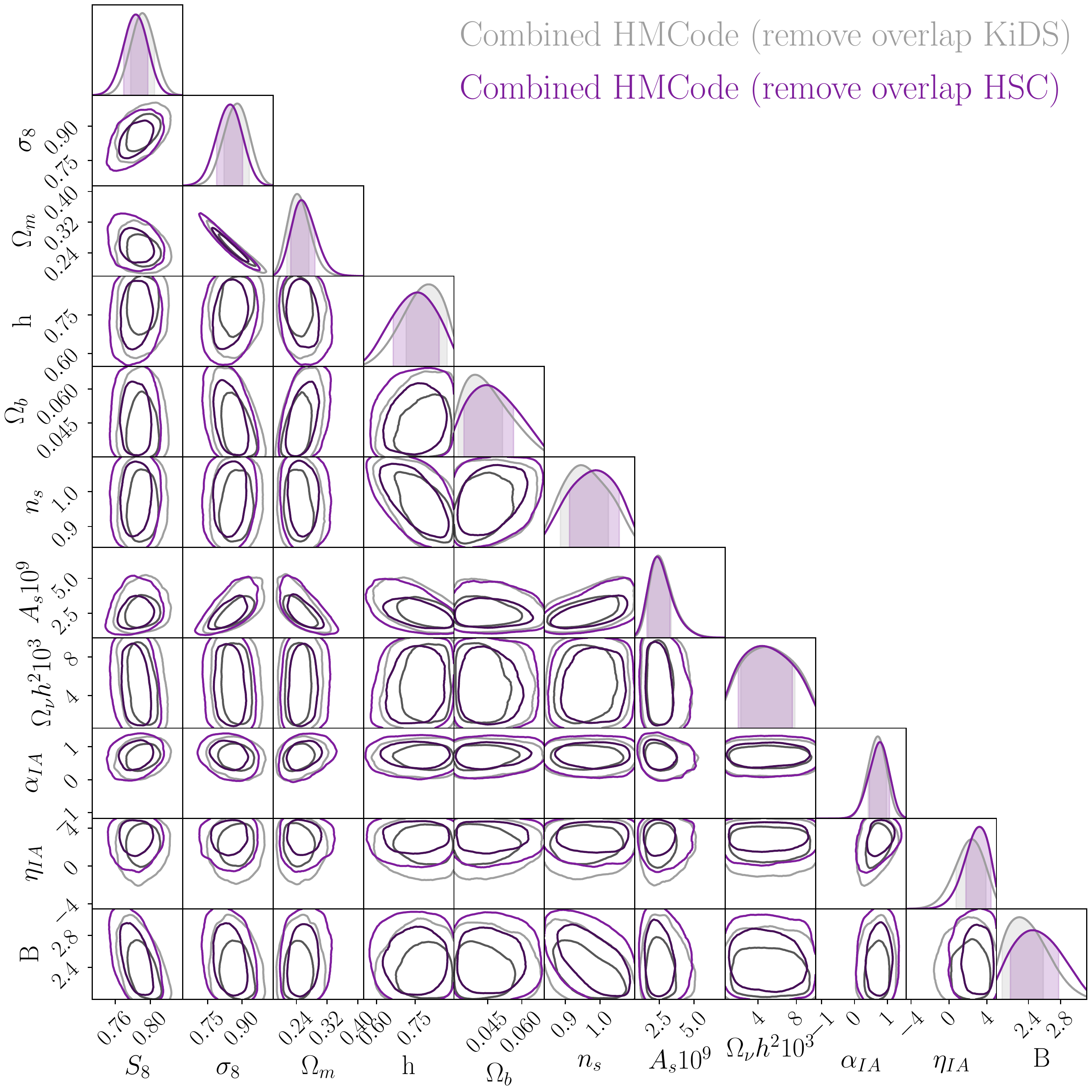}
	\caption{Posterior constraints for the full parameter space in the unified analyses; $\Delta\chi^{2}$ and HMCode case, and assuming fully overlapped footprint between the North half of KiDS-1000 and part of HSC-Y1. These correspond to the contours in Figure~\ref{fig:planck_compare_small}.}
    \label{fig:combinedsmallallpars}
\end{figure*}

In Figure~\ref{fig:combinedsmallallpars} we show the full parameter space in the unified analyses HMCode case, and assuming fully overlapped footprint between the North half of KiDS-1000 and part of HSC-Y1, corresponding to the contours in Figure~\ref{fig:planck_compare}. In general, we do not see any surprising discrepancies in the other parameters between the two approaches of the small-scale treatment.  We have similarly examined the full parameter space of the $\Delta\chi^{2}$ cut scenario and find similar results.

\section{Unified Intrinsic Alignment KiDS-1000} 
\label{appendix:IA}

\begin{figure}
	\includegraphics[width=0.49\textwidth]{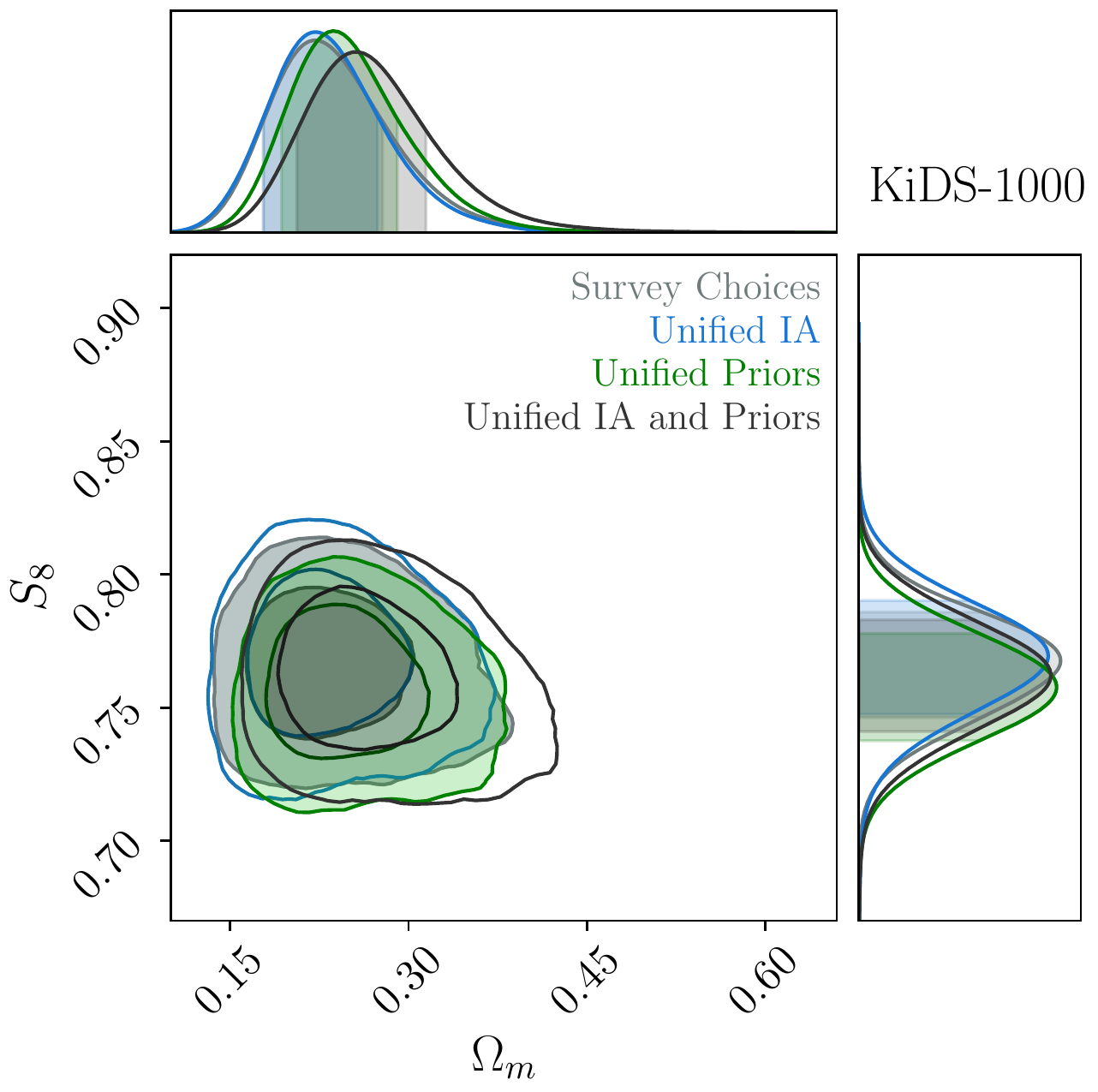}
	\caption{KiDS-1000's original fiducial analysis did not use the z-dependent power law term.  Here we isolate the results from just unifying the intrinsic alignment model, blue, and in green, the results from just unifying the priors.  The black results are the identical contour shown in \ref{fig:unified_priors}. This test demonstrates that the small shift is primarily driven by the change in priors, rather than the change in the IA model.}
    \label{fig:kids_IA_appendix}
\end{figure}

In this analysis we assess the change in the constraints for KiDS-1000 when unifying the cosmological priors and the IA modeling.  In Figure~\ref{fig:kids_IA_appendix} we show the effect of just changing one of these choices on the KiDS-1000 constraints.  We find a small shift upwards in $S_{8}$ of $\sim$0.15$\sigma$ when changing to a redshift dependent IA model.  There is a $\sim$0.28$\sigma$ downwards shift with the unified priors adopted. We find unifying the IA and prior choice gives an $S_{8}$ constraint that is lower than the survey choices by $\sim$0.13$\sigma$.

\section{Intrinsic Alignment Combined Constraint} 
\label{appendix:IA_combined}

\begin{figure}
	\includegraphics[width=0.49\textwidth]{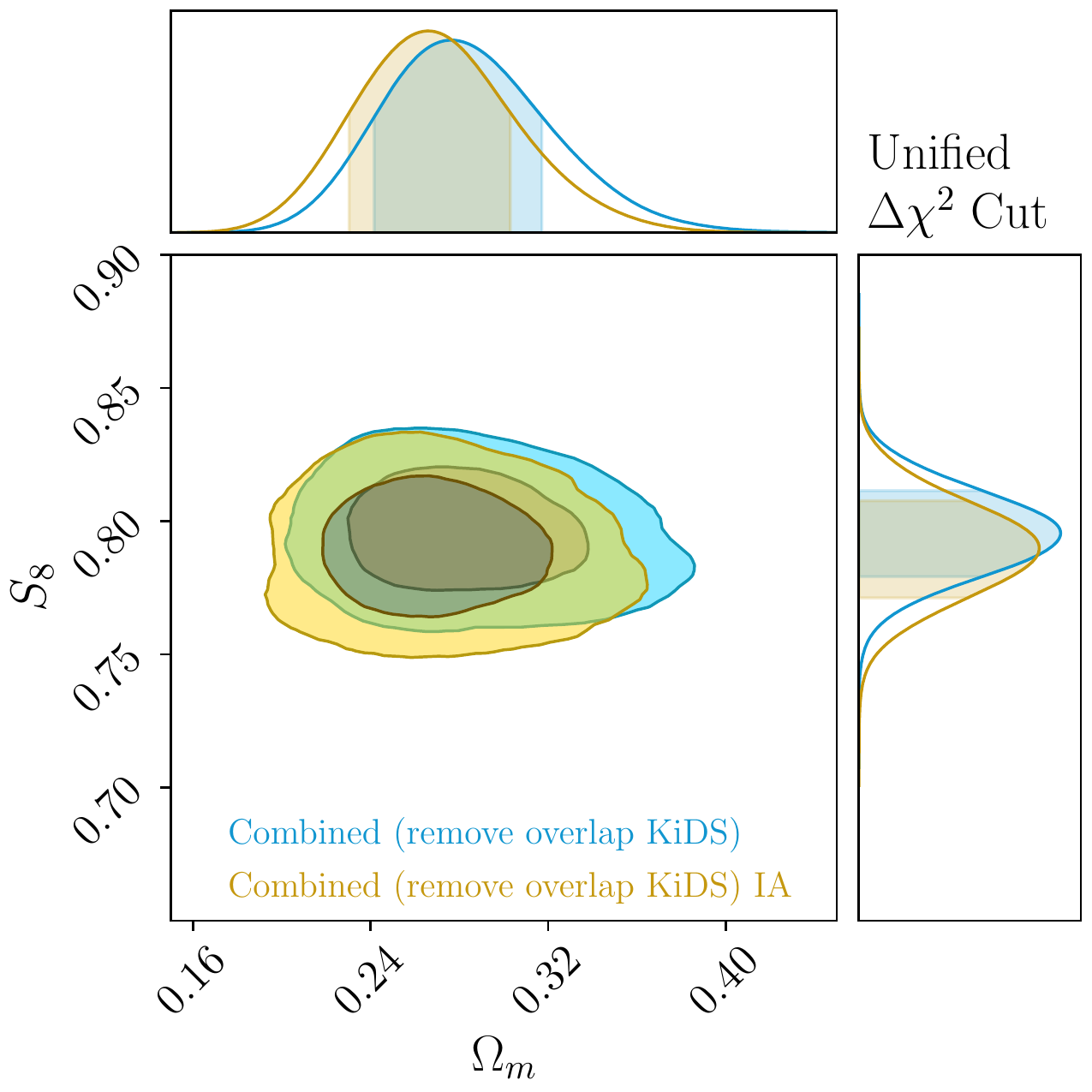}
	\caption{Here we show the results of the combined constraint when using combined IA parameters (blue) and when using different IA parameters for each survey (yellow).  We do not find a significant shift in the constraints, but do find slightly wider uncertainties.}
    \label{fig:IA_combined}
\end{figure}

 In Figure~\ref{fig:IA_combined} we show a comparison between the combined constraint when using combined IA parameters for each of the surveys, and when using different IA parameters for each survey.  We do not find a significant shift in the constraint ($<0.2\sigma$ for both $S_{8}$ and $\Omega_{\rm m}$) but find a slight increase in the uncertainty in both $S_{8}$ and $\Omega_{\rm m}$, (about $15\%$ and $5\%$ respectively) due to marginalizing over additional systematic parameters.  

\section{Linear and logarithmic $A_s$ priors} 
\label{sec:sample_as}

\begin{figure}
	\includegraphics[width=0.49\textwidth]{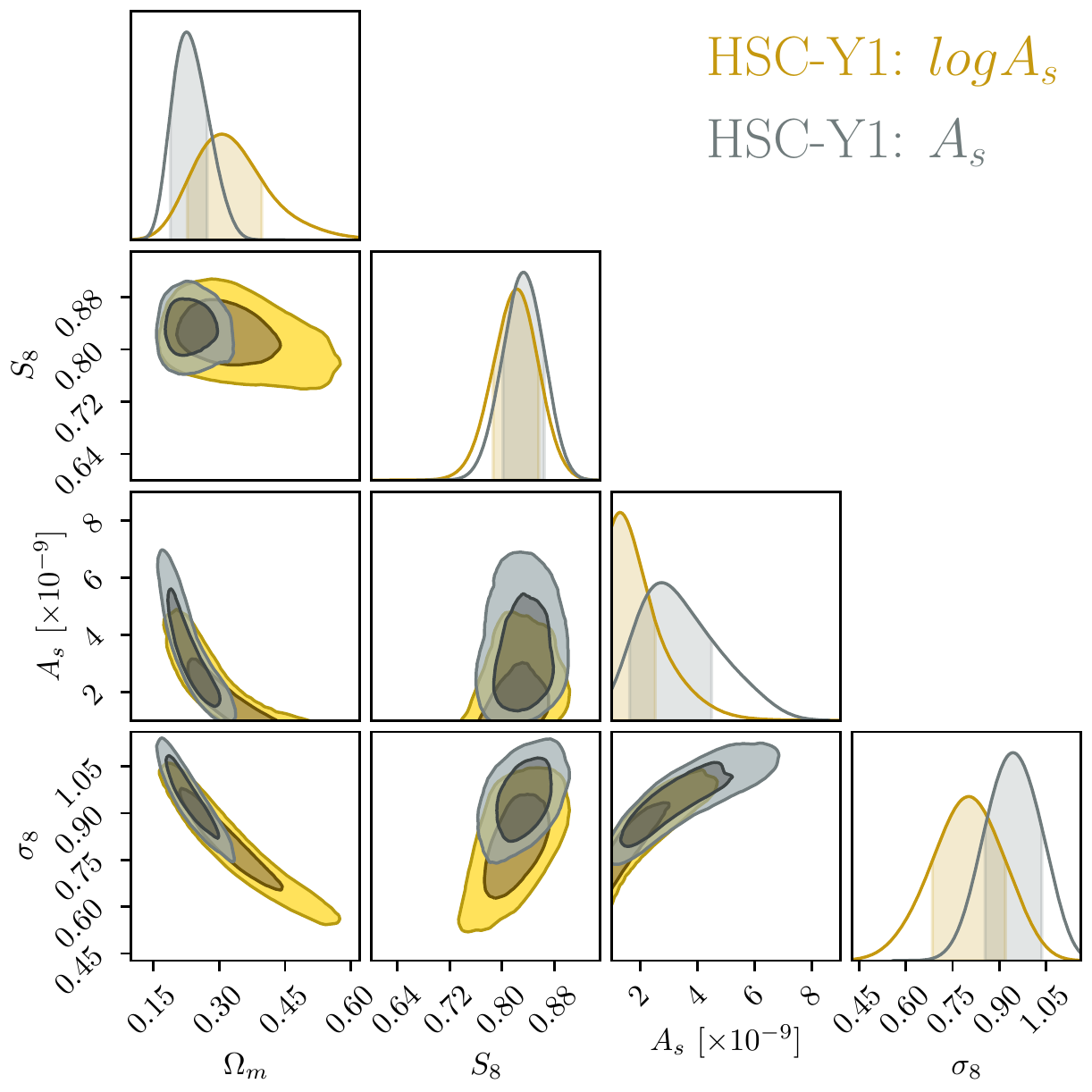}
	\caption{Posterior constraints for $S_{8}$, $A_{s}$, $\sigma_{8}$ and $\Omega_{\rm m}$ for HSC either sampling in $\mathrm{log}A_{s}$ or linear $A_{s}$ with the same upper and lower bounds.  The $\Omega_{\rm m}$ constraint in particular is heavily effected by this choice, with significantly tighter constraints dictated by the prior in the case of $A_{s}$ sampling.}
    \label{fig:log_lin_As}
\end{figure}

Previous analyses have differed in the sampling choice regarded $A_{s}$.  The prior chosen for the unified analysis corresponds to the widest range between the surveys, which is the fiducial choice for HSC-Y1.  Like their analysis we sample in $\log_{10}A_{s}$. DES-Y1, in comparison, sampled linearly in $A_{s}$. To test the effect of this choice we looked at the HSC-Y1 posteriors with identical range of $A_{s}$ sampling in logarithmic and linear space.  The results are shown in Figure~\ref{fig:log_lin_As}. 

\begin{figure}
	\includegraphics[width=0.49\textwidth]{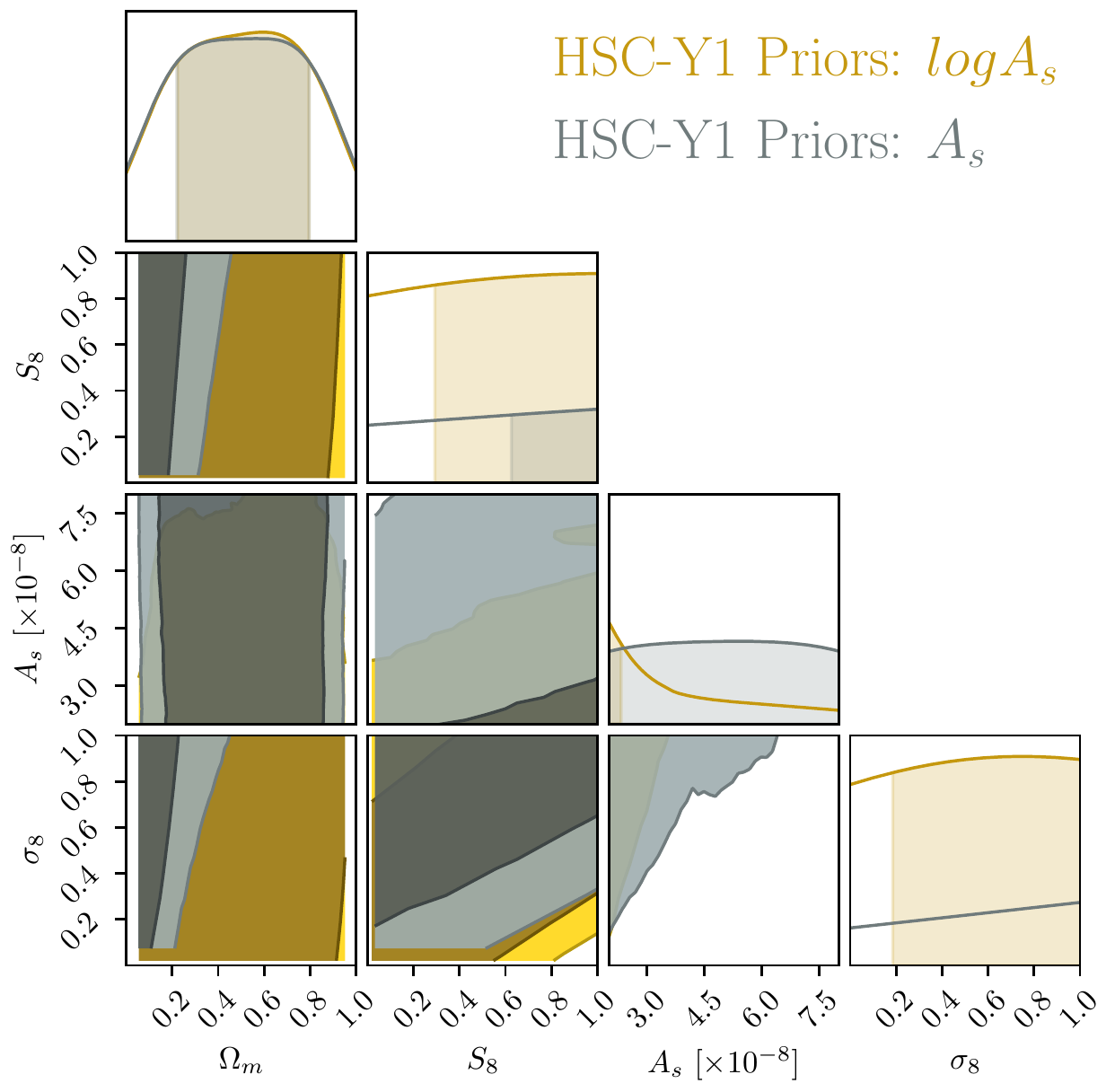}
	\caption{Priors in the full parameter space for HSC either sampling in $\textrm{logA}_{s}$ or linear $A_{s}$.  The primary parameters of interest in our analysis are $\Omega_{\rm m}$ and $S_{8}$.  The $\textrm{logA}_{s}$ parameterization gives the flattest priors for these parameters, so we adopt this prior in our unified analysis.}
    \label{fig:log_lin_As_priors}
\end{figure}

We find that $S_{8}$ is not sensitive to the choice of logarithmic versus linear sampling in for $A_{s}$, however other parameters are sensitive to the choice, in particular $\sigma_{8}$, $A_{s}$ and $\Omega_{\rm m}$.  We chose to keep the choice of $\mathrm{log}A_{s}$ for the unified analysis as it corresponds to the flattest prior in $S_{8}$, as shown in figure \ref{fig:log_lin_As_priors} which plots the priors for each parameter in each sampling space. In \citet{Sugiyama2020}, the authors explored an approach of reweighting the chains to achieve flatter priors in the parameters of interest ($\sigma_8$ in that work). KiDS-1000 avoids this issue in their analysis by sampling $S_{8}$ directly \citep{Asgari2021}, but found in \citet{Troester2021} that the choice of $A_{s}$ similarly does not affect the $S_{8}$ constraint.  Interestingly, the combined results, which yield a much tighter constraints on $\Omega_{m}$ compared to the prior, do not seem to be largely effected by the $A_{s}$ versus $\mathrm{log}A_{s}$ prior choice.

\section{Pairwise Constraints}
\label{appendix:pairwise}
In Figure~\ref{fig:pairwise_constraints} we show the $S_8$-$\Omega_{\rm m}$ constraints for each pairwise combination for the surveys, for each of our two small-scale treatments, compared with the primary CMB probes of {\it Planck}.  We have assumed independence between the surveys, which as mentioned in Section~\ref{sec:conclusion} is an approximation for the HSC-Y1 and KiDS-1000 combinations and we would like to emphasize the same caution as the results in Section~\ref{sec:conclusion} in the interpretation of the results, in particular refraining from quantitatively assessing tension with CMB results.  Similarly to the full combined constraints, we do not see a large shift between the two small scale treatments.  We find a slightly higher $S_{8}$ value for the DES-Y1/HSC-Y1 and HSC-Y1/KiDS-1000 constraints and slightly lower value when combining DES-Y1/KiDS-1000.

\begin{table}
  \caption{The pairwise parameter constraints for the $\Delta\chi^{2}$ cut case and \textsc{HMCode} case.}
  \label{fig:pairwise_constraints}
  \centering
  \begin{tabular}{l c c c}
\hline 
Dataset 1 & DES-Y1 & DES-Y1 & KiDS1000 \\
Dataset 2 & HSC-Y1 & KiDS-1000 & HSC-Y1 \\
\hline
$\Delta\chi^2$ cut & & \\
$S_{8}$ & $0.793^{+0.017}_{-0.017}$ & $0.764^{+0.018}_{-0.018}$ & $0.784^{+0.018}_{-0.018}$ \\ 
\hline
\textsc{HMCode} & & \\
$S_{8}$ & $0.801^{+0.015}_{-0.015}$ & $0.770^{+0.015}_{-0.014}$ & $0.789^{+0.015}_{-0.015}$ \\ 
 \hline
\end{tabular}
\end{table}

\begin{figure*}
	\centering
 	\includegraphics[width=0.95\columnwidth]{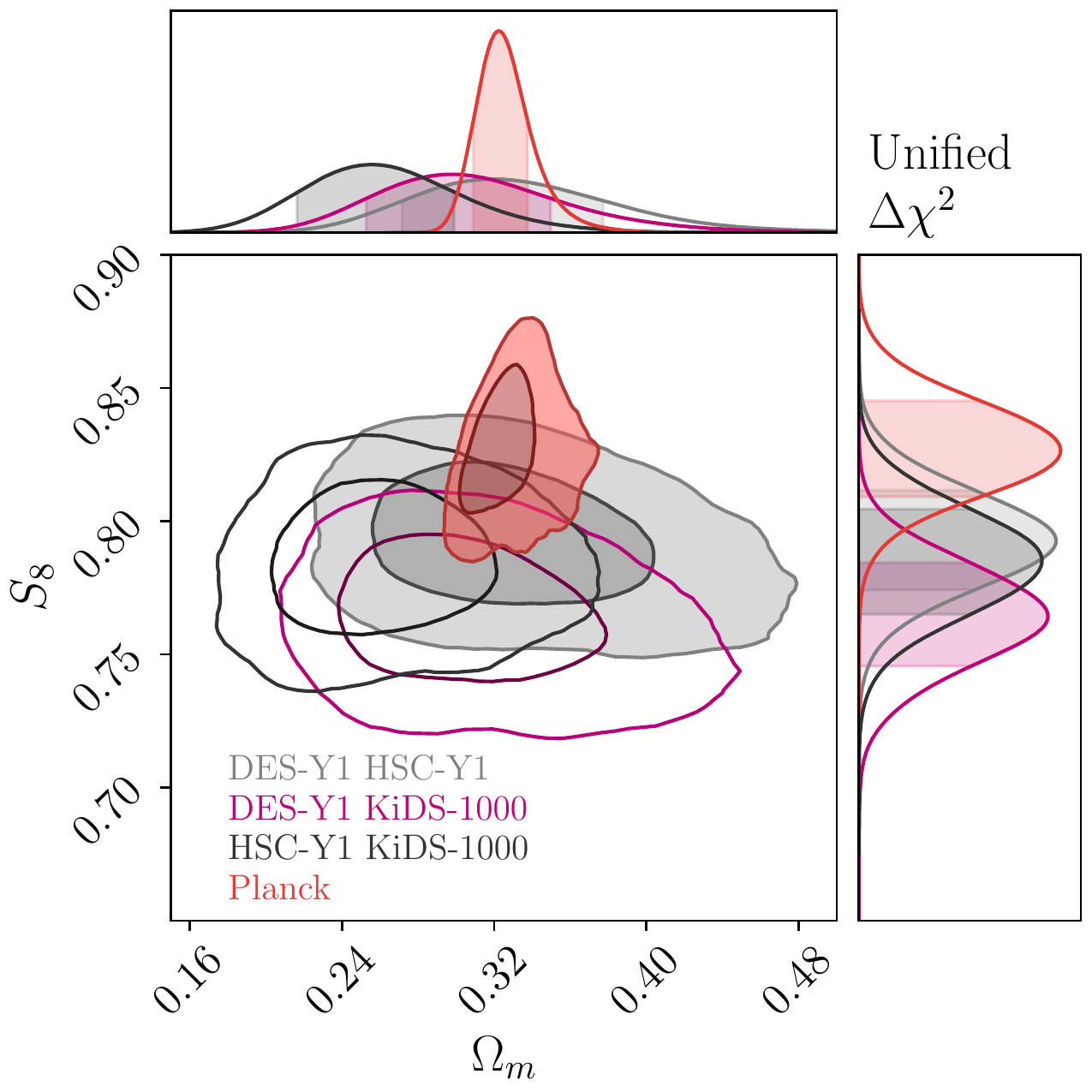}
	\includegraphics[width=0.95\columnwidth]{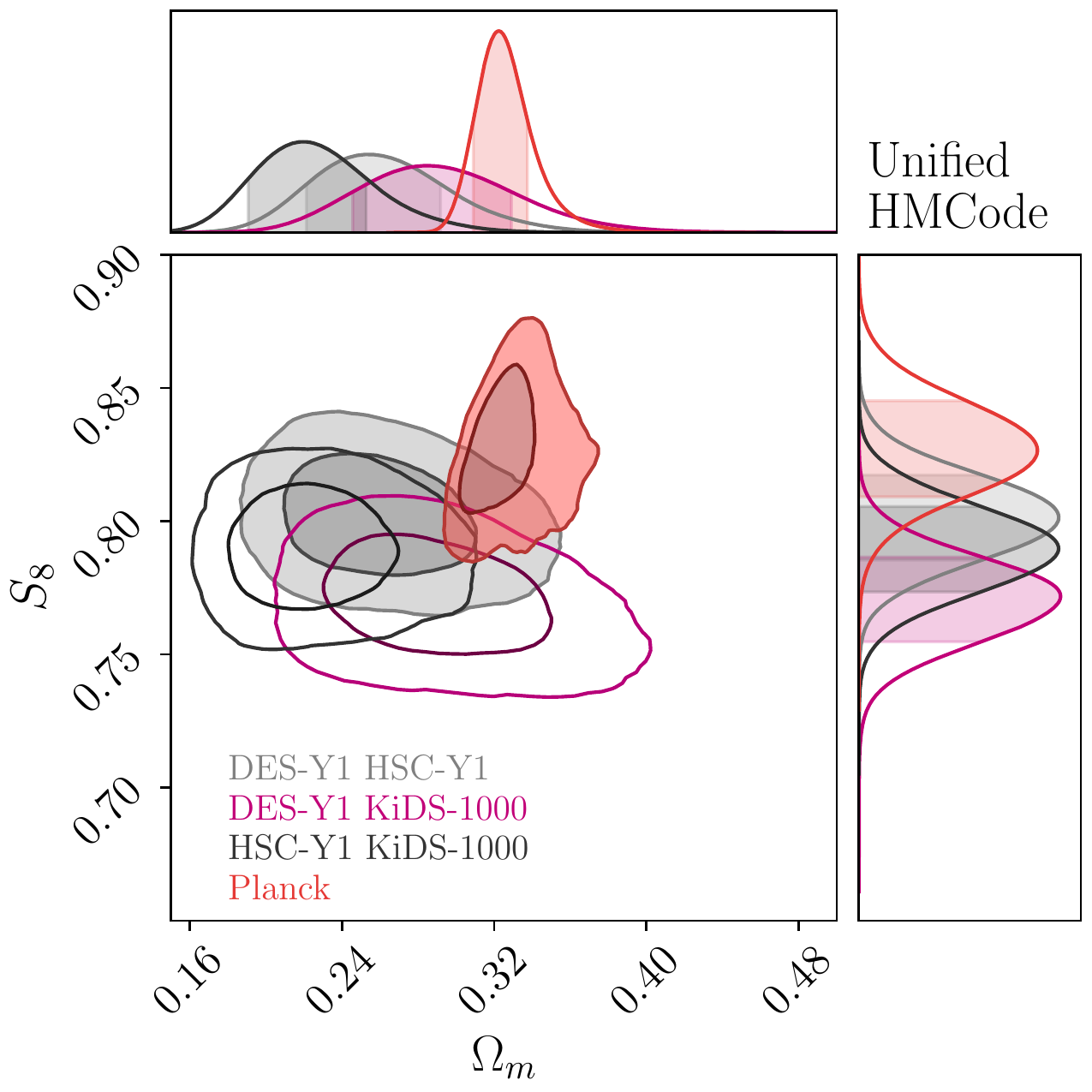}
	\caption{$S_8$-$\Omega_{\rm m}$ constraints from the combined constraints for each pair of surveys between DES-Y1, HSC-Y1 and KiDS-1000 assuming two different small-scale treatments.  The covariance for the joint constraint assumes independence between the surveys, as mentioned previously this is an approximation for the HSC-Y1 and KiDS-1000 combination.   These constraints are compared with the primary CMB probes of {\it Planck} \citep{Planck2018}.}
	\label{fig:pairwise_combined}
\end{figure*}

\section{$\Delta \chi^{2}$ Cut with HMCode} 
\label{appendix:chi2_hmcode}

\begin{figure*}
	\centering
	\includegraphics[width=0.95\columnwidth]{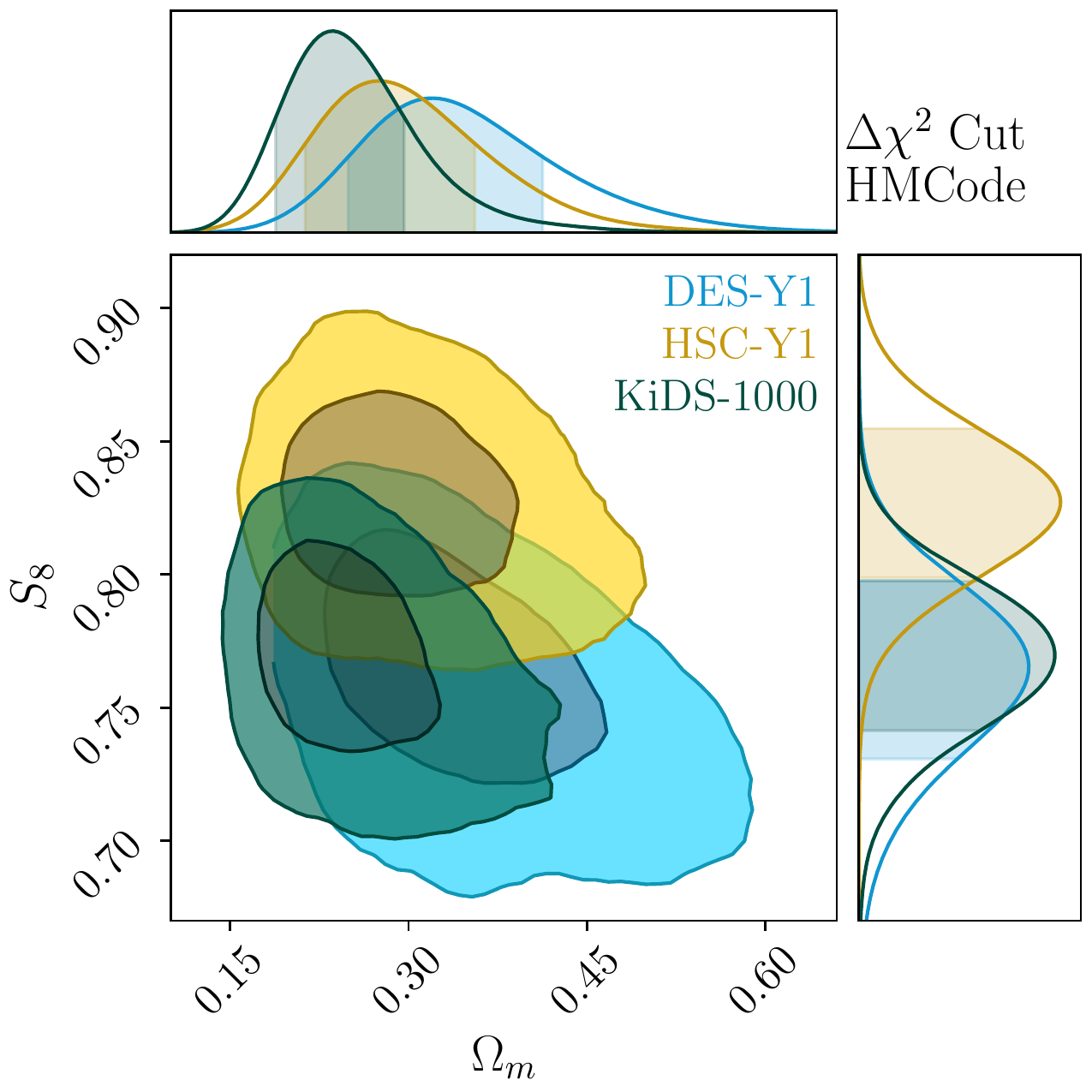}
	\caption{$S_8$-$\Omega_{\rm m}$ constraints for the unified constraints for each pair of surveys for DES-Y1, HSC-Y1 and KiDS-1000 adopting the $\Delta \chi^{2}$ cut and \textsc{HMCode}.}
	\label{fig:unified_chi2_hmcode}
\end{figure*}

In Section~\ref{sec:unify} we examine two different choices for small-scale modeling; a $\Delta \chi^{2}$ cut with \textsc{Halofit} and a small-scale cut with \textsc{HMCode}.  The 2D contour results of these choices for each survey are shown in Fig~\ref{fig:unified_chi2_hmcode}.  The $S_{8}$ constraints and goodness-of-fit results are summarized in Table~\ref{tab:constraints_chi2_hmcode}. In addition to these tests, we looked at the results of adopting \textsc{HMCode} with a $\Delta \chi^{2}$ cut.  In general, we did not find a significant change in the relative $S_{8}$, compared to the public analysis choices.  For DES-Y1, HSC-Y1 and KiDS-1000, there is $-0.43$, $0.06$ and $0.09$ $\sigma$ shift in the $S_{8}$ constraint.  The overall constraining power increases by roughly $10\%$ for DES-Y1 and $10\%$ for HSC-Y1 compared to the published analysis.  The KiDS-1000 constraint decreases by roughly $30\%$ compared to the published analysis.  

\begin{table*}
  \caption{Below we show, for the results for a unified analyses adopting the $\Delta \chi^{2}$ cut and \textsc{HMCode} across the three surveys, the $S_8$ constraints and the goodness-of-fit. We quote in terms of $\chi^{2}/$(d.o.f.-constrained parameters) and the resulting reduced $\chi^2$ (p-value).}
  \label{tab:constraints_chi2_hmcode}
  \centering
  \begin{tabular}{l c c c}
\hline 
Dataset & DES-Y1 & HSC-Y1 & KiDS-1000 \\
\hline
$\Delta\chi^2$ cut & & &  \\
$S_{8}$& $0.760^{+0.034}_{-0.026}$ & $0.827^{+0.025}_{-0.025}$  & $0.767^{+0.028}_{-0.024}$ \\
$\chi^{2}$/D.O.F. & 256.30/(258-11.07) & 253.10/(213-6.98) & 197.49/(164-6.23) \\
Reduced $\chi^{2}$ (p-value) & 1.04 (0.33) & 1.23 (0.014) & 1.25 (0.017)\\
 \hline
\end{tabular}
\end{table*}

\end{document}